\documentclass[a4paper,showkeys,floatfix,aps,prd,amsmath,amssymb,nofootinbib,preprintnumbers,superscriptaddress]{revtex4}

\usepackage{times,amsfonts}
\usepackage{natbib}
\usepackage[english]{babel}
\usepackage[T1]{fontenc}
\usepackage[latin2]{inputenc}
\usepackage{graphicx}
\usepackage{color}
\usepackage{tabularx}
\usepackage{mathrsfs}
\bibpunct{(}{)}{;}{a}{}{,}
\usepackage{aas_macros}

\usepackage{hyperref}
\usepackage{url} 

\makeatother

\renewcommand{\d}[1]{\ensuremath{\operatorname{d}\!{#1}}}

\newcommand{\lb}[2]{$(l^\circ,b^\circ)=(#1,#2)$}
\newcommand{\lmax}{\ell_{\rm{max}}}

\newcommand{\alms}{a_{\ell m}}
\newcommand{\onedegfield}{{\em SQDF}}
\newcommand{\PIsteradiansStr}{PISKY}
\newcommand{\PIsteradians}{{\em \PIsteradiansStr}}
\newcommand{\RTHdeepfield}{{\em RTHDF}}
\newcommand{\RTH}{{\em Hevelius} }

\newcommand{\Beginruledtabular}{\begin{ruledtabular}}
\newcommand{\Endruledtabular}{\end{ruledtabular}}

\begin{document}

\title{Simulations of cm-wavelength Sunyaev-Zel'dovich galaxy cluster and point source 
blind sky surveys 
and 
predictions for the RT32/OCRA-f and the Hevelius 100-m radio telescope.
}

\author{Bartosz Lew} \email[]{blew@astro.uni.torun.pl }
\affiliation{Toru\'n Centre for Astronomy, Nicolaus Copernicus University, ul. Gagarina 11, 87-100 Toru\'n, Poland}
\author{Mark Birkinshaw} 
\affiliation{HH Wills Physics Laboratory, University of Bristol, Tyndall Avenue, Bristol BS8 1TL, UK}
\author{Peter Wilkinson}
\affiliation{Jodrell Bank Centre for Astrophysics, The University of Manchester, Alan Turing Building, Manchester M13 9PL}
\author{Andrzej Kus} 
\affiliation{Toru\'n Centre for Astronomy, Nicolaus Copernicus University, ul. Gagarina 11, 87-100 Toru\'n, Poland}

\date{December 18, 2014}

\begin{abstract}
We investigate the effectiveness of blind surveys for radio sources
and galaxy cluster thermal Sunyaev-Zel'dovich effects (TSZEs) using
the four-pair, beam-switched OCRA-f radiometer on the 32-m radio
telescope in Poland. The predictions are based on mock maps that
include the cosmic microwave background, TSZEs from
hydrodynamical simulations of large scale structure formation, 
and unresolved radio sources. We validate the mock maps against
observational data, and examine the limitations imposed by simplified
physics. We estimate the effects of source clustering towards galaxy
clusters from NVSS source counts around \textit{Planck}-selected
cluster candidates, and include appropriate correlations in our mock
maps. The study allows us to quantify the effects of halo
line-of-sight alignments, source confusion, and telescope angular
resolution on the detections of TSZEs.

We perform a similar analysis for the planned 100-m Hevelius radio
telescope (RTH) equipped with a 49-beam radio camera and operating at
frequencies up to 22 GHz.

We find that RT32/OCRA-f will be suitable for small-field blind radio
source surveys, and will detect $33^{+17}_{-11}$ new radio sources
brighter than $0.87 \ \rm mJy$ at 30~GHz in a $1 \ \rm deg^2$ field at
$> 5\sigma$ CL during a one-year, non-continuous, observing campaign,
taking account of Polish weather conditions. It is unlikely that any
galaxy cluster will be detected at $3\sigma$ CL in such a survey.
A $60$-deg$^2$ survey, with field coverage of $2^2$ beams per pixel,
at~15 GHz with the RTH, would find $<1.5$ galaxy clusters per year
brighter than 60~$\mu$Jy (at $3\sigma$ CL), and would detect about
$3.4 \times 10^4$ point sources brighter than $1 \ \rm mJy$ at
$5\sigma$ CL, with confusion causing flux density errors
$\lesssim 2\%\, (20\%)$ in 68\% (95\%) of the detected sources.

A primary goal of the planned RTH will be a wide-area
($\pi$~sr) radio source survey at 15~GHz. This survey will
detect nearly $3 \times 10^5$ radio sources at $5\sigma$ CL down to
$1.3$ $\mathrm{mJy}$, and tens of galaxy clusters,
in one year of operation with typical weather conditions. Confusion
will affect the measured flux densities 
by $\lesssim 1.5\%\, (16\%)$ for 68\% (95\%) of the point
sources. We also gauge the impact of the RTH by investigating its
performance if equipped with the existing RT32 receivers, and the
performance of the RT32 equipped with the RTH radio camera.

\end{abstract}


\maketitle

\tableofcontents

\section{Introduction}
Dedicated wide-area and all-sky surveys over the next decade 
will boost our knowledge of large-scale structure formation, structure
evolution, and cosmology. Optical galaxy redshift surveys such as
LSST\footnote{\protect\url{http://www.lsst.org}}, 
EUCLID\footnote{\protect\url{http://sci.esa.int/euclid/45403-mission-status/}},
and 
4MOST\footnote{\protect\url{http://www.aip.de/en/research/research-area-ea/research-groups-and-projects/4most}}
will provide shapes and spectroscopic redshifts of millions of galaxies with
redshift $z\lesssim 2$. The 
eROSITA\footnote{\protect\url{http://www.mpe.mpg.de/eROSITA}} 
X-ray survey
will provide high angular resolution ($\lesssim 0.5'$)
maps and spectra of thousands of galaxy clusters. These optical and
X-ray data will trace the expansion history of the Universe and probe the
mass function of collapsed objects. This will constrain 
`dark energy' (DE) and/or improve our understanding of gravity on the largest scales.

Wide area cm-wavelength radio surveys will complement these
observations via measurements of the Sunyaev-Zel'dovich effects
\citep{Sunyaev1972} (hereafter SZE) from young galaxy clusters. SZE
data can combine with other data to constrain cluster masses and
abundances in redshift space.  Radio catalogues with sub-mJy
sensitivity will also provide a census of active galactic nuclei
(AGN), showing their distribution and activity in relation to forming
and evolving galaxies and clusters.  
Since inflation-based
cosmological models typically predict non-Gaussian curvature perturbations,
observations of the abundances of massive, high-redshift clusters provide a window into the
very early Universe via constraints on the shape and the amplitude 
of the primordial non-Gaussianity imprinted in matter distribution \citep{Dalal2008,Matarrese2008}. 

Wide area radio source surveys also have the potential to test DE by
measuring the expansion-driven decay of gravitational potential wells
-- the integrated Sachs-Wolfe effect \citep{Sachs1967} -- detected by
cross-correlation with the primary CMB at large scales
\citep{Afshordi2008}.

Galaxy clusters as observationally useful tracers of Mpc-scale mass distributions
have been used to constrain 
cosmological parameters 
\citep{White1993,Eke1996,Komatsu2002,Voit2005,Rapetti2010,Vanderlinde2010,Burenin2012,PlanckCollaboration2014b,Weinberg2013,Benson2013}, 
local non-gaussianity of primordial density perturbations 
\citep{Matarrese2000,Sadeh2007,Roncarelli2010} 
and departures from the standard cosmological model 
\citep{Schmidt2009,Schmidt2009a,Dunsby2010,Ferraro2011,Wyman2014,Giannantonio2014}. 
Clusters are also a promising source of information on the
dark energy equation of state parameter in the standard $\Lambda$CDM
model \citep{Sehgal2011,Kneissl2001,Alam2011},
and are crucial for attempts to solve the missing-baryon problem
\citep{Afshordi2007,PlanckCollaboration2014a,VanWaerbeke2014}.
Cool-core clusters provide laboratories for studying 
interplay between processes such as star
formation, AGN feedback, thermal emission, thermal conduction, plasma
magnetisation, photoionisation and metallicity-dependent molecular line cooling \citep{Voit2011,Voit2014}.
The most effective means of finding galaxy clusters to date has been
by X-ray detections of thermal emission from their hot intra-cluster
medium (ICM). However, complementary information on the 3D-temperature distribution of ICM is also
available through observations at radio frequencies\citep{Prokhorov2011}.  
With the advent of current and near-future blind radio surveys,
thousands of new galaxy clusters will be found via measurements of their
thermal Sunyaev-Zel'dovich effect (TSZE) at arcminute and
sub-arcminute scales  
\citep{Kneissl2001,DePetris2002,Vanderlinde2010,Sehgal2011,PlanckCollaboration2011,Muchovej2012,Mantz2014}.
On such scales, and at the frequencies of tens of GHz at which most
surveys are undertaken, unresolved radio sources are an
important source of contamination and confusion \citep{Vale2006}.
From another point of view, the data on sub-mJy radio sources that
will emerge from blind surveys is an important measure of 
the differential source count \citep{Muchovej2010}, another piece of
useful cosmological information.  
Recent important catalogues in this class have come from the
WMAP and \textit{Planck} satellites, which have surveyed Jy-level sources
over $3\pi$-sr, the 22-GHz Australia Telescope survey
\citep{Murphy2010}, which gives nearly $2\pi$-steradians coverage
with flux-density threshold of 40 mJy, the 15-GHz Ryle Telescope
survey \citep{Waldram2003}, which provides the differential source
count above $\sim 25$~mJy over $\sim 520$ $\mathrm{deg^2}$, and the
SZA survey to sub-mJy levels at 31-GHz in selected
fields with total area $7.7$ $\mathrm{deg^2}$ \citep{Muchovej2010}.

The OCRA-f radio array on the 32-metre telescope (RT32) in Poland is
another facility capable of cm-wave surveys. OCRA-f is the successor
to the OCRA-p single pair, dual beam, beamswitched radiometer
\citep{Browne2000}. The two OCRA receivers were constructed by the 
University of Manchester.
OCRA-f was designed to carry out targeted TSZE
observations \citep{Lancaster2011} and blind radio 
source surveys, and it is now about to start operation in small
northern sky regions. In the future the OCRA-f technology could be
developed into a wide-field, wide-band, multi-beam radio camera for
the Polish 100-m radio telescope \RTH (RTH) that is currently being
planned. 
The RTH project was signed into the Polish Roadmap for Research
Infrastructures in 2011. It has received strong scientific
support from the European VLBI Network Consortium, and funding 
from the European Regional Development Fund will be decided 
in the near future.
It is expected that the RTH will carry out blind,
large-area, radio surveys to mJy flux density levels at frequencies
around 15~GHz. 

Bearing in mind the importance of the TSZE for
cosmological studies, it is timely to investigate the possibility
that RTH and RT32/OCRA-f could find clusters in blind surveys, and
to examine the quality of the radio source counts that would result
from such surveys. This investigation is the main purpose of this paper.

Proper calibration and radio map reconstruction from survey data will
require excellent knowledge of the instrumental beams, the noise
properties of the radiometers, and the intervening foregrounds,
including the effects of atmospheric emission and absorption.
Simulations of atmospheric and receiver effects are in progress, but
in the present paper we focus on models of the centimetre-wavelength
radio sky. We have two main objectives: 
(i) to create reliable, calibrated, mock maps of the
astrophysical signals to be sought for in the planned surveys; and
(ii) to predict the number of objects that are likely to be found 
in these surveys.
This work can be used to inform specifications for future RT32
and RTH receivers. More specifically, for an assumed survey geometry
and duration we will estimate the number of radio sources that will be
detected, the likely flux density threshold, and the ability of the
systems to detect the TSZEs from galaxy clusters. These results,
combined with realistic simulations of atmospheric foregrounds and
receiver  performance (Lew 2015, in preparation), will provide tests
of the astrophysical signal reconstruction procedures 
that will be applied to the real data at the map-making stage.

The structure of this paper is as follows.
In section~\ref{sec:instruments} we outline the instrumental
properties. In section~\ref{sec:simulation} we describe simulations of
cosmic microwave background (CMB) intensity fluctuations,
the TSZEs, and the point source population.
In section~\ref{sec:comparizons} we test our hydrodynamic simulations
against available observational data and scaling relations
resulting from realistic high-resolution simulations of galaxy clusters.
The main results are gathered in section~\ref{sec:results}.
We discuss these results and conclude in sections~\ref{sec:discussion} 
and~\ref{sec:conclusions}.

\section{RT32 and RTH instruments}
\label{sec:instruments}

We derive predictions for the expected point source and galaxy cluster
detection rates principally for two instruments, the OCRA-f receiver
installed on the 32-m radio telescope in Poland (hereafter RT32), 
and the planned 49-beam receiver for the 100-m radio telescope \RTH
(RTH), which is planned for construction starting in 2017. We
highlight the impact of telescope and receiver size by also predicting
the survey outcomes with the receivers swapped between the telescopes.
For the 49-beam system we consider survey performance in three
sub-bands around 15 GHz. We also calculate predictions for the
existing 22-GHz RT32 receiver, since the planned 30-GHz RT32 survey
will be conducted with this receiver taking data in parallel with
OCRA-f. Again, we compare the effect of surveying with the
same 22-GHz receiver on the RTH. In this section we outline
the basic instrumental parameters of the three receivers. 

\subsection{The OCRA-f receiver}
OCRA-f is a 30-GHz, four-pair, beam-switched, secondary focus,
receiver array installed at the fully-steerable, 32-metre, Cassegrain
radio telescope in Toru{\'n} (Poland). The receiver pairs are
identical and run independently. In simple terms, the receiver
operates as follows. Signals from the two arms of each receiver pair
are mixed in a $90^\circ$, four-port Lange coupler and amplified at cryogenic
temperatures in first-stage low-noise amplifiers (LNAs). 
Next, they are phase shifted by $180^\circ$, mixed again, in a second,
identical, $90^\circ$ coupler, and amplified in second-stage
room-temperature amplifiers. The resulting signal 
is passed to a square-law detector, filtered, amplified in
video-amplifiers, and digitised in an analog-to-digital data
acquisition card installed in a PC-class computer. 

The technical design and laboratory tests of individual OCRA-f components 
have been described in \cite{Kettle2007} and also in \cite{Peel2010}.
By design OCRA-f mitigates fluctuations in atmospheric radio power by
differencing the total power signals detected in the two arms of the
receiver, which view slightly different directions on the sky through
corrugated feeds. The observable cm-wavelength signals from
astrophysical sources are contaminated by 
atmospheric emission (and reduced by variable atmospheric absorption),
mostly from water vapour and water droplets (in clouds). These signals
fluctuate in time and space due to the turbulent nature of atmospheric
flows. The single difference mitigates these fluctuations because
the near-field telescope beams largely overlap at the altitudes of
clouds. Internal gain instabilities in the receiver are mitigated by
differencing pairs of single difference signals phase shifted with
respect to each other by $180^\circ$. Phase switching (between
$0^\circ$ and $180^\circ$) is introduced at a rate which is adjustable
up to a few kHz. In its design and operation, each OCRA-f receiver
pair is similar to the receivers used by the WMAP satellite.

In preparation for the planned surveys we numerically simulate the
whole receiver chain. This allows us to generate realistic 
time-domain noise, and to test map reconstruction techniques that can
cope with non-uniform and incomplete sky coverage (Lew 2015, in
preparation). For the purpose of the present work, the details of
these simulations can be characterised by a few 
parameters describing the system noise and antenna sensitivity.
These are estimated based on in-lab measurements, or astronomical
calibrations, and are gathered in table~\ref{tab:RTspecs}.

Details of the data preparation and reduction pipelines that
will lie behind map-making and signal reconstruction from the raw data are
beyond the scope of the present work and will be discussed separately.
Here we note only that the raw data from each OCRA-f detector, as well as from
the K-band receiver, are digitised at an adjustable rate -- typically
about $10^5$ samples per second per data channel. The data are time-tagged
according to a 1-second pulse signal from a hydrogen maser. Next, they
are averaged within a switching state, and the double-difference (DD)
is calculated and referenced to sky coordinates by linear
interpolation from the RT32 pointing datastream. At the end of this
real-time process, the DD data are samples with time resolution 
compatible with the switching time, which is typically set to 3~ms, so
that fast-scanning strategies are supported. 
Occasional calibration pulses are injected into a single switching
state, so as to appear in the DD data, and absorber on-off sequences
are introduced periodically during the scan to track the stability of
the calibration diode. Prototype map-making and source-extraction
pipelines are currently being tested using simulations of the
astronomical sky and the sources of noise: the astrophysical part
of these simulations is described in the present paper.

Over the last few years RT32 control, OCRA-f data
acquisition, and the data processing pipelines have been 
developed and improved to provide good calibration and to support
sky-scanning strategies that can optimise between field size, 
scan completeness, and the elevation-dependence of atmospheric and
astronomical signals.

\subsection{A 100-metre Hevelius radio telescope}
The construction of a fully-steerable, $\sim 100$-m, radio
telescope is based on a number of science projects, but its operation
will be centred on an all-purpose radio camera, providing 49 broad-band
horns, with four 2-GHz bands per channel, and full polarisation, for a
total of 784~independent data channels. The polarisation and  
spectroscopic data from all horns will be recorded at
high time resolution. The large collecting area of the primary antenna 
and wide instantaneous frequency coverage ($6 - 22$~GHz) will
result in high sensitivity, and the multiple feeds will provide a
wide instantaneous field of view (FOV), allowing for 
effective mitigation of atmospheric foregrounds and fast sky mapping. 
It is expected that the analyses of 
OCRA-f simulations and data acquired from its early observations 
will provide software pipeline that will be readily adapted to
the data from the RTH.

The preferred location of the RTH is deep within a forested part 
of a natural reserve in northern Poland. This site is sufficiently
remote that the present-day radio frequency 
interference (RFI) environment is benign, and this 
should persist several decades into the future, even with 
increasing urbanisation and density of telecommunications signals.

It is anticipated that the RTH will be an instrument uniquely 
suitable for wide-area, blind, radio surveys in the
frequency ranges accessible 
to ground-based, sea-level radio astronomy. Its operations will be
constrained mostly by the typical weather conditions in eastern
Europe, which dictate the high-frequency limit for effective
operation, and define the fraction of the year for which sensitive
observing is possible. In addition to its stand-alone capabilities,
RTH will provide a dramatic enhancement to the European Very Long
Baseline Interferometry Network (EVN), and open new opportunities for
the international scientific community to carry out a wide range of
radio observations including:
(i) molecular studies of star-forming regions and circumstellar
envelopes;
(ii) searches for new molecules in the interstellar medium and 
solar system objects; 
(iii) measurements of the redshifted, 115-GHz emission line of
$^{12}\mathrm{C}^{16}\mathrm{O}(1-0)$ from sources with redshift of $z\gtrsim 4.2$;
(iv) blind surveys for discrete sources;
(iv) continuum surveys of Galactic emission;
(v) measurements of the Sunyaev-Zel'dovich effect in galaxy clusters;
(vi) observations of transient sources; and
(vii) studies of radio pulsars.

An important scientific goal for the planned telescope is a new
radio survey, at around 15~GHz, that 
will detect and characterise the spectra of a few $\times 10^5$~radio
sources in the northern hemisphere down to mJy flux densities (point
iv above). Combinations of different but overlapping field sizes will
result in tiered surveys of varying depths limited only by source
confusion, and, potentially, the detection
of new TSZEs. The most important parameters of the RTH and
the 49-beam receiver are collected in table~\ref{tab:RTspecs}.

\begin{table}
\caption{\label{tab:RTspecs}
Selected parameters of the 32-m and the planned 100-m
radio telescopes and their receivers.  We compare the existing K- and
Ka-band receivers installed on the 32-metre telescope in 
Toru{\'n} with the 49-receiver array of the RTH. We highlight
the effect of antenna size by including the performance that would
be expected if the existing receivers were to operate on the 100-m
telescope, or the radio camera was to be sited on the RT32.
}
\Beginruledtabular
\begin{tabular}{rcccccc}
& \multicolumn{3}{c}{32-m radio telescope (RT32)}& \multicolumn{3}{c}{100-m {\it Hevelius} radio telescope (RTH)}  \\
\hline
Location\footnotemark[1] [dms]&\multicolumn{3}{c}{$18^\circ\, 33'\, 50.6''\, \mathrm{E},\, 53^\circ\, 05'\, 43.7''\, \mathrm{N}$}&\multicolumn{3}{c}{$18^\circ\, 21'\, 42''\, \mathrm{E},\, 53^\circ\, 39'\, 32''\, \mathrm{N}$}\\
Aperture [m]		& 32 & 32 & 32 & 100 & 100 & 100 \\
Geometric area [$\mathrm{m}^2$]		& 804 & 804 & 804 & 7854 & 7854 & 7854\\
Surface rms error [mm]	&$\sim 0.6$ & $\sim 0.6$ & $\sim 0.6$ & 0.5 &0.5 &0.5 \\
\hline
Band name & Ku & K & Ka & Ku & K & Ka\\
Central frequency [GHz] &15 &22 & 30\footnotemark[2] &15  &22 & 30\footnotemark[2] \\
Technology & \multicolumn{6}{c}{HEMT/MMIC radiometers}\\
Bandwidth [GHz] & 6\footnotemark[4] & 4\footnotemark[3] & 8  & 6\footnotemark[4]  & 4\footnotemark[3] &  8 \\
Wavelength [cm] & 2.00	& 1.36 & 1.00	&2.00	&1.36	& 1.00 \\
HPBW ($\theta_b$) [$'$]\footnotemark[5] & 2.46	&1.68&	1.23&	0.79&	0.54&	0.39 \\
$\Omega_B$ [$10^{-7}$ sr]\footnotemark[6] & 5.81 &	2.70&	1.45&	0.59&	0.28&	0.15 \\
Surface efficiency\footnotemark[7]&$0.42^{+0.07}_{-0.06}$&	$0.35^{+0.06}_{-0.05}$&	$0.26^{+0.05}_{-0.04}$&	$0.44^{+0.07}_{-0.07}$&	$0.40^{+0.06}_{-0.06}$&	$0.33^{+0.05}_{-0.05}$ \\
Effective area [$\mathrm{m}^2$]	&337&	282&	212&	3479&	3102&	2592 \\
Sensitivity ($\Gamma$)\footnotemark[7] [K/Jy]	&	$0.12^{+0.02}_{-0.02}$&	$0.10^{+0.02}_{-0.02}$&	$0.08^{+0.01}_{-0.01}$&	$1.26^{+0.20}_{-0.20}$&	$1.12^{+0.18}_{-0.17}$&	$0.94^{+0.15}_{-0.15}$\\
$T_{\mathrm{rec}}$ [K]		&16	&16	&30	&16&	16&	30\\
$T_{\mathrm{atm}} (z=0^\circ,\mathrm{Jun})$\footnotemark[8]  [K]	&8      &31     &	17&	8&	31&	17\\
$T_{\mathrm{atm}} (z=0^\circ,\mathrm{Dec}) $$^[\footnotemark[8]^]$ [K]	&7      &	16&	13&	7&	16&	13\\
$T_{\mathrm{sys}}(z=0^\circ)$ [K]	&24     &	47&	47&	24&	47&	47\\
Air mass $z=(30^\circ, 60^\circ)$ 	&$(1.15,2.00)$ & $(1.15,2.00)$ & $(1.15,2.00)$ & $(1.15,2.00)$ & $(1.15,2.00)$ & $(1.15,2.00)$ \\
$T_{\mathrm{sys}}(z=45^\circ)$ [K]\footnotemark[11]&		$28^{+5}_{-4}$&	$60^{+18}_{-25}$&	$54^{+10}_{-10}$&	$28^{+5}_{-4}$&	$60^{+18}_{-25}$&	$54^{+10}_{-10}$\\
Number of receivers&		49&	1&	4&	49&	1&	4\\
RMS noise [$\mathrm{mK/s}^{1/2}$]\footnotemark[10]$^,$\footnotemark[11]&		$0.36^{+0.06}_{-0.05}$&	$0.94^{+0.27}_{-0.40}$&	$0.86^{+0.16}_{-0.16}$\footnotemark[9]&	$0.36^{+0.06}_{-0.05}$&	$0.94^{+0.29}_{-0.40}$&	$0.86^{+0.16}_{-0.16}$\footnotemark[9]\\

\end{tabular}
\Endruledtabular
\footnotetext[1]{Geodetic coordinates --- approximate
coordinates of the planned location for the RTH.}
\footnotetext[2]{Central frequency for the effective bandwidth
resulting from the OCRA-f LNA gain characteristics.}
\footnotetext[3]{Target Ka-band bandwidth to be achieved at RT32 in
Toru{\'n}. Currently only 500 MHz bandwidth is supported.} 
\footnotetext[4]{The effective bandwidth, centred at 15~GHz, is
assembled from three 2-GHz sub-bands.} 
\footnotetext[5]{Half power beamwidth. Cassegrain optical system with 12~db taper on the
edge of secondary mirror is assumed with a primary to secondary
mirror size ratio of 10.}  
\footnotetext[6]{The main beam solid angle ($\Omega_b$) estimated value based on the half power
beamwidth, which should be accurate to 5\% when modelling a
Gaussian beam profile \citep{Wilson2009}.}
\footnotetext[7]{68\% confidence range estimate (which includes
systematic and random errors resulting from the seasonal
efficiency variations, taking account of snow and ice during
winter) was obtained from C2-band sensitivity measurements multi-season campaign. 
The surface efficiency spectral dependence for RT32 is obtained by fitting 
Ruze's formula surface RMS errors parameter ($\epsilon$) 
to match the C1 and K-band surface efficiency measurements. With such
calibrated spectral dependence RTH surface efficiency for each band was calculated
by assuming $\epsilon=0.5 \mathrm{mm}$.}
\footnotetext[8]{The atmospheric brightness temperature (for zenith
distance $z$) is based on the average atmospheric conditions during
$2000 - 2010$ at the RT32 in June [December] using 
a radiative transfer code (developed at the Smithsonian
Astrophysical Observatory) for clear sky conditions (no droplets)  
for a model with a standard air oxygen, nitrogen and ozone mixture
and using measurements of vertical pressure, and  
temperature and relative humidity profiles extracted from (i)
weather balloon data, (ii) International Reference
Atmosphere data, and (iii) satellite measurements.  
The adapted atmospheric model pipeline was developed 
within the RadioNet-FP7 Joint Research Activity "APRICOT"
(All Purpose Radio Imaging Cameras On Telescopes)
and  
will be described in Lew (2015).}
\footnotetext[9]{OCRA-f is a double-difference radiometer and the RMS
noise estimate is increased by a factor $\sqrt{2}$ with respect to
a single feed radiometer.} 
\footnotetext[10]{The RMS noise uncertainties include the seasonal
variations of system temperature due to changes in atmospheric
brightness temperature. Similarly the telescope sensitivity
($\Gamma$) uncertainties cover the variations due to changing
seasons.} 
\footnotetext[11]{The uncertainties include variations due to air
mass varying within the considered range of elevations.} 
\end{table}

\section{Simulations}
\label{sec:simulation}
We perform simulations within the framework of the standard
$\Lambda$CDM cosmological  model with cosmological parameters
\begin{equation}
\{\Omega_{b0},\Omega_{m0},\Omega_{\Lambda0},n_s,\sigma_8, h\} = \{0.0463, 0.279, 0.721, 0.972, 0.821, 0.7 \}
\label{eq:cosmoParams}
\end{equation}
which are
compatible with the nine-year WMAP data
\citep{Hinshaw2013}. 
For a scale-free primordial power spectrum this setup implies the
amplitude of the scalar comoving curvature perturbation
$\Delta_{\mathcal{R}}^2 = 2.405\cdot 10^{-9}$. While the \textit{Planck}
(CMB+lensing) estimates on $\sigma_8$ 
are slightly higher \citep{PlanckCollaboration2014a}, this 
will not significantly alter our main results. We do not address the
reported tension between the calibrations of the matter power spectrum 
inferred using high-redshift (CMB) and low-redshift (cluster) data \citep{PlanckCollaboration2014},
and henceforth assume the CMB-calibrated cosmology (see
section~\ref{sec:discussion} for discussion of the anticipated
impact of a slightly modified value of the $\sigma_8$ parameter).

\subsection{Large scale structure simulations}
\label{sec:lss}
We used a publicly-available version of the smooth particle
hydrodynamics (SPH) code Gadget-2 \citep{Springel2005} to simulate 
large scale structure evolution in the past light cone. All
calculations were carried out on the 2048-CPU shared memory
supercomputer at the Pozna{\'n} Supercomputing and Networking Centre.
We created a number of large scale structure simulations that we stacked
together in comoving space in order to generate a deep field of view
that mimics 
the observational fields to be scanned in the planned surveys. For the 
primary results we generated 11 simulations, each of which
contains $N_{\rm{CDM}}=512^3$ dark matter and  
$N_{\rm{gas}}=512^3$ gas particles within a comoving
volume of $512^3 \mathrm{Mpc}^3$. 
The simulations were recorded at 12 evolutionary stages, covering
redshift range $z \approx (0.06,2.25)$. 
We used comoving gravitational softening
lengths of 15 kpc/$h$ for CDM and baryon particles, and a mesh grid of
size of $M=N^{1/3}$ was used for long-distance gravity force
computations. The initial conditions were generated 
with the N-GenIC program \citep{NgenIC} working within the Zel'dovich
approximation \citep{Zeldovich1970}, at an initial redshift 
$z_{\mathrm{ini}} = 50$. 
The gas temperature at the initial redshift was calculated using the
mean value between the coupled and adiabatic cases 
\citep{Shapiro1994}, $T_{\mathrm{ini}} \approx 73\, \mathrm{K}$.
The CDM and gas particles masses are $M_{\rm{CDM}}\approx2.2148 \cdot
10^{10} M_\odot h^{-1}$, and  $M_{\rm{gas}}\approx0.4407 \cdot 10^{10}
M_\odot h^{-1}$. The simulations together contain nearly 3 billion
particles and cover comoving volume 
$V_{\mathrm{tot}} \approx 1.476\, \mathrm{Gpc}^3$. 

In order to test the stability of our results and to understand
their sensitivity to variations of key parameters,
we created a number of smaller test simulations, altering the selected
parameters such as the initial redshift $z_{\mathrm{ini}}=\{50,100\}$,  
mass resolution (number of particles for a fixed comoving simulation
volume) $N^{1/3}=\{128, 256, 512\}$, gravitational softening length
$\{66,33,15,5\}$ kpc/$h$, number of neighbours used for smoothing
length calculations (multiple values $N_{\mathrm{neig}}$
between $5$ and $66$),
initial condition generator (N-GenIC and Grafic++), 
and cosmological parameter $\sigma_8$. 
The details of these tests are beyond the scope of the current paper,
but we found that our central choice is 
both stable and in good agreement with observational data. The
consistency checks that we made are described
in Section~\ref{sec:comparizons}. 

\begin{figure}[!t]
\centering
\includegraphics[width=\textwidth]{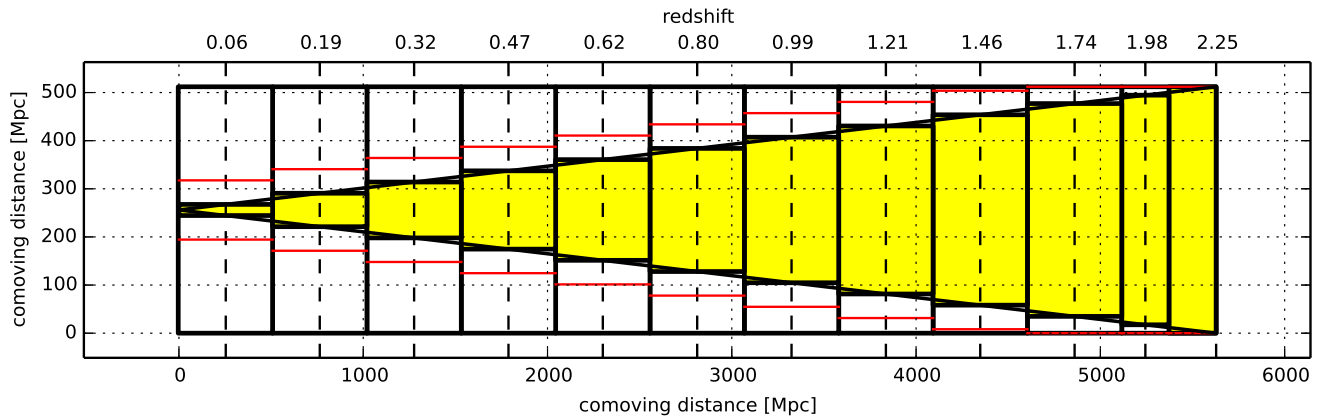}
\caption{Deep field geometry: the $z$-$y$ plane. The horizontal,
$z$, axis represents field depth
and the vertical axis represents the $y$ coordinate. The $x$
coordinate, which is orthogonal to $y$ and $z$, is suppressed. 
The heavy black boxes indicate adjacent simulation box cubes.
The shaded (yellow) rectangles represent slices used 
to approximate the past light cone for an observer
located at $(x,y,z)=(L/2,L/2,0)$
where $L = 512$~Mpc is the length of the side of the simulation
box. Vertical dashed lines indicate the redshifts at which
simulations were recorded. The horizontal red lines
represent the 5-Mpc margin (expanded
by a factor of 10 for visualisation) 
that define the cut-away domain used in SPH interpolations.
The opening angle of the field of view in the assumed flat geometry
is $\sim 5.2^\circ$.
}
\label{fig:deep_field}
\end{figure}

Every deep field realisation consists of 11 simulation boxes stacked 
as shown in figure~\ref{fig:deep_field}. Since the selected field of
view cuts only about a third of the total simulation volume, in
different FOV realisations we introduce random permutations of the
simulation box ordering, apply random coordinate switches between
$y$-$z$ and $x$-$z$, and apply random periodic particle shifts in
three dimensions within the simulation box. These operations make
better use of the total simulation volume and so allow better
assessment of cosmic and sample variance, although the resulting
samples are not entirely independent. 

As explained in the caption of figure~\ref{fig:deep_field}, we define
a ``slice'' as a parallelepiped cut-away region  
corresponding to the inside of the shaded rectangles
in figure~\ref{fig:deep_field} (marked in yellow).
The actual region used for the SPH interpolations (see
section~\ref{sec:cluster_search}) is larger in the $x$-$y$ directions 
as indicated by the red lines in the figure. We refer to this larger
region as a ``domain''. Some particles within domains
influence the density estimates inside the associated slice.

We use slice depths of 512~Mpc (the full size of the simulation box) 
in the $z$-direction. Our methodology allows us to use 
arbitrarily smaller values, which would result in a larger number of
domains, but would also require more simulation snapshots to
sample fully the evolution of the density field.
Our choice causes the most distant simulation box, at the
high-redshift extreme of the light cone, to contain
at least two slices.

\subsection{Galaxy clusters: search and properties}
\label{sec:cluster_search}
We run a Friends-of-Friends (FOF) algorithm with linking length
parameter $b=0.2$ to identify groups of particles 
either in the hypersurface of the present for the whole simulation volume or
as observed on the light cone, on a box-by-box.
The corresponding comoving linking length, $L_l = b \langle
l_{pp}\rangle$, depends on the mean inter-particle separation,
$\langle l_{pp}\rangle$, which we crudely approximate as 
$\langle l_{pp}\rangle = (V/N)^{1/3}$ where $N$ is the 
total number of particles of a given type and $V$ is the simulation
volume. The search for halos is done using both CDM and gas particles, 
and we require halos to contain at least $N_h = 600$ particles. This
number typically corresponds to halo mass 
$> M_{h,\rm min} \approx 1.3 \cdot 10^{13} M_\odot h^{-1}$.
$M_{h,\rm min}$  depends on gas mass fraction, and so slowly evolves
during cluster formation. 
This is because halos are defined in terms of the
FOF linking length and slightly different ratios of gas to DM
particles can appear in halos of the same $N_h$, depending on their
merging history.

At our mass resolution, galaxy-cluster-sized halos are resolved
with tens of thousands of particles. 
The most massive systems, with halo masses 
$M_{h} > 2 \cdot 10^{15} M_\odot h^{-1}$, are 
comprised of $N_h > 1.6 \cdot 10^5$ particles. 
We find tens of thousands of galaxy-group or galaxy-cluster size
halos in each deep-field simulation, out to the limiting redshift
that we used, $z_{\rm max}=2.25$. For each
identified halo we 
calculated a set of properties including location in the comoving
space, peculiar velocity, velocity dispersion, redshift
corresponding to comoving distance (which typically 
does not exactly correspond to the redshift resulting from the stage
of evolution), kinetic and potential energies, 
angular size, gas mass fraction, shape parameters (such as
eccentricity), the location of the gravitational potential minimum,
the mass centre, etc.
In order to test halo properties at certain overdensity thresholds 
we reconstruct the three-dimensional (3D) distributions of mass
density, gas temperature, and gas density-weighted temperature 
using SPH interpolations (see section~\ref{sec:SPHinterpolations}) on a
regular grid that resolves the halo and maintains a guard region of
adjustable size around it. For all halos we use a constant
grid resolution of 50~kpc (comoving), which we find sufficient at our
mass resolution. The innermost regions of galaxy
clusters (with overdensities above 2500) are not well
resolved, but in that regime our adiabatic simulations are also
deficient. In this paper we are concerned with regions of much lower
overdensity, $\delta=\{\mathrm{vir}, 200,500\}$ (see below).
Using the 3D distributions we reconstruct direction-averaged radial profiles 
of (i) overdensity (with respect to the critical density of the
Universe at the corresponding redshift); 
(ii) mass; (iii) temperature; and (iv) gas mass fraction. We use
these functions to derive the halo properties at the selected  
overdensity thresholds. 

The radial distance from the halo centre for a chosen overdensity
($r_\delta$) and the corresponding quantities calculated at that
distance are sensitive to the choice of centroid. 
We define the halo centroid as the location of the minimum of the
gravitational potential derived from the halo SPH particle
distribution. Profiles are calculated with respect to that
point because it is much more stable to the presence of sub-halos
than, for example, the mass centroid. We make an exception for
the integrated Compton Y-parameter $Y^{\mathrm{INT}}$
(Eq.~\ref{eq:Yint}) which we always calculate within $r_\delta$ of
the maximum of the line-of-sight
integrated, two-dimensional (2D),
map of the Compton $y$-parameter, since this matches the way
that observers measure $Y^{\rm INT}$.

Throughout this paper we denote ``virial'' quantities (such as
virial radius $r_{\mathrm{vir,c}}$) by subscript ``vir'' which
formally corresponds to overdensity $\delta_{\mathrm{vir}} \approx
100.2$ calculated for the assumed cosmological parameters at the
lowest considered redshift of ($z=0.06$) \citep{Eke1998} and with
respect to the critical density as usually indicated by subscript
``c''. 
Throughout this paper we work only with overdensities calculated with
respect to the critical density of the Universe at the measurement
redshift and henceforth we will skip the explicit
subscripting with ``c''.

The details of the halo finder used are irrelevant for our main
results, but our choice of the FOF algorithm allows us to generate a 
relatively high-resolution (compared to the size of the simulation)
grid with SZE relevant quantities (such as density weighted
temperature). The fact that FOF binds multiple halos into a single
system (be it gravitationally bounded or not) does not worsen the  
resolution since we use a fixed resolution for all halos. The
definition of what constitutes a halo (i.e., details of the
criteria for finding a halo) may have some impact on the scaling
relations constructed for selected overdensity thresholds, and our use
of a non-hierarchical FOF (with a single linking length) will also
impact the reconstructed mass functions
(section~\ref{sec:mass_function}), but investigating the impact of
these differences is beyond the scope of this paper.

\begin{figure}[!t]
\centering
\includegraphics[width=\textwidth]{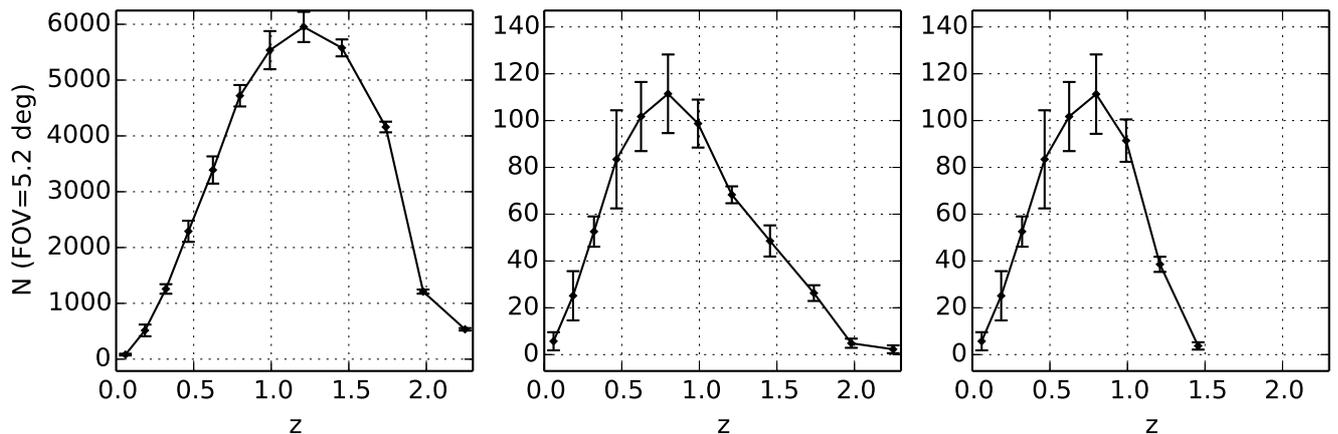}
\caption{Halo counts as a function of redshift in the simulated field
of view. 
{\it Left}: All FOF halos at $z < 2.25$
with $N_h\geq 600$.
{\it Middle}: All halos with FOF-derived mass $M_{\mathrm{FOF}} >10^{14}
M_\odot\, h^{-1}$. 
{\it Right}: Halos with $M_{\mathrm{FOF}}
>10^{14} M_\odot\, h^{-1}$ and $\theta_\mathrm{vir}>1'$.  
Error bars indicate sample variance as estimated from multiple
realisations of the deep field.
}
\label{fig:cluster_redshift_abundance}
\end{figure}

A combination of geometry and the history of structure
formation defines the distribution in redshift of the number 
of galaxy clusters per square degree. At low redshifts there will be
few clusters in the FOV because of the small comoving volume of the  
simulation box slices. At large redshifts, although the comoving
volume is large, massive clusters are not yet abundant.  
This effect is shown in figure~\ref{fig:cluster_redshift_abundance}.
Most cluster halos are found in redshifts $z\approx [0.5,1]$.
The strong suppression in the halo count at $z > 1.5$ (left plot in 
figure~\ref{fig:cluster_redshift_abundance}) results from the low-mass
halo cut-off due to the mass resolution of the simulation and 
the assumed minimal number of particles ($N_h\geq 600$) required to
identify an FOF halo. 

\subsection{SPH interpolations}
\label{sec:SPHinterpolations}
For each FOF identified halo, we calculate the 3D distribution of
quantities like mass, temperature, etc. on a grid using a generic SPH
interpolation algorithm \citep{Gingold1977, Lucy1977} 
referred to as the ``scatter'' scheme in \cite{Hernquist1989}. 
Thus the local density estimate at location $\mathbf{x}$ due to a set
of particles can be obtained from 
\begin{equation}
\rho(\mathbf{x}) = \sum_i m_i W(|\mathbf{x}-\mathbf{x_i}|, h_i)
\label{eq:SPHdensity}
\end{equation}
where $m_i$ is the mass of $i$'th SPH particle, situated at
$\mathbf{x}_i$. $W(|\mathbf{x}-\mathbf{x_i}|, h_i)$ is the 
smoothing kernel which we take to be the same as in the
Gadget-2 code 
(eq.4 of \cite{Springel2005}) with $W(|\mathbf{x}-\mathbf{x_i}|\geq h_i,
h_i) = 0$. $h_i$ is the local smoothing length of the $i$'th particle,
and is calculated  
according to the criterion that each particle should have a constant
number of neighbours. For interpolations within the same 
particle species, this is the same as using the criterion 
requiring a constant mass within the sphere of radius $h_i$ since
all particles of a single species have the same mass. The latter criterion
is used for all massive particles in density calculations 
in the SPH code Gadget-2.
For our main results we use a
fixed number of neighbours, $N_{\mathrm{neigh}}=33$.
See the sections~\ref{sec:discussion} and~\ref{sec:MY} for a 
discussion of this choice for $N_{\rm neigh}$.
Given a density estimate, the mass overdensity 
$\delta(\mathbf{x}) = \frac{\rho(\mathbf{x})}{\rho_c(z)}$, 
where $\rho_c(z) = \frac{3 H^2(z)}{8 \pi G}$ is the critical density
of the Universe at redshift $z$.
The local value of any other quantity, $f(\mathbf{x})$, is given by 
\begin{equation}
f(\mathbf{x}) = \sum_i \frac{f_i}{\rho(\mathbf{x_i})} m_i 
W(|\mathbf{x}-\mathbf{x_i}|, h_i),
\label{eq:SPHinterpolation}
\end{equation}
where the summation spans all particles, but only those with
$|\mathbf{x}-\mathbf{x_i}|<h_i$ contribute. 

We calculate scaling relations by reconstructing the 3D mass
density distributions
$\rho(\mathbf{r})$ of both dark matter and gas. The 3D
distributions 
of temperature ($T(\mathbf{r})$), and density-weighted temperature
($T_\rho(\mathbf{r})$), (see eq.~\ref{eq:TSZ3}) 
are reconstructed using only gas particles. From these we calculate
the radial profiles of density $\rho(r)$, 
temperature $T(r)$, density-weighted temperature $T_\rho(r)$, and
baryon gas-mass fraction $f_{\mathrm{gas}}(r)$, and hence the scaling
relations. 

We process the TSZE signal on a halo-by-halo basis for reasons of
efficiency. High spatial resolution is required
to resolve intra-cluster structure, but large cosmological volumes are
required to develop sample statistics. The combination
is not possible on a single grid. In consequence we create a 
high-density grid only around the locations that will generate a
significant TSZE signal. Cool particles and low-density regions will
not contribute significantly and they are neglected in the analysis. 
The simple creation of TSZE maps could be done by processing each SPH
particle independently, as is often done in simulations. However, our
present approach allows us to measuring scaling relations and
assessing the effects of TSZE flux boosting due to line-of-sight (LOS)
overlap of halos.

\subsection{Thermal Sunyaev-Zel'dovich effect}
The thermal Sunyaev-Zel'dovich effect (TSZE) in galaxy clusters alters
the CMB specific intensity\footnote{Through the rest of this paper we
refer to the spectral radiance -- according to the more
traditional nomenclature -- as specific intensity. }  
towards a cluster as compared to a reference
direction. This effectively is seen as a frequency dependent
black-body thermodynamic temperature variation of amplitude  
\begin{equation}
\delta T\equiv\Delta T/T_{\mathrm{CMB}}= f_\nu y,
\label{eq:TSZ1}
\end{equation}
where 
$y$ is the Comptonisation parameter and $f_\nu$ encodes the spectrum
of the effect. In the non-relativistic limit, and with the standard
redistribution 
function, $f_\nu=x \frac{e^x+1}{e^x-1}-4$,
where $x=\frac{h \nu}{k_B T_{\mathrm{CMB}}}$, and $h$ and $k_B$ are the \textit{Planck} and the Boltzmann constants respectively.
The Compton $y$-parameter is proportional to the LOS integral of the
product of the electron number density, $n_e$, and electron temperature,
$T_e$, 
\begin{equation}
y = \sigma_0 \int n_e \frac{k_B T_e}{m_e c^2} \d l,
\label{eq:TSZ2}
\end{equation}
where the integral is in physical distance.
Following the notation of \cite{Refregier2000} this formula can also be expressed as the integral over comoving 
distance via $\d l=a d\chi = d\chi/(1+z)$, and by introducing
the number of electrons per proton $\mu_e^{-1} = n_e /(\rho/m_p)$, where $\rho$ is the gas mass density, and $m_p$ is the 
mass of the proton, the integral can be converted to
\begin{equation}
y =  y_0 \int  T_\rho (1+z)^2 \d\chi,
\label{eq:TSZ3}
\end{equation}
where $T_\rho \equiv T \frac{\rho}{\bar\rho}$ is the density weighted temperature and 
$\bar\rho = \rho_b = \rho_{b0} (1+z)^3 = \rho_{c0} \Omega_{b0} (1+z)^3$ 
is the average baryon mass density at redshift $z$.
The constant 
\begin{equation}
y_0 = \frac{\sigma_T \rho_{c0}\Omega_{b0}k_B}{\mu_e m_p m_e c^2} 
\approx 7.76 \cdot 10^{-17}\, [\mathrm{K}^{-1}\,\mathrm{Mpc}^{-1}]
\label{eq:y0}
\end{equation}
for the assumed cosmology, where $\sigma_T$ is the Thomson
scattering cross-section, $m_e$ and $m_p$ are the electron and proton
masses respectively, and $c$ is the vacuum speed of light.
We assume a standard chemical composition, compatible with Big-Bang nucleosynthesis 
(helium mass fraction $Y_p=0.24$) in which case $\mu_e=1.136$.
We calculate the SPH gas particle temperatures as
\begin{equation}
T_e(u) = \frac{m_p}{k_B} \mu(Y_p) (\gamma-1) u,
\label{eq:Te}
\end{equation}
where $u$ is the internal energy per unit mass associated with a given SPH particle and 
$\gamma=5/3$ is the ratio of specific heats for a monatomic gas. The mean molecular weight 
$\mu = \frac{4}{8 X_p +3 Y_p} \approx 0.588$ for a fully-ionised
hydrogen/helium mixture, which is the case for the intra-cluster medium.
$X_p = 1- Y_p$ is the mass fraction of hydrogen, and we neglect the
minor contribution from heavier elements.
Using eq.~\ref{eq:TSZ1} and first-order expansion of the black body specific intensity
\begin{equation}
B_{\nu,\mathrm{CMB}}(x)=\frac{2\nu^2}{c^2}\frac{h\nu}{e^x-1},
\label{eq:BBraiance}
\end{equation}
about the black body temperature, it is easy to show that 
the TSZE-induced radiance variation $\Delta I= I_\nu(T)-I_\nu(T_{\mathrm{CMB}})$ for a cluster is
\begin{equation}
\Delta I = \frac{2 (k_B T_{\mathrm{CMB}})^3}{(h c)^2} \frac{x^4 e^x}{(e^x-1)^2} f(x) y = I_0 g(x) f(x) y
\label{eq:TSZ4}
\end{equation}
where $g(x) = x^4 e^x/(e^x-1)^2$ and $I_0=\frac{2 (k_B T_{\mathrm{CMB}})^3}{(h c)^2}$.
Then the TSZE flux density per beam can be calculated as 
\begin{equation}
\Delta S_{\mathrm{TSZ}}(\nu,P_b) = \int \Delta I_\nu(\mathbf{\hat{n}}) P_b(\mathbf{\hat{n}}) \d\Omega 
\label{eq:TSZ5}
\end{equation}
where $P_b(\mathbf{\hat{n}})$ is the instrumental beam profile.

Following \cite{Nagai2007}, we also define 
the solid-angle integrated Compton $y$-parameter as: 
\begin{equation}
Y^{\mathrm{INT}} = d_A^2(z) \int y \d\Omega
\label{eq:Yint}
\end{equation}
to test the compatibility of our simulational procedures with similar simulations via the $M$-$Y$ scaling relations.

\begin{figure}[!t]
\centering
\includegraphics[width=\textwidth]{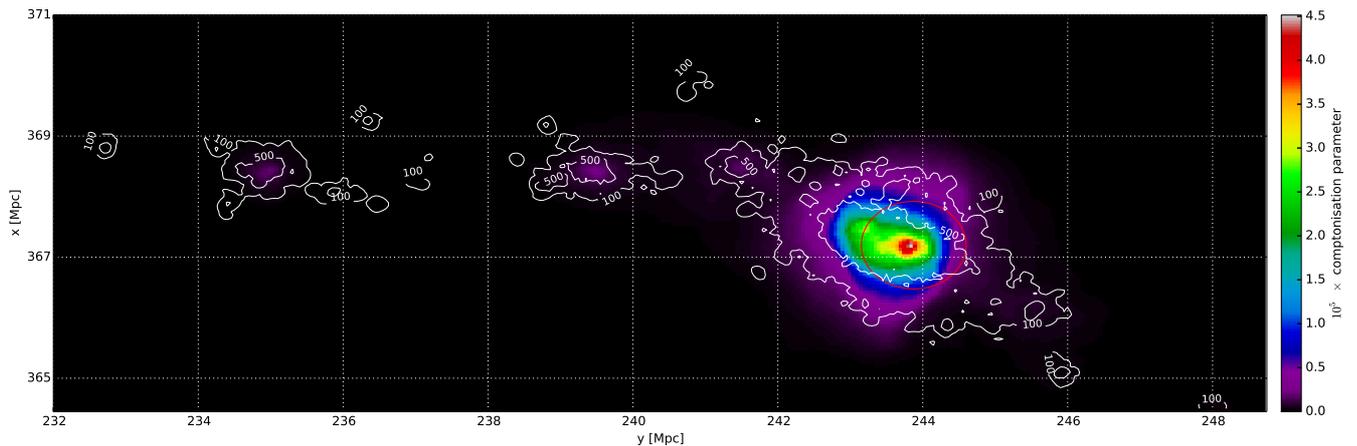}
\caption{One of the most massive simulated galaxy cluster systems
found by FOF at $z \approx 0.624$. The colour scale indicates the
value of the Compton $y$-parameter, while the contours indicate the
iso-overdensity regions for $\delta=\{100,500\}$ in the projected,
line-of-sight--maximised mass overdensity distribution.  
The red circle marks a projected 
sphere of radius $r_{500}$ centred at the minimum of the
gravitational potential of the  
whole system. The total halo mass is $M_{\mathrm{FOF,tot}} \approx
7.4\times 10^{14} \mathrm{M}_\odot h^{-1}$, and this halo is
resolved with over 
$5.4 \times 10^4$ dark matter and gas particles.
}
\label{fig:halo34582}
\end{figure}
In figure~\ref{fig:halo34582} we plot an example of a system of
halos. This example indicates how the FOF algorithm connects different
halos into a single system if they form a filamentary structure. It is
likely that other methods (such as those based on spherical
overdensity) or a hierarchical approach would identify this system  
as a set of nested individual systems, however it is not obvious that
processing the individual halos separately for calculation of
(e.g.) density profiles 
would provide a better $r_\delta$ estimate because this could represent
a gravitationally bounded system during a merger event, when the
(over)density profiles should include the substructures. The presence
of sub-halos also undermines the usefulness of radial profiles, at
least at low overdensities, by the implicit assumption of spherical
symmetry.

\begin{figure}[!t]
\centering
\includegraphics[width=.75\textwidth]{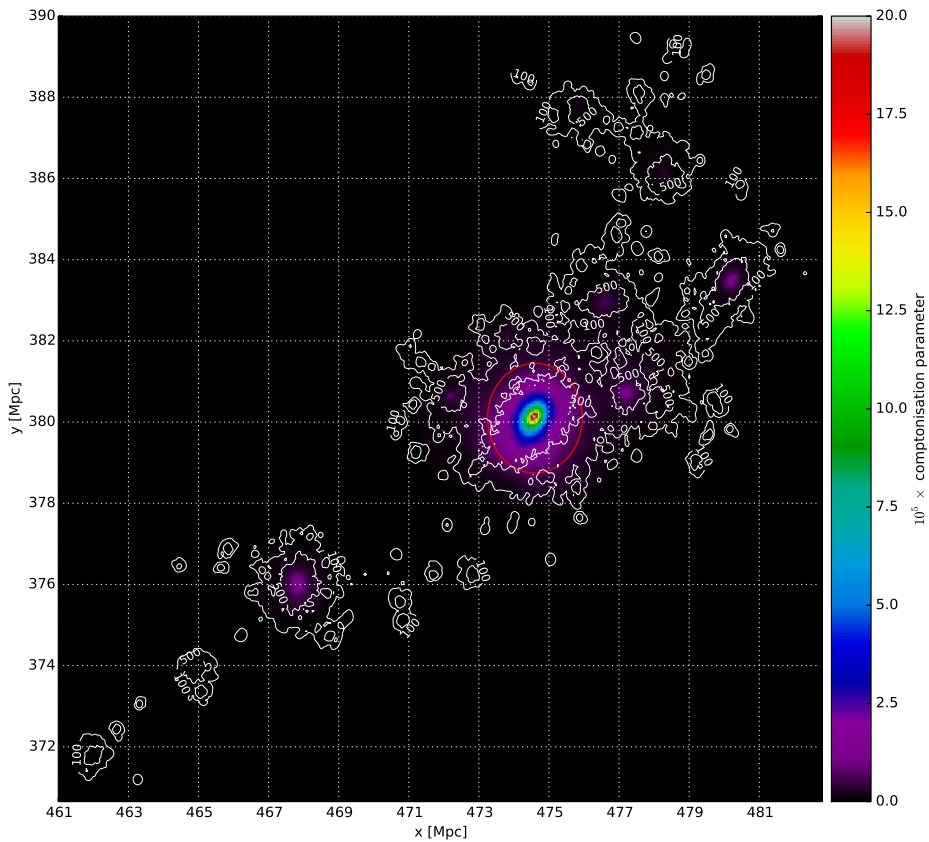}
\caption{Same as figure~\ref{fig:halo34582}, but for a halo of mass 
$M_\mathrm{FOF,tot} \approx 2.5\times 10^{15} \mathrm{M}_\odot
h^{-1}$ which is resolved with more than 
$1.8\times 10^5$ dark matter and gas particles, and is evolved up to
$z\approx 0.06$. 
As indicated by the colour scale, the TSZE signal is more than four times
larger than in figure~\ref{fig:halo34582}. 
However the chance of finding halos such as this in 
small-field surveys is low, due to the small survey volume and the
low number density of halos of such high mass
(figure~\ref{fig:massFn}).  
}
\label{fig:halo10801}
\end{figure}

The reliability of temperature and density profile reconstructions
from the point of the minimum of gravitational potential is
illustrated by Figures~\ref{fig:halo34582} and~\ref{fig:halo10801}.
In these figures
it is clearly seen (i) that the Compton $y$-parameter traces the
cluster mass distribution, and (ii) that the location of the maximum
$y$ (relative to which the flux density is calculated)  
coincides well with the minimum of the gravitational potential (the
centre of the red circle). 
The radius of the red circle, $r_{500}$, is calculated within the
spherical region enclosing the mass that yields overdensity
$\delta=500$. It is evident that this $r_{500}$ roughly coincides
with the contour tracing the same overdensity value.

In dynamic situations, such as those shown in
figures~\ref{fig:halo34582} and~\ref{fig:halo10801}, 
the kinematic SZ effect (kSZE) increases towards regions of the
infalling gas lying close to the line of sight. While the kSZE is much
smaller than the TSZE, it can dominate in regions of cold and
fast-moving gas. Because the kSZE tends to be averaged out by
superpositions along the line of sight, since the sign of the kSZE
reverses with the sign of gas velocity, in the current paper we
neglect kSZE contributions to the change of CMB intensity.

\subsection{CMB simulations}
We generate the CMB background fluctuation field starting from a
random realisation of a lensed CMB power spectrum, generated with CAMB
\citep{Lewis2000}. A random Gaussian realisation of
the $\alms$ coefficients in spherical harmonic space up to $\lmax =
4096$ was then made, and a spherical harmonic transformation then
converted this to the sky map of CMB temperature fluctuations. All
cosmological parameters are as in section~\ref{sec:simulation}.
For the selected field we project CMB fluctuations from the Healpix
grid \citep{Gorski2005} to a tangent plane, and interpolate 
(using the SPH interpolation algorithm) onto the regular Cartesian
grid coincides our map. The Healpix map used for interpolation is 
significantly larger than the final map to allow for interpolation to
the edge of the required area. Using this 
approach we can generate maps of arbitrary resolution without 
producing spurious signals, since
the CMB fluctuations are sufficiently smooth and well resolved by the
initial Healpix grid, which has resolution parameter $n_s=2048$.
The instrumental beam is introduced (when needed) by convolving with
the beam transfer function over the full sky in spherical harmonic
space, rather than by FFT (Fast Fourier Transform) performed in the
selected field, so that there are no induced aliasing effects.

In the flat field limit a set of $(\Delta_\lambda , \Delta_b)$ offsets
in coordinate directions $(\lambda,b)$ approximate a spherical
square. 
While this approximation is useful for plotting two-dimensional maps of small fields, for large FOVs
it necessary to work on a grid defined over the full
celestial sphere (such as Healpix).
We choose the $b$ coordinate to coincide with galactic latitude, but
$\lambda$ is taken to be a great-circle coordinate 
orthogonal to $b$, so its orientation depends on direction.

We prefix tangent plane coordinates by ``$\Delta$'' in order to
indicate that these are the offsets with respect to the chosen field
centre. 

\subsection{Field of view projections}

The TSZE due to a given halo is calculated using its properties at the
epoch closest, in terms of the redshift, to the redshift inferred from
the location of the halo in our light-cone construction (also referred
to as a ``deep field''). 

The comoving coordinates of the halo in the deep field $(x_h,y_h,z_h)$ 
are related to the coordinates in the 2D FOV $(\Delta_\lambda$, $\Delta_b)$ by
\begin{equation}
\begin{array}{ccc}
\Delta_\lambda(x_h,z) &=& \mathrm{atan}\bigl[\frac{(x_h-L/2)/(1+z)}{d_A(z)}\bigr]\\\\
\Delta_b(y_h,z) &=& \mathrm{atan}\bigl[\frac{(y_h-L/2)/(1+z)}{d_A(z)}\bigr],
\end{array}
\label{eq:xyz2FOV}
\end{equation}
where the observer is located at $(x,y,z)=(L/2,L/2,0)$\footnote{In the
$(x,y,z)$-tuple, $z$ indicates comoving coordinate along the central projection line of sight,
and not redshift.}, 
$L$ is the comoving size of the simulation box, 
and $d_A(z)$ is the angular diameter distance at redshift $z$, 
corresponding to the comoving distance at redshift $z_h$. 
For TSZE signal map-making we consider every halo if it at least
partially overlaps with the FOV. Partial overlaps are common for halos
with low $z_h$ and large angular sizes.

\subsection{Point source simulations}

\subsubsection{Flux density}
\label{sec:flux_density}
We simulate contributions to the integrated flux density 
due to unresolved extra-galactic non-thermal sources based on source
counts from AT20GB, a 20-GHz southern equatorial hemisphere 
blind survey \citep{Murphy2010}, and the SZA (Sunyaev-Zel'dovich Array) 31-GHz blind survey
of about 4.3 $\rm{deg}^2$ \citep{Muchovej2010}.
AT20GB is complete at 78\% above flux density 50 mJy. The
SZA survey is complete at 98\% above flux density 1.4 mJy.
Although we do not directly use the 15-GHz 9C
results\citep{Waldram2003}, we find that counts based on the 9C survey
are consistent with those from the other two surveys within the
uncertainties, when corrected for the frequency and completeness.

In the low flux-density limit we use the SZA survey, which probes the
flux density range (0.7,15)~mJy, and extrapolate from 15~mJy to 50~mJy, based on 
the best fit power-law model of the form
$\d{N}/\d{S} = A (S/S_0)^{-\gamma}$, with parameters as given in table~\ref{tab:alphaPDF}.
We calculate the spectral index distribution required for our
extrapolation only for SZA sources fainter than 30 mJy.

In the flux density range above 50 mJy we use the AT20G survey
which covers a wide area and hence is well-suited to
probe the strong radio source population.
Our comparison of the two surveys, at a single frequency based on our
spectral model, indicates that the SZA survey predicts slightly more
faint sources than AT20G.

We simulate the $\d N/\d S$ relation for the three target frequencies, 15, 22, and 30~GHz,
using Monte-Carlo realisations of a skew-Gaussian--fitted probability distribution function
(PDF) (eq.~\ref{eq:skewGauss}) of spectral indexes $\alpha_{8-20}$ for
the AT20GB survey and $\alpha_{5-31}$ for the SZA survey respectively, 
where $\alpha_{\mathrm{f_1-f_2}}$ is the spectral index of a source
between frequencies $f_1$ and $f_2$. Our convention for the sign of
spectral index is that 
\begin{equation}
S(\nu) = S_0 \Bigl(\frac{\nu}{\nu_0}\Bigr)^{-\alpha},
\label{eq:Snu}
\end{equation}
and the flux density measurements at the reference frequencies 
were provided with the catalogue data.\footnote{\protect\url{http://heasarc.gsfc.nasa.gov/W3Browse/radio-catalog/at20g.html}}$^,$\footnote{\protect\url{http://heasarc.gsfc.nasa.gov/W3Browse/radio-catalog/sza31ghz.html}}
The fitting function for the spectral index distributions $\d
N/\d\alpha$ take the form
\begin{equation}
\phi(\alpha,A_\alpha,m_\alpha,\sigma_\alpha,S_\alpha) = 
A_\alpha \exp\Bigl(-\frac{(\alpha-m_\alpha)^2}{2 \sigma_\alpha^2}\Bigr)  \Bigl(1.0+\mathrm{erf}\bigl(S_\alpha\frac{\alpha-m_\alpha}{\sigma_\alpha \sqrt{2}}\bigr)\Bigr)
\label{eq:skewGauss}
\end{equation}
where $\mathrm{erf}$ is the error function.

\begin{figure}[!t]
\centering
\includegraphics[width=0.79\textwidth]{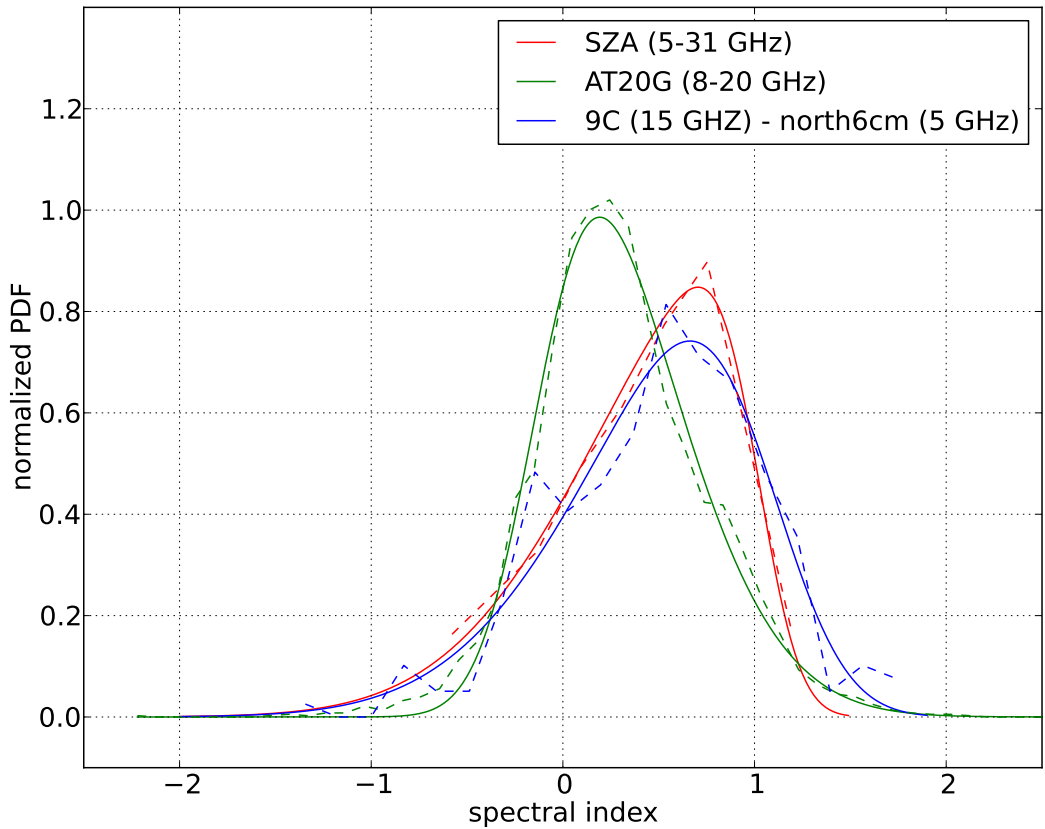}
\caption{Spectral index distributions based on
the SZA 31-GHz survey (dashed red), the AT20G survey (dashed green), and
the 9C survey data cross-correlated with the North6cm survey (dashed blue), with
corresponding skew-Gaussian distributions (solid lines) as in
equation~\ref{eq:skewGauss}. The fitted parameters are given in table~\ref{tab:alphaPDF}.
}
\label{fig:alphaPDF}
\end{figure}

We use the Levenberg-Marquardt (LM) algorithm \citep{Levenberg1944,Marquardt1963}
to find the best-fit parameter values. In order to avoid getting trapped in the local minima of the likelihood function,
we use a Monte-Carlo approach to generate initial parameter guesses
from within a parameter space that extends beyond the range of
plausible parameter values. We impose a flat prior on each of the
initial parameters values in every LM run. We fit
spectral index distributions as shown in Figure~\ref{fig:alphaPDF}, with
parameters $(A_\alpha,m_\alpha,\sigma_\alpha,S_\alpha)$ as given in 
table~\ref{tab:alphaPDF}.

\begin{table}
\caption{\label{tab:alphaPDF} 
Best-fit parameters for the skew-Gaussian distributions
(equation~\ref{eq:skewGauss}) of spectral indexes for the  
9C, AT20GB, and SZA radio catalogues. The last four columns give the
amplitude and the tilt of the power-law $\d N/\d S$ relation (for
pivot flux density $S_0=1\, \mathrm{mJy}$), the assumed catalogue
completeness, and frequency respectively. 
}
\Beginruledtabular
\begin{tabular}{rrrrrrrrr}

Catalogue name & $A_\alpha$ & $m_\alpha$ & $\sigma_\alpha$ & $S_\alpha$ & $A$ $[\mathrm{Jy}^{-1}\mathrm{sr^{-1}}]$ & $\gamma$ & $f_c$ & $\nu$ $[\mathrm{GHz}]$\\
9C& $0.471 \pm 0.058$ & $1.075 \pm 0.073 $ & $0.816 \pm 0.109 $ & $-2.49 \pm 1.04 $ & 51.0 & 2.15 & 1.00 & 15.2\\
AT20GB & $0.641 \pm 0.055$ & $-0.120\pm 0.040 $ & $0.603 \pm 0.055 $  & $2.26 \pm 0.42$ & 31.0& 2.15& 0.78 & 20.0\\
SZA & $0.475 \pm 0.070$ & $1.018 \pm 0.063$ & $0.809 \pm 0.138$ & $-4.61 \pm 2.83$ & 30.4& 2.18 & 0.98 & 31.0
\end{tabular}
\Endruledtabular

\end{table}

The fit to the AT20G survey slightly underestimates the number of
sources with inverted spectra. The 9C survey suggests the presence of
a population of inverted-spectrum sources at a level not present in
the other two distributions. Since statistical significance of this
feature is poor, we prefer to use the simple skew-Gaussian models,
rather than the superposition of two spectral populations of sources,
in constructing our model skies.

The differences in the estimated spectral index distributions and 
in differential source counts resulting from different surveys make the predictions
at the target frequencies inaccurate at the level of several percent,
with the degree of mismatch being a function of 
the flux density range of interest, the survey sensitivity, and the
estimated coverage completeness.

For the assumed field size of $5.2\,\rm{deg}^2$  we 
generate 100 Monte-Carlo realisations of point source flux density
distributions at the original catalogue frequencies.\footnote{For
practical reasons we actually generate sources for a much larger
field (400 deg$^2$) and rescale the resulting counts to the field
size of interest.} 
The flux density PDFs are probed within the range from 1 $\mu$Jy to 
1 Jy. We then generate random realisations of the 
spectral indexes according to the spectral index PDF
(eq.~\ref{eq:skewGauss}) generated for a given catalogue and create a
Monte-Carlo realisation of the flux densities at the target frequency
of interest using eq.~\ref{eq:Snu}. We then combine the
realisations from the AT20GB and SZA catalogues by
removing sources below 50 mJy for the AT20GB simulation and above 50 mJy
for the SZA simulation. 

The result is plotted in Figure~\ref{fig:ptSrcCounts} as a cumulative
distribution, is discussed in Section~\ref{sec:results}.

In order to avoid extrapolations below the measurement-probed flux
densities ($S$) for simulations of point source flux density embedded
in our mock maps, we only use the sources with $S>100\,\mu \mathrm{Jy}$.

While we treat all unresolved sources as sources of a constant 
pixel-size angular extent, a more realistic simulation
could account for the fraction of the radio sources that will be
resolved by the telescope beam. However, the vast majority of sources
of interest for our purposes will be unresolved, and we 
ignore this possibility for the present.

\subsubsection{Spatial correlations}
\label{sec:spatial_correlations}
Radio sources tend to cluster towards galaxy clusters
(e.g. \cite{Coble2007}). We assess this for the specific case of
radio sources and clusters with strong TSZ effects by
cross-correlating the NVSS catalogue \citep{Condon1998}
of radio sources with the early \textit{Planck}-SZ cluster candidates sample \citep{PlanckCollaboration2011}.

\begin{figure}[!t]
\centering
\includegraphics[width=0.45\textwidth]{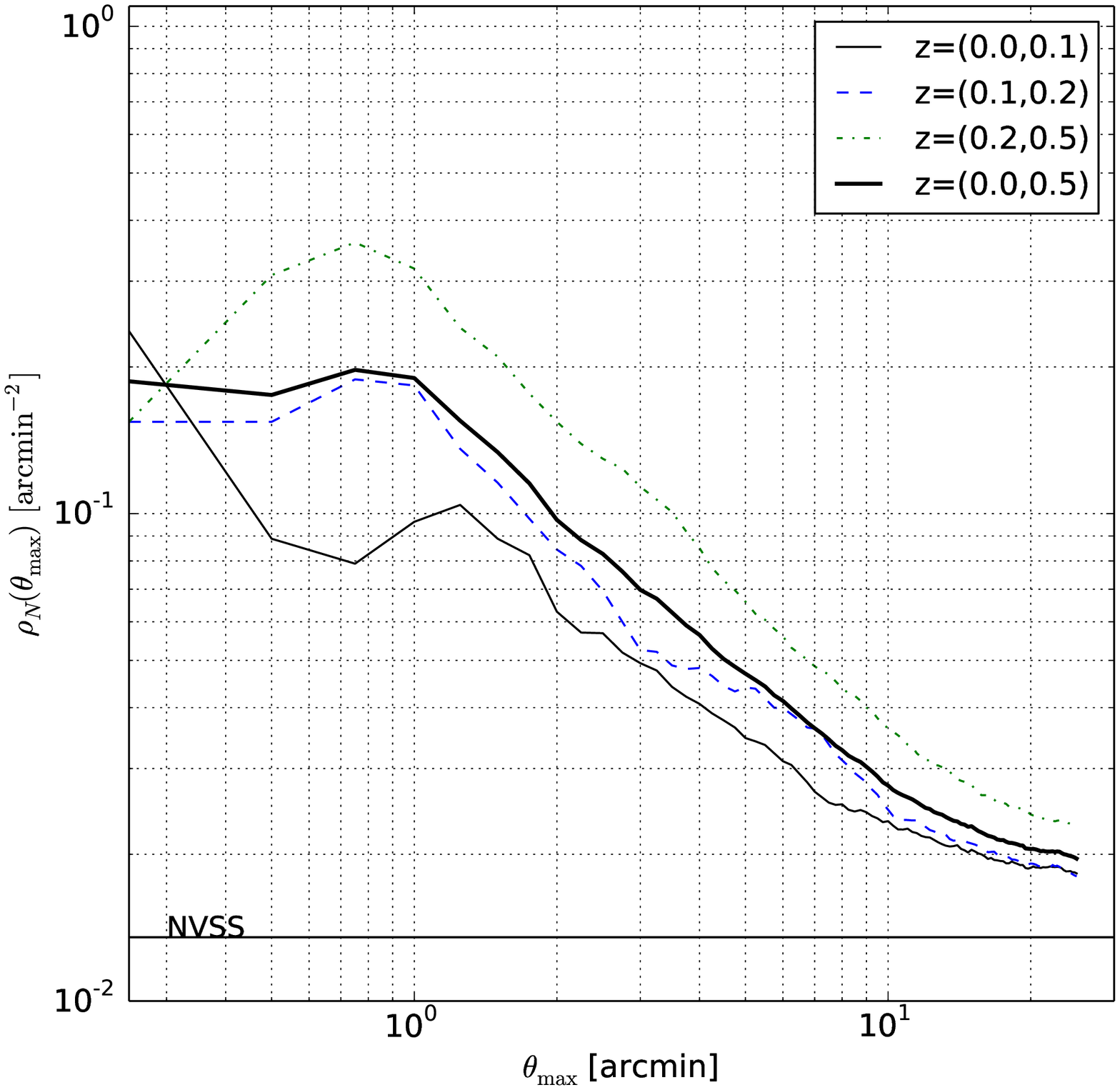}
\includegraphics[width=0.45\textwidth]{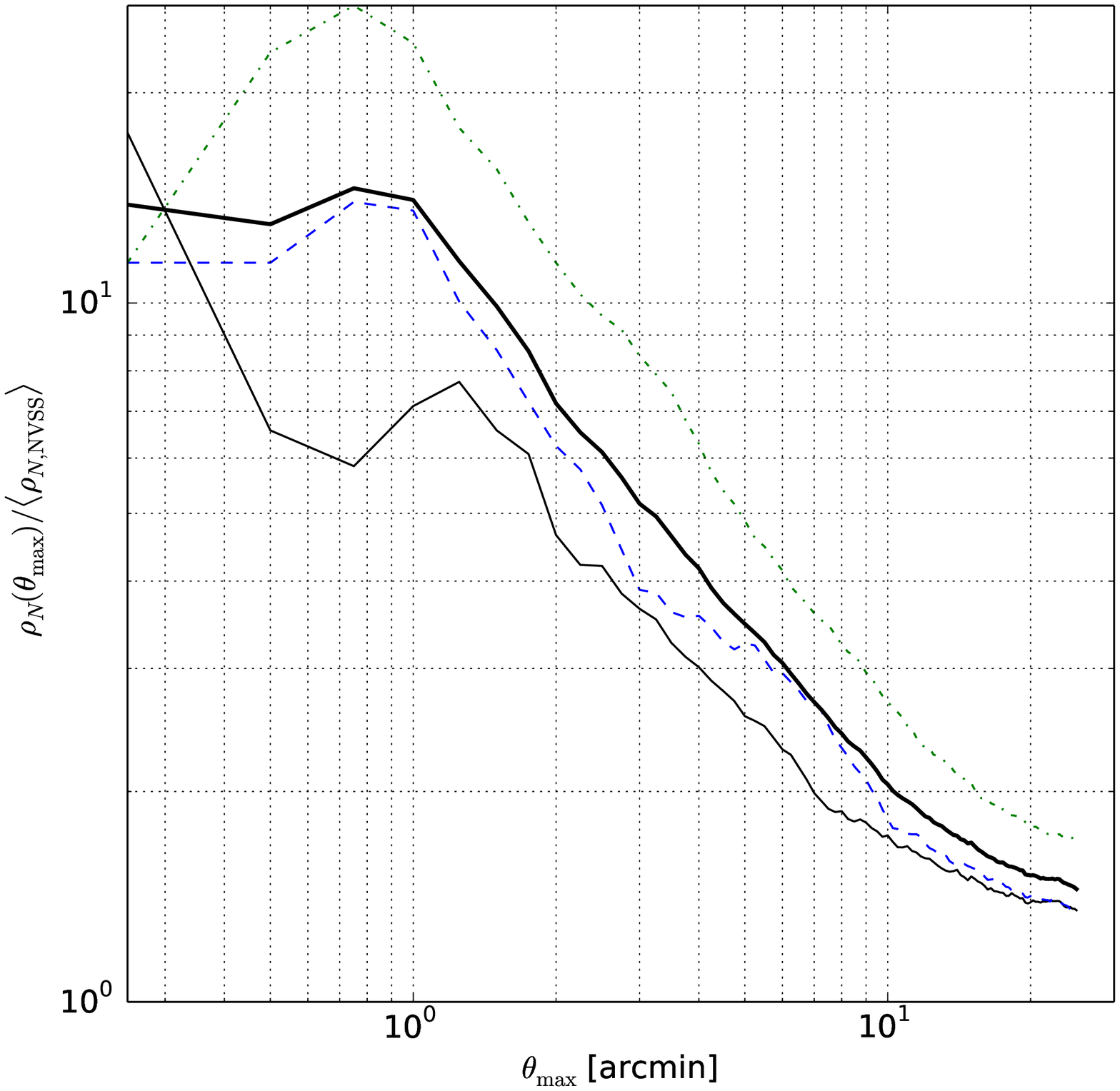}
\caption{({\emph left}) The average NVSS survey point source number
density per unit solid angle around 
the centroids of \textit{Planck} TSZE-selected cluster candidates as a
function of the angular distance from the suspected cluster
centres. The solid-thin (black), dashed (blue), dot-dashed (green)
and solid-thick (black) lines represent the densities for
sub-samples selected according to the 
redshift range given in the plot legend. The horizontal line
indicates the average number density of point sources
for the NVSS survey. ({\emph right}) The relative overdensity of
point sources for the same redshift sub-samples with respect to the
NVSS survey. 
}
\label{fig:ptSrcCorr}
\end{figure}

We calculate the average
cumulative number density of radio sources per unit solid angle  $\rho_N(\theta_{\mathrm{max}})$
as a function of angular distance from the cluster centre
\begin{equation}
\rho_N(\theta_{\mathrm{max}}) = \frac{1}{\pi \theta_{\mathrm{max}}^2 N_0} \int_0^{\theta_{\mathrm{max}}} \frac{\partial N(\theta)}{\partial \theta} \d\theta \approx \frac{1}{\pi \theta_{\mathrm{max}}^2 N_0} \sum_i A_1(\theta_i)
\label{eq:rhoNtheta}
\end{equation} 
where $A_1(\theta_i)=1$ if the radio source is at angular distance
$\theta_i < \theta_{\mathrm{max}}$ from its associated cluster's centre 
and $A_1=0$ otherwise. The summation extends over all radio sources. 
$\partial N(\theta)/\partial\theta$ is the differential radio source count as 
a function of radial angular distance from the cross-correlated cluster centre, and
the $N_0$ factor is taken as the total number of clusters in the
sub-sample selected for a given redshift range. It therefore
calibrates the relation allowing for comparison of data sub-samples of different sizes.
The result is shown in the figure~\ref{fig:ptSrcCorr}.
If point sources around galaxy clusters were not clustered but rather uniformly distributed,
$\rho_N(\theta_{\mathrm{max}})$ would be a constant function, and 
equal the average number density 
($\langle \rho_{N,\mathrm{NVSS}}\rangle$) defined by the survey sensitivity
threshold. Figure~\ref{fig:ptSrcCorr} clearly shows that this is not the case. 
For the NVSS survey $\langle \rho_{N,\mathrm{NVSS}}\rangle \approx 0.0135\,
\mathrm{arcmin}^{-2}$, which is simply the total number of sources in
the catalogue ($1\,810\,672$) divided by the total survey area ($2326
\times 4^2$ $\mathrm{deg}^2$).\footnote{$N_0=1$}

We find that the $\rho_{N}(\theta_{\mathrm{max}})$ relation is redshift dependent and exhibits somewhat stronger clustering at higher redshifts.
The redshift dependence results in part 
from the larger angular sizes of cluster of a given mass at lower
redshifts, but also from the \textit{Planck} cluster selection function. 
A full analysis of the 
statistical properties of point source
clustering around TSZE-detected clusters
would be complicated, 
but for
the purpose of this work, we use our result to infer that radio
sources are roughly 10 times more abundant in galaxy cluster centres
than in their peripheries (figure~\ref{fig:ptSrcCorr} right panel).
The \textit{Planck}-SZ sample \citep{PlanckCollaboration2014a} yields 
lower central clustering values by several per-cent, hence our choice may be conservative
when inferring the effective SZ-cluster counts.

Thus we simulate the direction-dependent radio source density by
constructing a two-dimensional probability distribution  
function (PDF) for a source's appearance 
using the locations of CDM halos extracted from the deep field simulations.
The shape of an individual PDF component due to a single halo is assumed to be a two-dimensional
Gaussian with full width at half maximum defined by the halo virial diameter.
Its amplitude is proportional to the halo virial mass. This is
motivated by the heaviest halos being the richest in galaxies, and
hence likely to contain more radio sources.
If the virial quantities are not defined (due to too small overdensity)
we use instead the halo total FOF extents along the directions tangential to the LOS and the total FOF masses.
The location of individual PDF maximal values coincide with the directions towards the minima
of the gravitational potential.
We then use Monte-Carlo realisations to create a point source
distribution with fluxes generated as described in
section~\ref{sec:flux_density}. The resulting map is  
converted to units of specific intensity for the assumed angular
resolution of the map and added to the simulated maps of the TSZE
component. 

The choice of a Gaussian shape for the correlated components of the
PDF could, in principle, be replaced by the $\beta$-like profiles. 
The shape of a Gaussian differs from that of a $\beta$~profile, in
particularly by showing a faster decrease at large angular
distances from the halo centre. However, since the background source
density is about 10\% of the peak source density, we expect the shape
assumption to have little effect on generic results from our
simulations.
Furthermore, the observational data (figure~\ref{fig:ptSrcCorr}) do
not give strong evidence for either shape, at present.

In addition to the correlated component of radio sources, we also
insert a uniformly-distributed source population in the field 
adding a constant density to the PDF. 
This simulates high-redshift sources outside the redshift range of the
simulation, ignoring the (usually weak) effects of gravitational
lensing.\footnote{This also captures radio sources hosted in galaxies
in the local neighbourhood that that were actively star forming a few
billion years ago, although they are still assumed to be unresolved.}
The constant source density used depends on the survey flux density
threshold, which for the NVSS is about $\sim 2.5\, \mathrm{mJy}$ -- the flux density
corresponding to the weakest detected sources
\citep{Condon1998}. Although this constraint is similar to independent
estimates reported in \cite{Coble2007} one would obtain a slightly
different radio source overdensity values if the clusters sample was
cross-correlated with a survey of a different sensitivity.

We neglect the contributions from the thermal sources as they are not dominant in the considered
frequency range.

\section{Comparison to other simulations, observations and theoretical predictions}
\label{sec:comparizons}
In the following sections we test our simulations of structure
formation by comparing selected scaling relations with those
from other simulations and with observations provided by the Chandra
and the XMM-Newton satellites. We also test the consistency of the
extracted mass functions with the theoretical predictions of the
Press-Schechter theory and fitting functions fixed by other numerical
simulations.

\subsection{Halo mass function}
\label{sec:mass_function}
\begin{figure}[!t]
\centering
\includegraphics[width=0.75\textwidth]{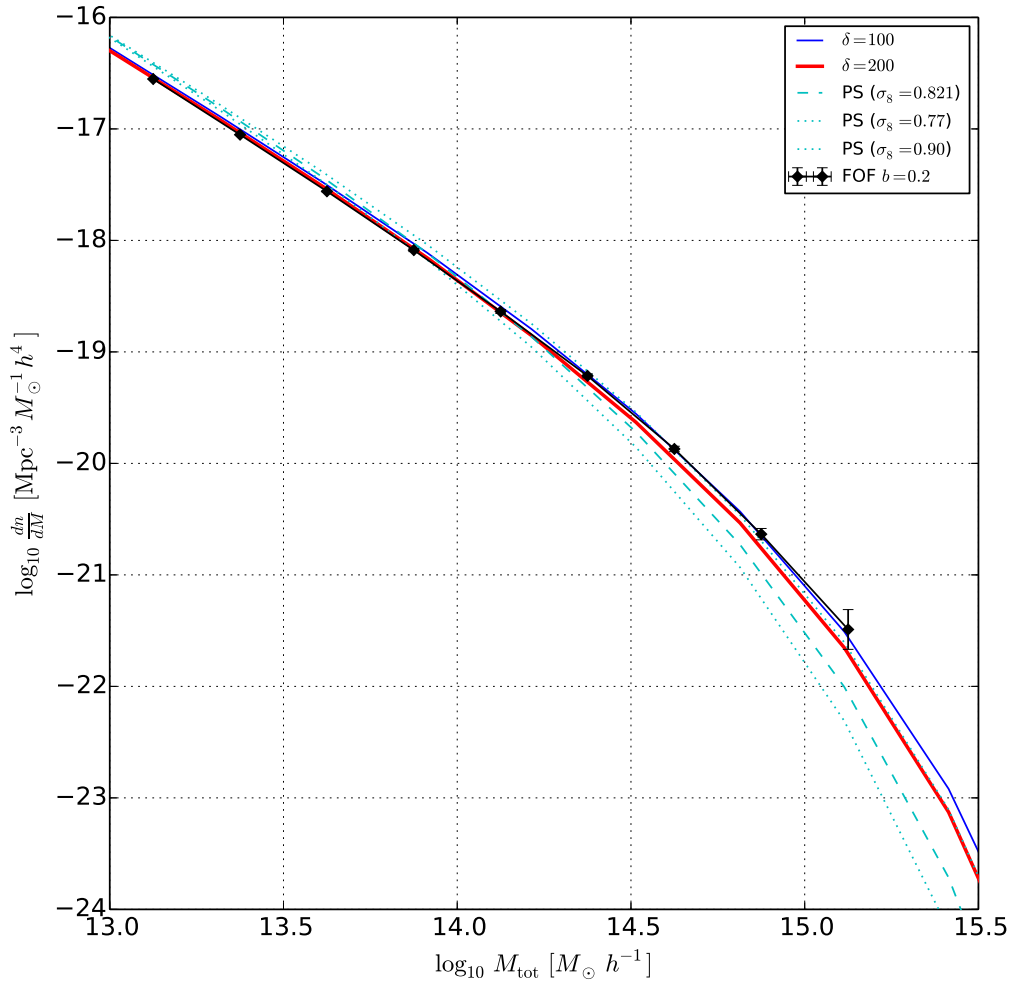}
\caption{Average halo mass function resulting from several
realisations of the simulation box comoving volume (black diamonds),
compared with the Press-Schechter (dashed, cyan) and Tinker (solid)
mass functions for a LCDM model. Results are shown for $\delta=100$
(blue,thin) and for $\delta=200$ (red,thick) overdensity
thresholds. The error bars correspond to $\pm 1\sigma$ from sample
variance in our simulations. The dotted (cyan) lines are the
PS predictions that show the impact of a different $\sigma_8$ normalisations.
}
\label{fig:massFn}
\end{figure}

In figure~\ref{fig:massFn} we plot the mass function  $\d n/\d M$ of the FOF halos identified 
in our deep-field simulations, sliced at the hypersurface of the
present. We also show the Press-Schechter (PS) mass function
\citep{Press1974} calculated with the same cosmological parameters 
\begin{equation}
\frac{\d n_{\mathrm{PS}}}{\d M} = \sqrt{\frac{2}{\pi}}  \frac{\rho_m}{M}  \frac{\delta_c}{\sigma(M)}   \frac{\d\ln\sigma^{-1}(M)}{\d M} \exp\Bigl(-\frac{\delta_c^2}{2 \sigma^2(M)}\Bigr)
\label{eq:PSmf}
\end{equation}
where $\sigma(M)=\sigma(M(R))$ with $M(R)=\frac{4}{3}\pi R^3 \rho_m$ and 
\begin{equation}
\sigma^2(R) = \int P_g(k) W^2(kR) \frac{\d k}{k},
\label{eq:sigmaM}
\end{equation}
is the variance of the matter density fluctuations smoothed at scales $R$.
\begin{equation}
P_g(k) = \frac{4}{25} f^2(\Omega_m,\Omega_\Lambda) \Bigl(\frac{k}{a_0 H_0}\Bigr)^4 T^2(k)  \mathcal{P}_{\mathcal{R}}(k)
\label{eq:Pg}
\end{equation}
is the power spectrum of the mass density distribution (in units where $c=1$). 
In equation~\ref{eq:Pg} 
$\mathcal{P}_{\mathcal{R}}(k) = \Delta_{\mathcal{R}}^2 (k/k_0)^{n_s-1}$ is the power spectrum of primordial curvature perturbations
with pivot scale $k_0=0.002\, \mathrm{Mpc^{-1}}$, 
$T(k)$ is the matter transfer function, and 
$f(\Omega_\Lambda,\Omega_m) \equiv \frac{g(\Omega_\Lambda,\Omega_m)}{\Omega_m}$ is 
the $\Lambda$CDM linear perturbation growth factor relative to the Einstein-de-Sitter case.
The $g(\Omega_\Lambda,\Omega_m)$ quantity is model dependent, and for the flat $\Lambda$CDM model is given 
by the fitting formula as in eq. 29 of \cite{Carroll1992}.
In that formula the variable names are different so it's not a direct reference.
The kernel function $W(kR)$ is chosen to be the Fourier transform of the top-hat function 
and its exact form depends on the assumed Fourier transform convention.
The quantity $\delta_c = 1.69$ is the present linear-theory overdensity required for a uniform spherical region to 
collapse into a singularity and it has a weak dependence on cosmological parameters \citep{Eke1996}.
We use the CAMB software to obtain a high resolution $\Lambda$CDM matter transfer function, tabulated 
up to $k \eta_{\mathrm{max}}=1.5\times 10^5$ (where $\eta_{\mathrm{max}}$ is the comoving scale of particle horizon). 
This is required for accurate predictions 
for the lowest-mass halos. 

We also compare our results with the Tinker mass function
\citep{Tinker2008} at redshift $z=0$
\begin{equation}
\frac{\d n}{\d M} = f(\sigma(M))  \frac{\rho_m}{M} \frac{\d\ln \sigma^{-1}(M)}{\d M},
\label{eq:tinkerMF}
\end{equation}
for the overdensities $\delta=\{100,200\}$ which
should enclose the range statistically probed by the FOF algorithm with linking length parameter $b=0.2$.
The Tinker mass function is parametrised by 
\begin{equation}
f(\sigma(M))=A \Bigl[ \Bigl(\frac{\sigma(M)}{b}\Bigr)^{-a} + 1 \Bigr] \exp\bigl[-\frac{c}{\sigma^2(M)}\bigr],
\label{eq:tinkerParam}
\end{equation}
where the four parameters $(A,a,b,c)$ are linearly interpolated (and extrapolated for the case of $\delta=100$) 
based on the values tabulated in table~2 of \cite{Tinker2008}.

Figure~\ref{fig:massFn} shows that there is a good consistency between
the recovered FOF mass function and the theoretical expressions at
$z\approx 0$, confirming that our numerical calculations are robust,
and supporting the validity of the halo abundances in our light-cone
realisations.
Apart from the known underestimate of the abundance of the heaviest
halos made by the Press-Schechter mass function,
the extracted mass function seems to overestimate the abundance 
of heavy halos with respect to the 
Tinker mass function, and slightly underestimate it (by a factor
$\approx 1.02$ with respect to the Tinker 
prediction at $\delta=200$) for 
the lightest halos, which have $N_h$ only slightly above the
minimum of $600$ (section~\ref{sec:cluster_search}).

This effect could be explained by the tendency of 
the FOF algorithm to connect multiple light halos into single, heavier
systems (as depicted in  figure~\ref{fig:halo34582} and
figure~\ref{fig:halo10801}).  
For the same reason, and because of the limited mass resolution of our
simulations, for low-mass halos the 
reconstructed mass function slightly deviates from the Tinker mass function
prediction. 
Details of the MF redshift evolution do not have impact on most of the presented
results (but see section~\ref{sec:PIsteradiansFieldPrediction} where an
order-of-magnitude estimates are obtained by neglecting the MF redshift
evolution up to redshift $z=1$).

We find that the creation of the low-mass halos is sensitive to the
settings of the softening length in the gravity computations. 
Too large a softening length may significantly suppress light halo
abundance and underestimate baryon temperature. 
Decreasing the softening length significantly increases the
computational time. We experimented with various softening lengths to
check that the value of $\sigma_8$ calculated in the simulations is
consistent with the input, linearly-predicted, value. 
We found that a softening length of 15 kpc/$h$ provides
reasonably-converged $\sigma_8$ values for our mass resolution with
acceptable execution times. The overall consistency between our mass
function and that from the Tinker mass function assures us that our
halo abundances are realistic.

\subsection{The M-$\sigma_{v}$ scaling relation}
\label{sec:Msigmav}
\begin{figure}[!t]
\centering
\includegraphics[width=0.85\textwidth]{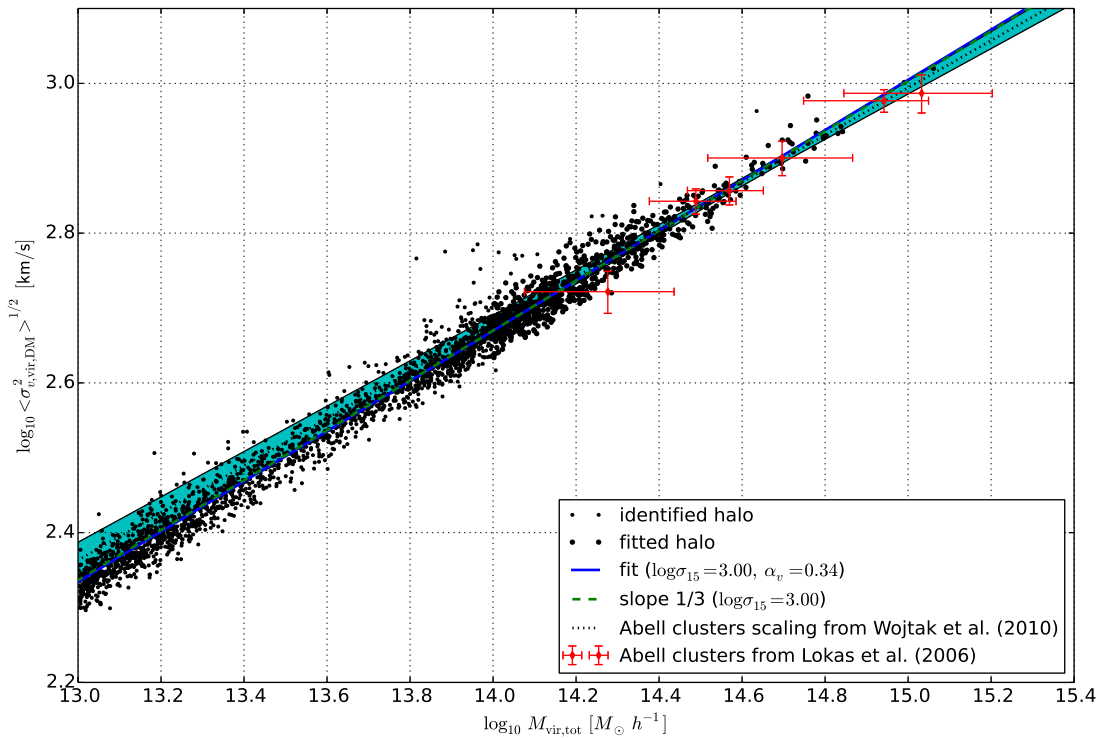}
\caption{The M-$\sigma_v$ scaling relation for the sample of simulated
halos (black dots) is compared against the measurements of the line-of-sight 
velocity dark matter dispersion in Abell clusters as derived in \cite{Lokas2006} (red diamonds).
The solid line is the fitted power law scaling relation, 
equation~\ref{eq:Msigma}, and the dashed line is the fitted scaling
relation for power-law slope $\alpha_v=1/3$. The fitting is done
only for halos indicated by large dots, with $M_{\mathrm{vir,tot}} >
10^{14} \ M_\odot h^{-1}$. For clarity, we plot only 10\% of the 
halos with $M_{\mathrm{vir}} < 10^{14} \ M_{\odot} h^{-1}$.
The dotted line with shaded area is the scaling relation reported in \cite{Wojtak2010}
using a larger sample of Abell clusters.
}
\label{fig:Msigmav}
\end{figure}

In order to further test the initial conditions and the compatibility
of the assumed cosmology with observations we compare the velocity
dispersions measured in the simulated halos with the LOS velocity
dispersions measured in Abell clusters. On the simulation side as a
measure of the velocity dispersion we choose the square root of the
mean dark matter halo coordinate-velocity variance: $(\langle
\sigma_{v_i,\mathrm{vir,DM}}^2\rangle_i)^{1/2}$, calculated within spheres of
virial radius 
We plot these estimates against the total viral mass of halos
$M_{\mathrm{vir}}$  to construct the $M-\sigma_v$
scaling relation, which for the case of virialised clusters should
follow 
\begin{equation}
\sigma_v(M_\delta) = \sigma_{15} \Bigl(\frac{M_{\delta}}{10^{15} M_\odot/h }\Bigr)^{\alpha_v} [\mathrm{km/s}],
\label{eq:Msigma}
\end{equation}
with $\alpha_v=1/3$, where $\sigma_{15}$ is the calibration of the
relation at the mass scales of $10^{15} M_\odot/h$. 
In figure~\ref{fig:Msigmav} we plot the cluster masses and
line-of-sight velocity dispersions of our simulated halos together
with the reconstructed cluster masses and velocity dispersions
from \cite{Lokas2006}. This compilation provides the best-fit masses
of six nearby ($z<0.06$) galaxy clusters\footnote{A0262, A0496, A1060,
A2199, A3158, A3558} within their virial radii, and were found by
fitting solutions of Jeans equation to the reconstructed
cluster velocity variance and velocity kurtosis profiles,
assuming the validity of their NFW density profiles to the virial
radii. The cluster velocity dispersion profiles, as probed by large
galaxies, were derived from the NASA/IPAC Extragalactic Database 
(NED) with a selection including a test to reject likely merging
systems, as described by \cite{Lokas2006}. 
We also plot the reconstructed $M-\sigma_v$ scaling relation from 
a projected phase-space analysis for the sample of nearby Abell
clusters reported in \cite{Wojtak2010}, where different 
methods of mass estimation are discussed. We ignore the small
difference in the virial overdensity definition between the value
assumed for this analysis and that of \cite{Lokas2006}.

The reconstructed velocity dispersion follows the virialised cluster
scaling relation well, and the simulated clusters have realistic
velocity dispersions, with no significant mass resolution effect.

\subsection{The M-T scaling relation}
\label{sec:MT}
In this section we test our simulations against the mass-temperature scaling relation.

\begin{figure}[!t]
\centering
\includegraphics[width=0.75\textwidth]{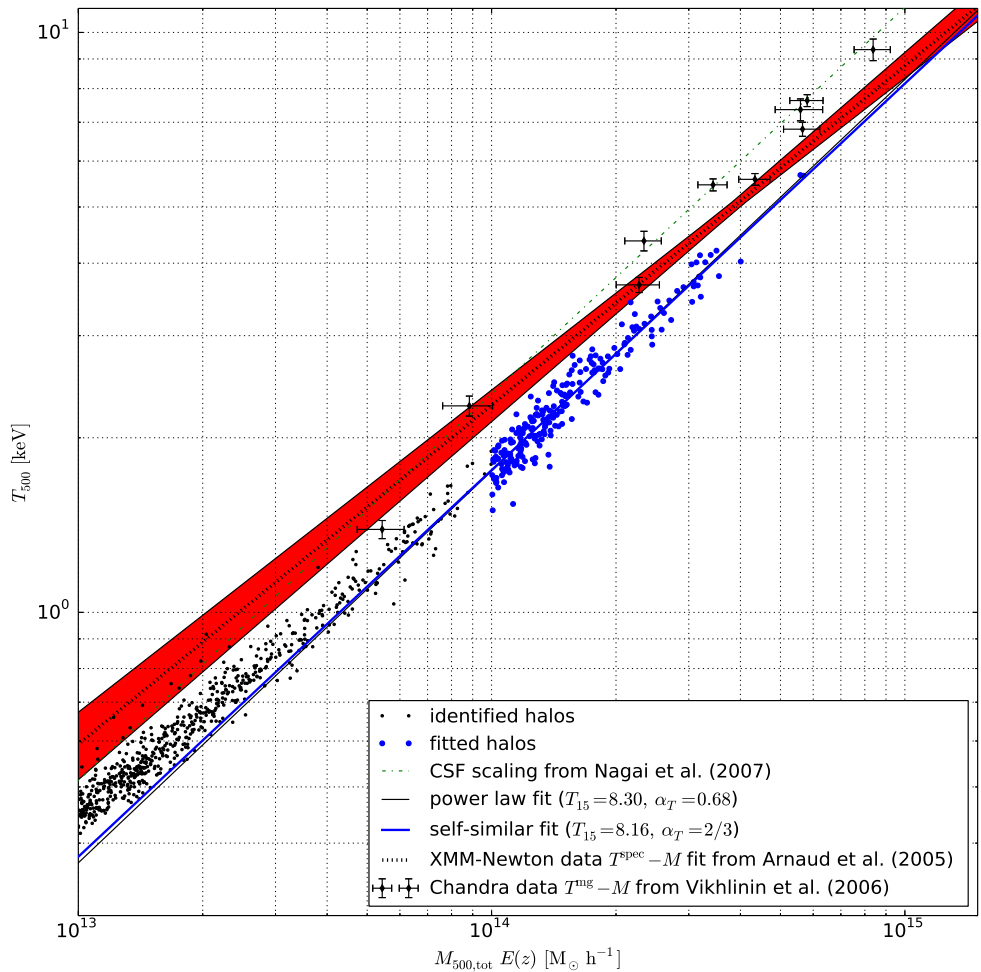}
\caption{Halo-mass--temperature scaling relation for halos
identified in the lowest-redshift simulation  
($z\sim 0.06$) for $\delta=500$.
Small (black) dots indicate the FOF identified halos and large
(blue) dots indicate the halos used for fitting the scaling relation
defined in equation~\ref{eq:MT}.  
We discard halos with masses $M_{500}<10^{14} M_{\odot} h^{-1}$
and outliers identified as points deviating from the initial
whole-sample power-law fit 10\%. For clarity, we plot only every
tenth halo with $M < 10^{14} M_{\odot} h^{-1}$.
The solid lines indicate the power law two parameter fit
(thick,blue) and the self similar fit (thin,black) to the filtered
data. The scaling relation of \cite{Nagai2007}, from 
high mass-resolution simulations that include radiative cooling and 
star formation effects, and assuming a self-similar slope, is shown
by the dash-dotted line. Data for the scaling relation for gas-mass
weighted temperature are taken from \cite{Vikhlinin2006} (black
diamonds). The best-fit scaling relation using spectroscopic temperature 
from \cite{Arnaud2005} is shown as the dotted line with 
the 1-$\sigma$ uncertainty band (shaded red) under assumption of
zero covariance between amplitude and power-law index.
The tens of percent discrepancy between the adiabatic simulations
and the observations is known to be the result of gas physics
not included in our simulations. 
}
\label{fig:MT}
\end{figure}

Our simulation framework relies on the adiabatic gas approximation (AD) and 
the resulting temperature profiles therefore deviate from those
extracted from simulations that include radiative cooling, star
formation, and AGN or supernova feedback (hereafter CSF). Such
processes play an important role in explaining cool-core clusters, and
may be required to fine-tune simulations to observational data,
especially in cluster central regions. In the central parts of
clusters the strong radiative cooling due to thermal processes
dominates over the heating from processes associated with star
formation, while in cluster outskirts the net effect results in an
increase of the ICM temperature \citep{Nagai2007}. This difference
alters ICM density profiles to some extent, and should affect the 
profiles of cluster X-ray emission and Compton $y$-parameter. The
difference between the AD and CSF cases  is estimated to be
a few tens of per-cent. This is seen in figure~\ref{fig:MT}. In our AD
simulations we find gas temperatures systematically lower than as
reported from CSF simulations, by a factor $\sim 1.3$ at $\delta=500$.

As in section~\ref{sec:Msigmav}, 
we examine the $M$-$T$ scaling relation for our simulated sample of
heavy halos and fit a scaling relation to those with
$M_{500,\mathrm{tot}}>10^{14} M_\odot/h$. A low-mass cut-off is required 
to avoid the biases resulting from our limited mass resolution. Mass
resolution effects also become increasingly important at large
$\delta$, and lead to artificial deviations from self-similar scaling.  
For the adiabatic case the cluster gas temperature is expected to
scale self-similarly with the cluster mass \citep{Kaiser1986}
\begin{equation}
T(M_{\delta}) = T_{15} \Bigl(\frac{M_{\delta} E(z)}{10^{15} M_\odot/h }\Bigr)^{\alpha_T}  [\mathrm{keV}],
\label{eq:MT}
\end{equation}
where $\alpha_T=2/3$, $T_{15}$ is the scale temperature at mass $10^{15} M_\odot/h$, and
$E(z) = \sqrt{\Omega_{m0}(1+z)^3 + \Omega_{\Lambda0}}$.\footnote{
Note that this is not a conventional definition of the scaling
relation, which is usually given in terms of the
$M(T)$ function, but throughout this paper we consistently use the
total halo mass as an independent variable, although observationally
it is the quantity to be sought using a scaling relation.}

We observe that the scaling relation 
(as probed by the simulated halos) clearly follows the self-similar prediction for high-mass
clusters, but apparently  underestimates the cluster temperature at
$\delta=500$. The low-mass tail deviates from self-similar scaling.
We verified that this is due to the limited mass resolution in our
$2\times 512^3$ particles simulation, and is the main reason why
we neglect lighter halos in fitting the scaling relations. 
In order to exclude outliers we perform the fitting in two
stages. First we fit a straight line in the space of logarithms of
masses and temperatures, calculate relative residuals, and exclude
halos that deviate by more than 10\%. Then we perform the second fit
on such pre-filtered data. Halos ignored in the fitting process
are marked with small dots in figure~\ref{fig:MT}. In the particular
case of the scaling relation calculated from the simulation presented
in the figure~\ref{fig:MT} there are no outliers within the fitted
mass range. In that figure we also provide a power-law fitting function 
parameters for the cases of (i) fixed slope (according to the self-similar case) and 
(ii) with fitted slope.

In our 3-D gas temperature reconstructions (section~\ref{sec:cluster_search})
it is assumed that the degree by which our adiabatic approximation simulations 
deviate from the actual temperature distributions in the
forming clusters is similar throughout the whole range of the considered redshifts (section~\ref{sec:lss}).

\subsection{The M-Y scaling relation}
\label{sec:MY}
As a final consistency test we compare the M-Y scaling relation from our simulations with that from a similar, but higher mass-resolution, adiabatic simulation 
performed in \cite{Nagai2006}. 

\begin{figure}[!t]
\centering
\includegraphics[width=0.49\textwidth]{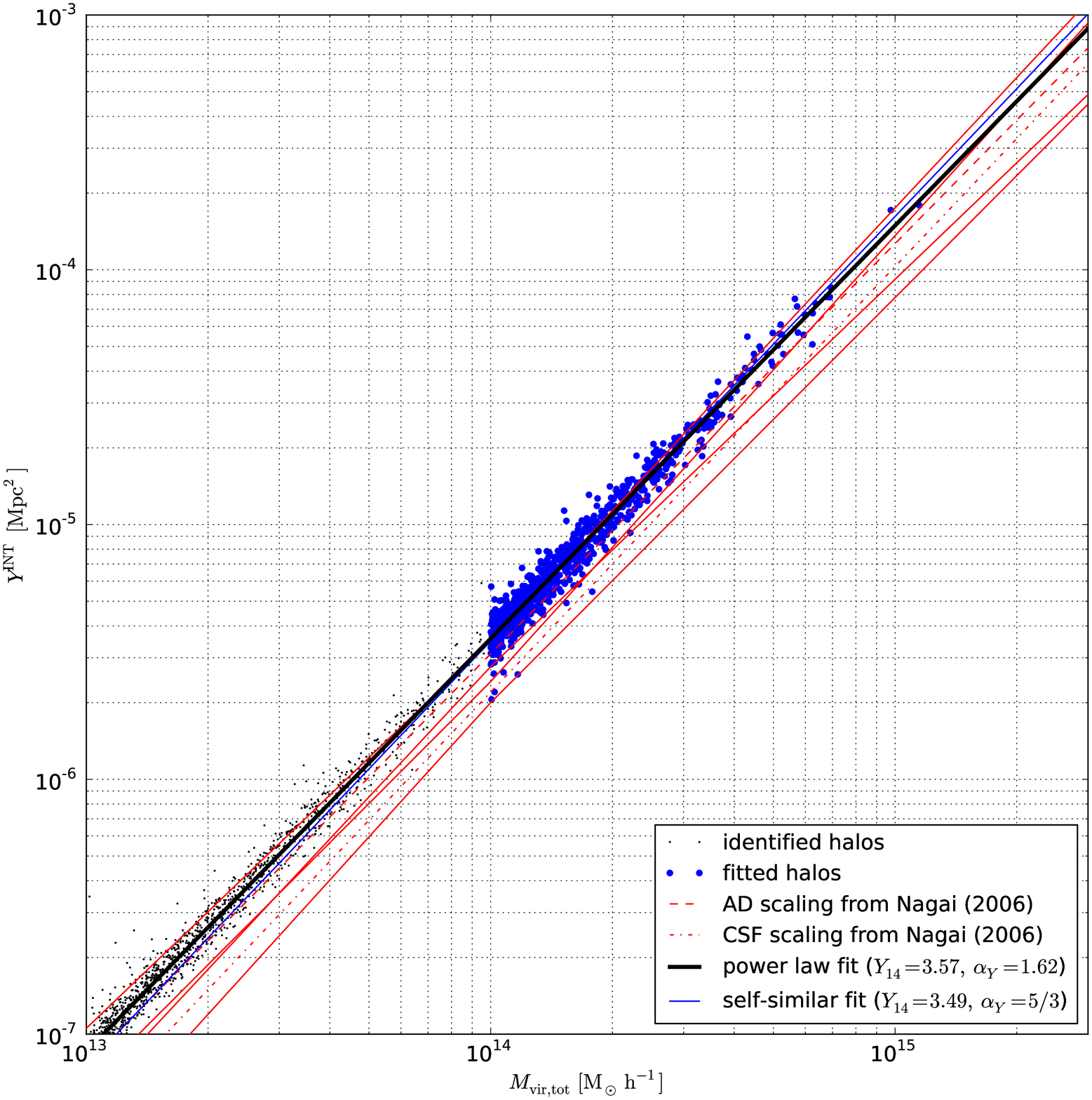}
\includegraphics[width=0.49\textwidth]{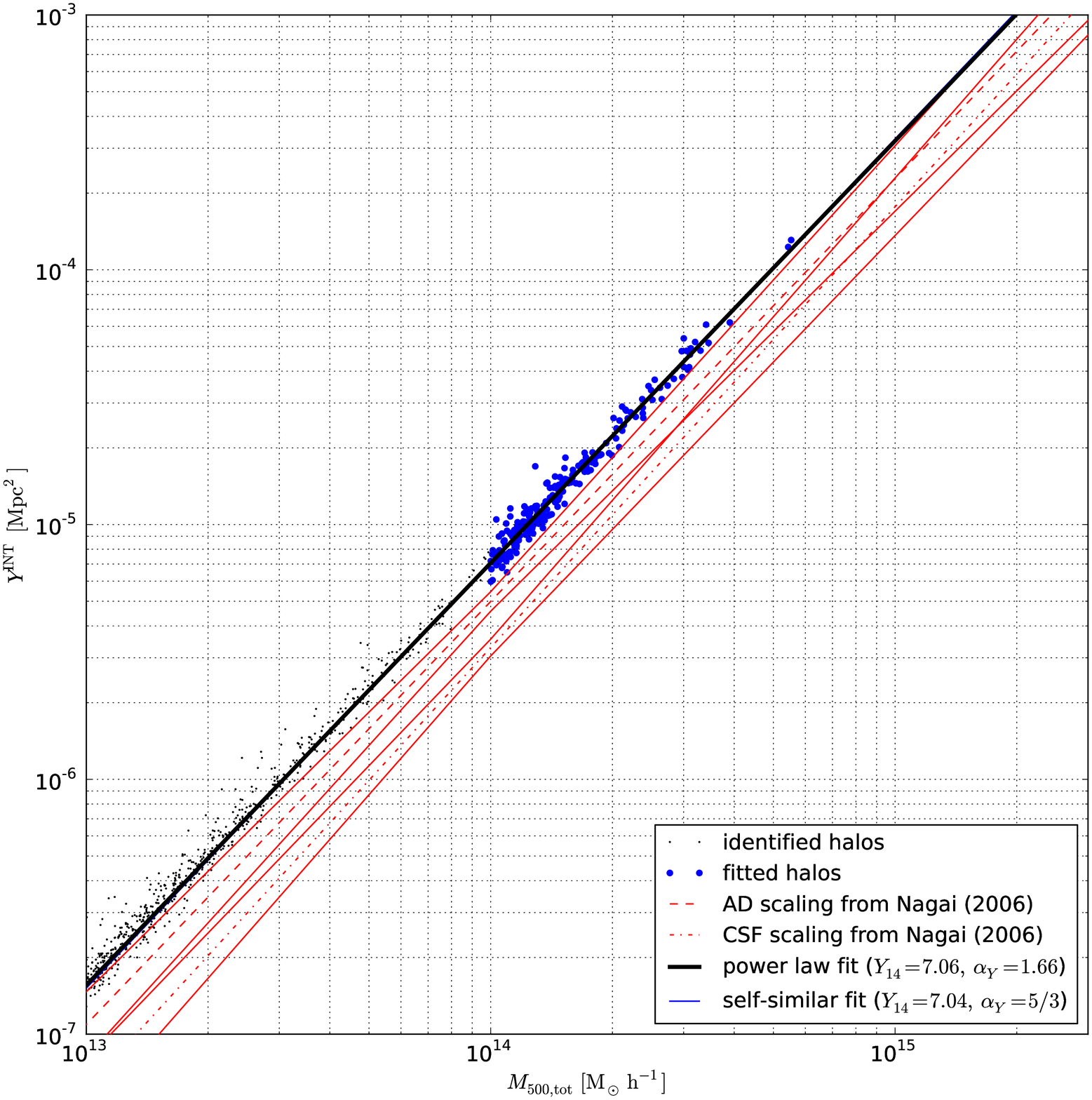}
\caption{Halo-mass--integrated Comptonisation-parameter scaling relation   for halos identified in the lowest redshift simulation 
($z\sim 0.06$) for the virial (left) and $\delta=500$ (right) mass definitions.
The solid lines indicate power law two-parameter fits (thick) and self-similar fits   (thin) to halos with outliers removed and with total estimated FOF masses   $> 10^{14}\mathrm{M_{\odot}\,h^{-1}}$.
For clarity, we plot only every tenth halo with mass
$< 10^{14}\mathrm{M_{\odot}\,h^{-1}}$.
We also show the scaling relations of \cite{Nagai2006}   from high-resolution adiabatic simulations (dashed) and   simulations that include radiative cooling and 
star formation (dash-dotted). 
}
\label{fig:MY}
\end{figure}

Figure~\ref{fig:MY} shows over 2000 of the most massive halos found in one of the simulations evolved up to the 
redshift of $z=0.06$.
The $M-Y^{\mathrm{INT}}$ relation for self-similar evolution is given by 
\begin{equation}
Y^{\mathrm{INT}}(M_{\delta}) = \Bigl(\frac{Y_{14}}{10^6}\Bigr) \Bigl (\frac{M_{\delta} }{10^{14} M_\odot/h }\Bigr)^{\alpha_Y} E(z)^{2/3}  [\mathrm{Mpc^2}],
\label{eq:MY}
\end{equation}
with $\alpha_Y=5/3$ and where $Y_{14}$ is scale Comptonisation parameter at mass scale $10^{14} M_\odot/h$.
This relation follows from equations~\ref{eq:MT} and ~\ref{eq:Yint}.
As before, we fit the scaling relation using the LM algorithm supported Monte-Carlo selection of initial parameters for individual LM minimisations. The data filtering and fitting strategy is as described in section~\ref{sec:MT}.

We find that the $M-Y$ scaling relation is consistent with the results 
from the AD simulations of \cite{Nagai2006} to within a factor of order unity for low $\delta$ values. The $Y_{14}$ scaling, however, depends
on a number of factors, particularly on the
choice of the number of neighbours used in the smoothing kernel for the SPH interpolations and density calculations. We find that 20\% changes in $Y_{14}$ occur as we change the number of neighbours from~5 to~66, with larger values of $Y_{14}$ arising from a larger neighbour count.  
Even with our highest mass resolution ($N_{\mathrm{tot}}=2\times 256^3$ and $L=128$ Mpc) we find that the amplitude of the scaling relation 
for 
$\delta=500$ is larger by a factor of $\sim 1.1$ for $N_{\mathrm{neigh}}=15$ than in \cite{Nagai2006}.
For $N_{\mathrm{neigh}}=33$ (used for figure~\ref{fig:MY}) the calibration is larger by a factor of $\sim 1.40$ 
than the value $Y_{14}^{\mathrm{AD}}(\delta=500)=4.99$, and by $\sim 2.1$ from the value $Y_{14}^{\mathrm{CSF}}(\delta=500)=3.29$), from \cite{Nagai2006}.
However, the scaling relation is much more similar at lower overdensity thresholds, $\delta=\{200,\mathrm{vir}\}$.
In the case of  $\delta=\delta_{\mathrm{vir}}$, the scaling calibration is larger only by factors $\sim 1.1$ and
$\sim 1.6$ for the AD and CSF cases of \cite{Nagai2006}.

Mass resolution also has an important impact on the value of
$Y_{14}$. As the mass resolution is improved, the amplitude of the
scaling relation approaches that of the AD simulation of
\cite{Nagai2006}.  A 64-fold mass resolution increase, from
$N_{\mathrm{tot}}=2\times 256^3$ and $L=512$ Mpc to
$N_{\mathrm{tot}}=2\times 512^3$ and $L=512$ Mpc to
$N_{\mathrm{tot}}=2\times 256^3$ and $L=128$ Mpc induces changes of
order 20\% in the value of $Y_{14}$.

This parameter is also affected by the choice of redshift used to
generate the initial conditions. The reason probably stems from the
challenge of preserving high numerical accuracy during time evolution
of structure formation. This is more difficult as perturbations are
followed through a broader range of density contrast (i.e., from
earlier epochs).

We experimented with different numbers of neighbours at fixed mass
resolution to optimise the interpolation for TSZ signals. For each SPH
particle that comprises a part of a halo we investigated how precisely
we could perform interpolations, weighted according to the
density-weighted temperature, at grid cell centres.  We used the
density-weighted temperature to ensure that the interpolation error is
dominated by regions that contribute most to the SZ signal and not by
the cluster peripheries, where the density weighted temperature is
small.  We found that the relative interpolation accuracy is or order
1\% as a function of number of neighbours, and that the values of
$Y_{14}$ change by less than 1\% as the grid resolution is improved
from 50~to 25~kpc.

In figure~\ref{fig:MY} ({\em right}), we find that at overdensity
$\delta=500$ there is a small departure from self-similarity at low
mass, because of the limited mass resolution.  The effect becomes
worse for simulations with worse mass resolution, and is more
noticeable at larger $\delta$, as expected.

The net effect of ignoring cooling and star formation in our
simulations is an overestimation of the TSZE signal. This may result
in our simulations overestimating the number of objects that would be
detected for a given flux density threshold, although the amplitude of
the difference between AD and CSF simulations is also a function of
redshift. Thus our estimates of the number of TSZE detections may be
high by a modest factor ($< 2$).

\section{Results}
\label{sec:results}
We now describe the main results from this study. We split these into four categories: 
(i) mock maps; (ii) survey sensitivity limits; (iii) predictions for TSZE blind surveys;
and (iv) predictions for blind radio-source surveys.

\subsection{Mock maps}
\label{sec:mockMaps}
\begin{figure}[!t]
\centering
\includegraphics[width=0.35\textwidth]{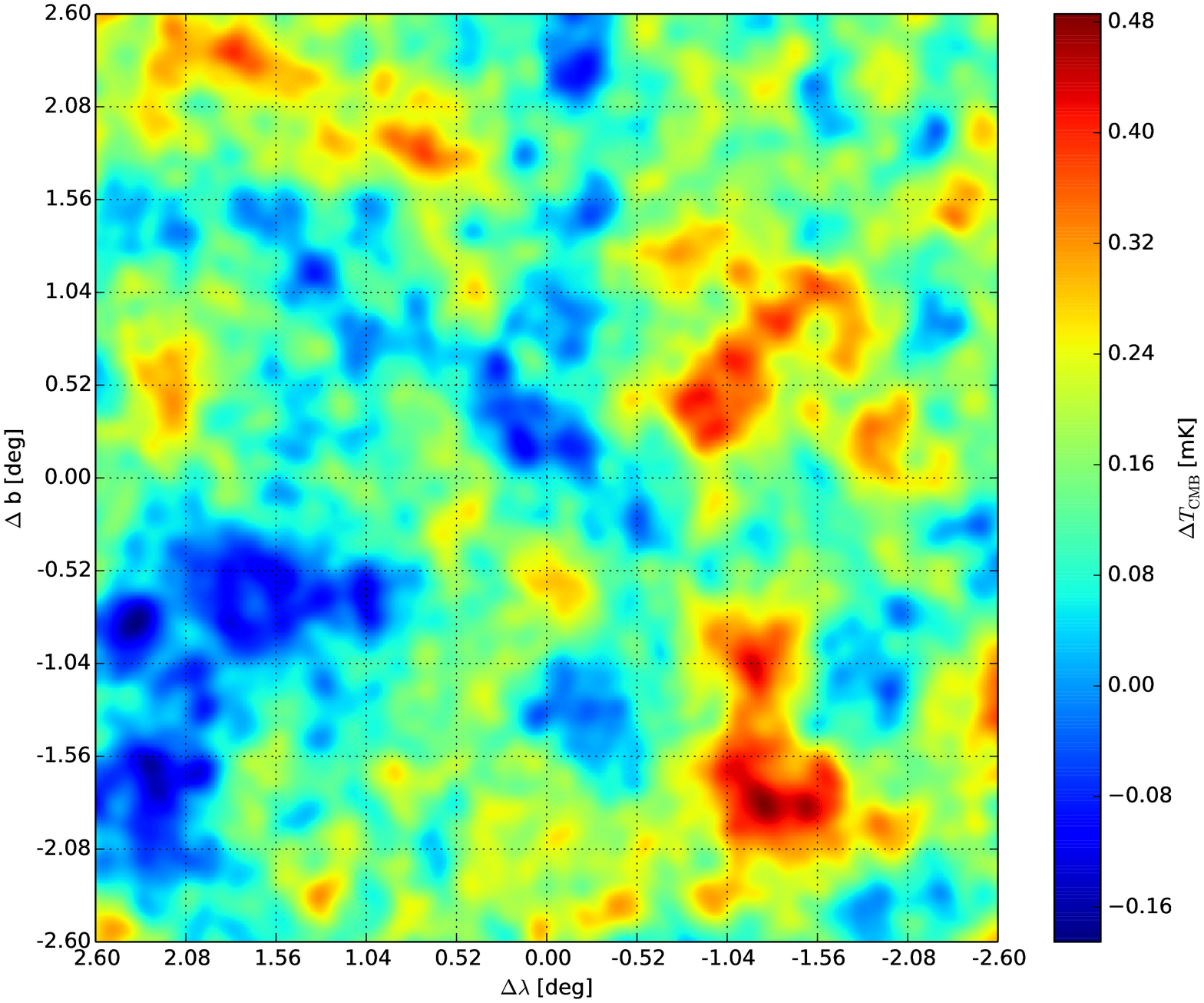}
\includegraphics[width=0.35\textwidth]{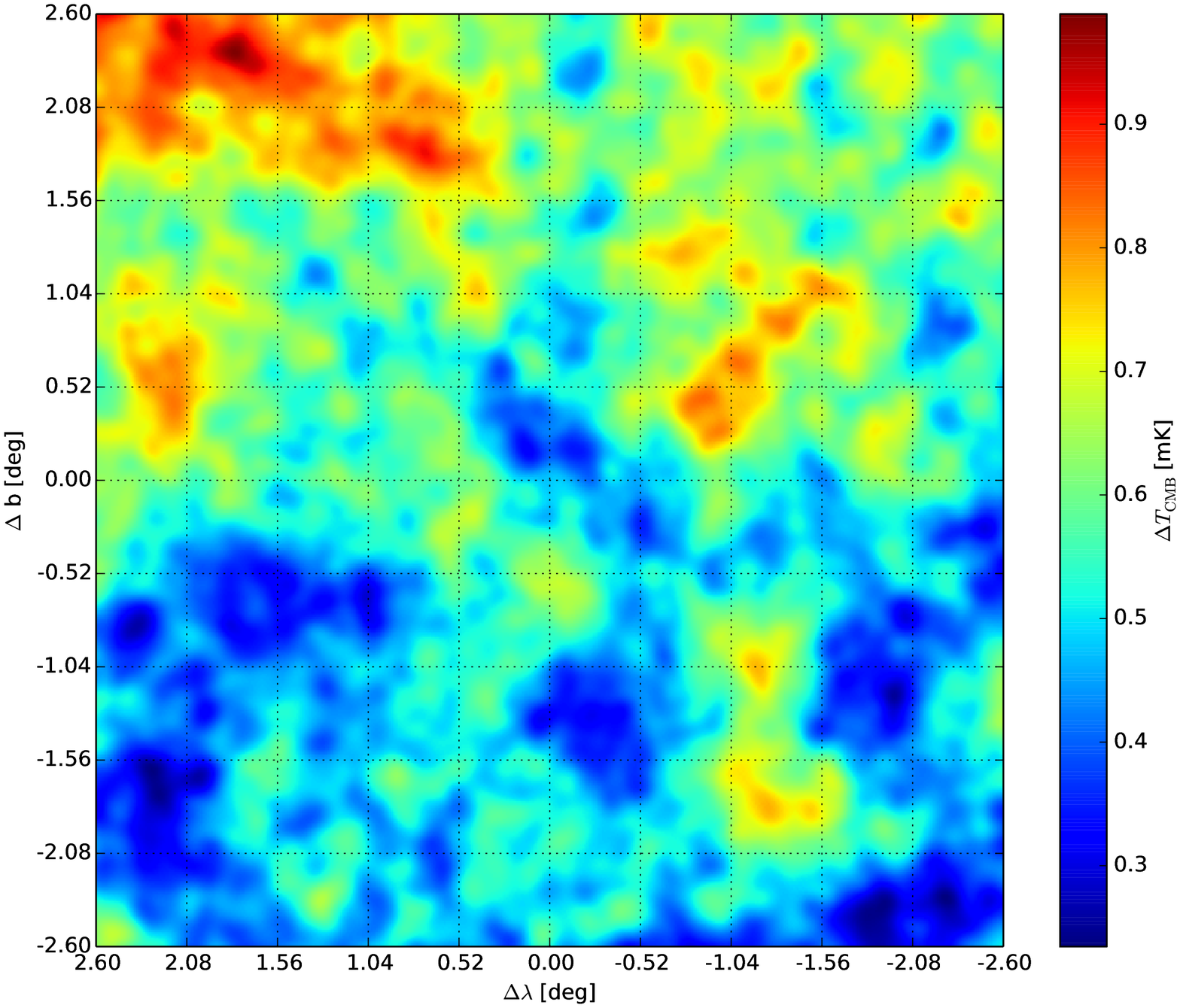}
\includegraphics[width=0.35\textwidth]{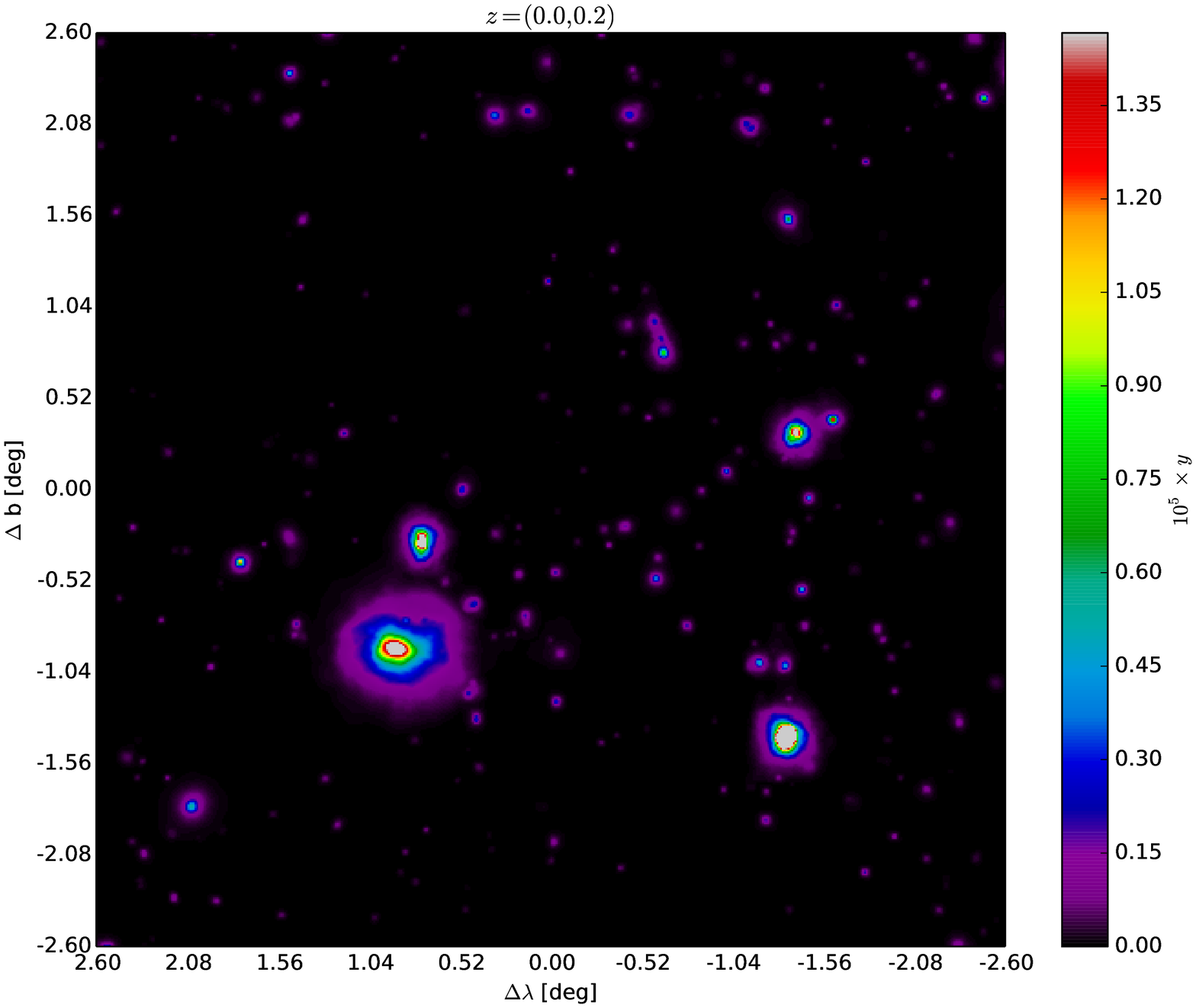}
\includegraphics[width=0.35\textwidth]{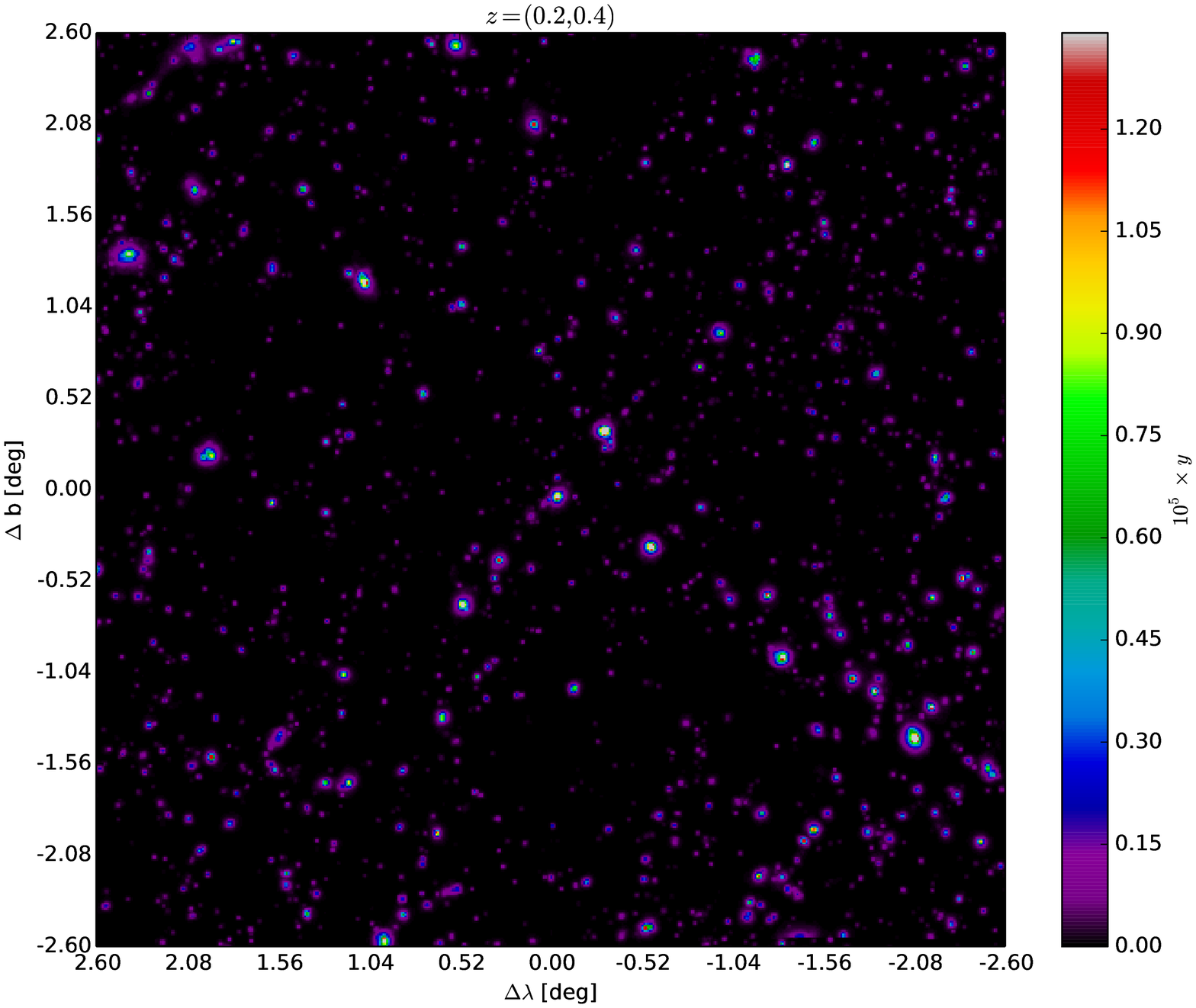}
\includegraphics[width=0.35\textwidth]{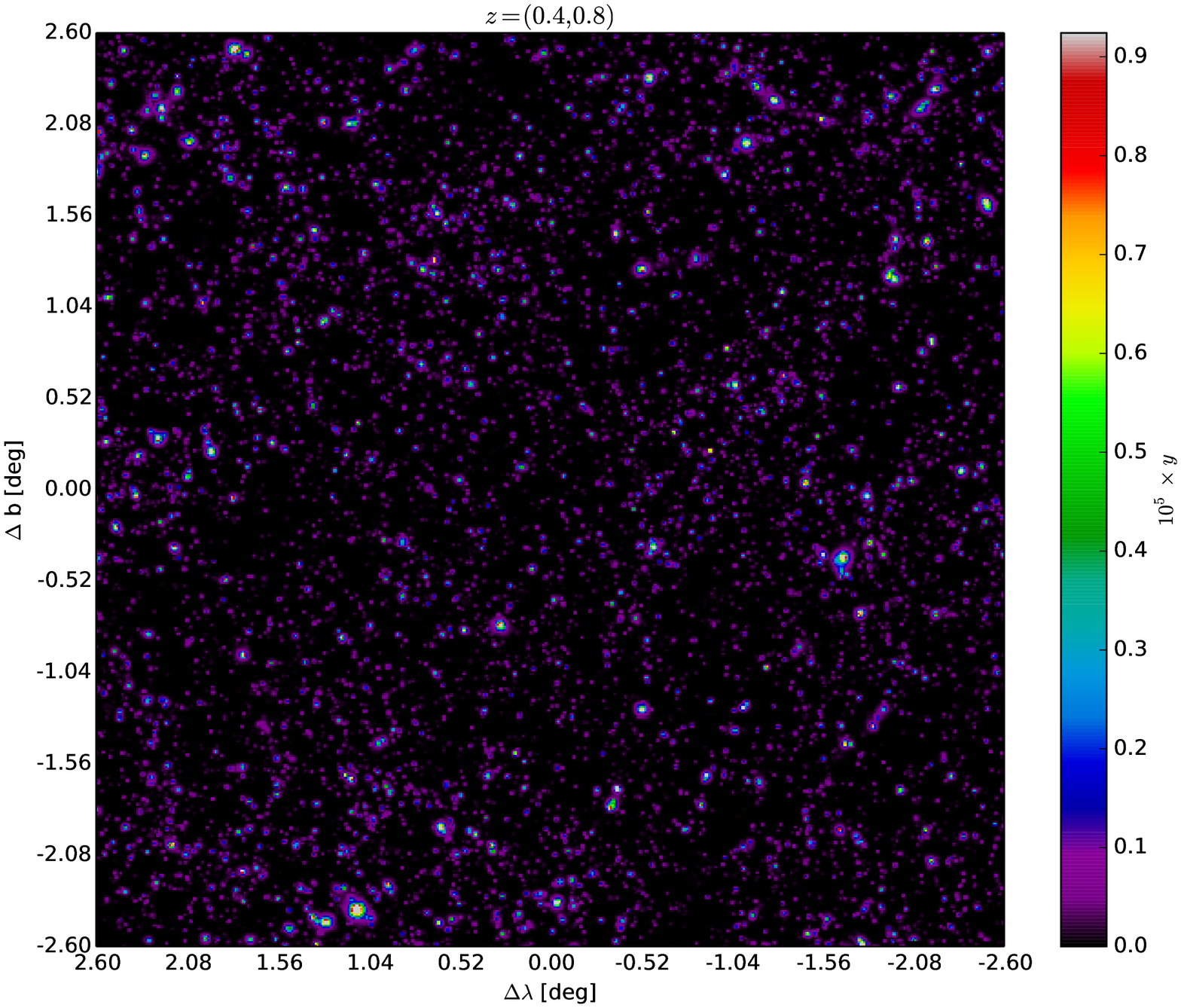}
\includegraphics[width=0.35\textwidth]{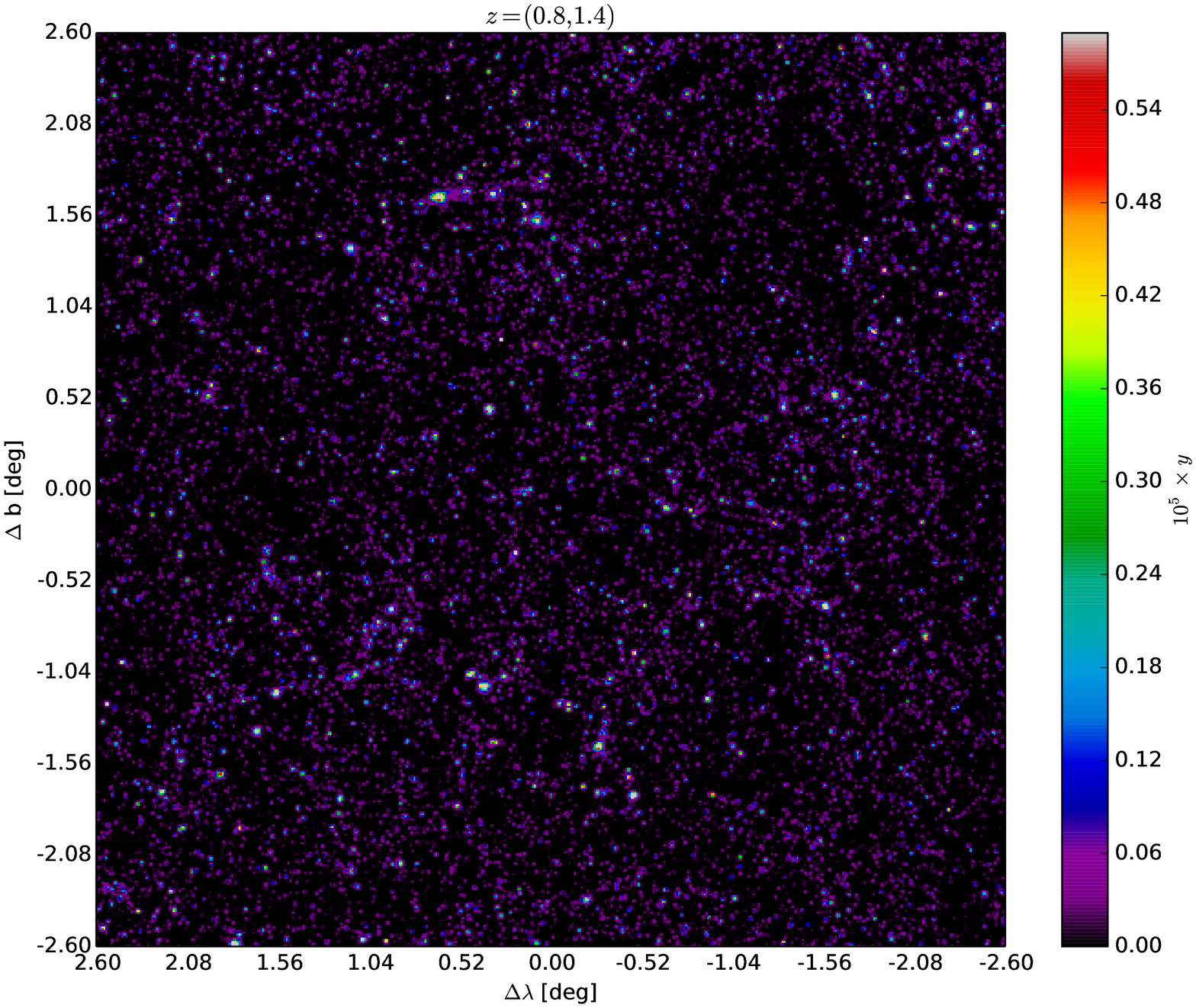}
\includegraphics[width=0.35\textwidth]{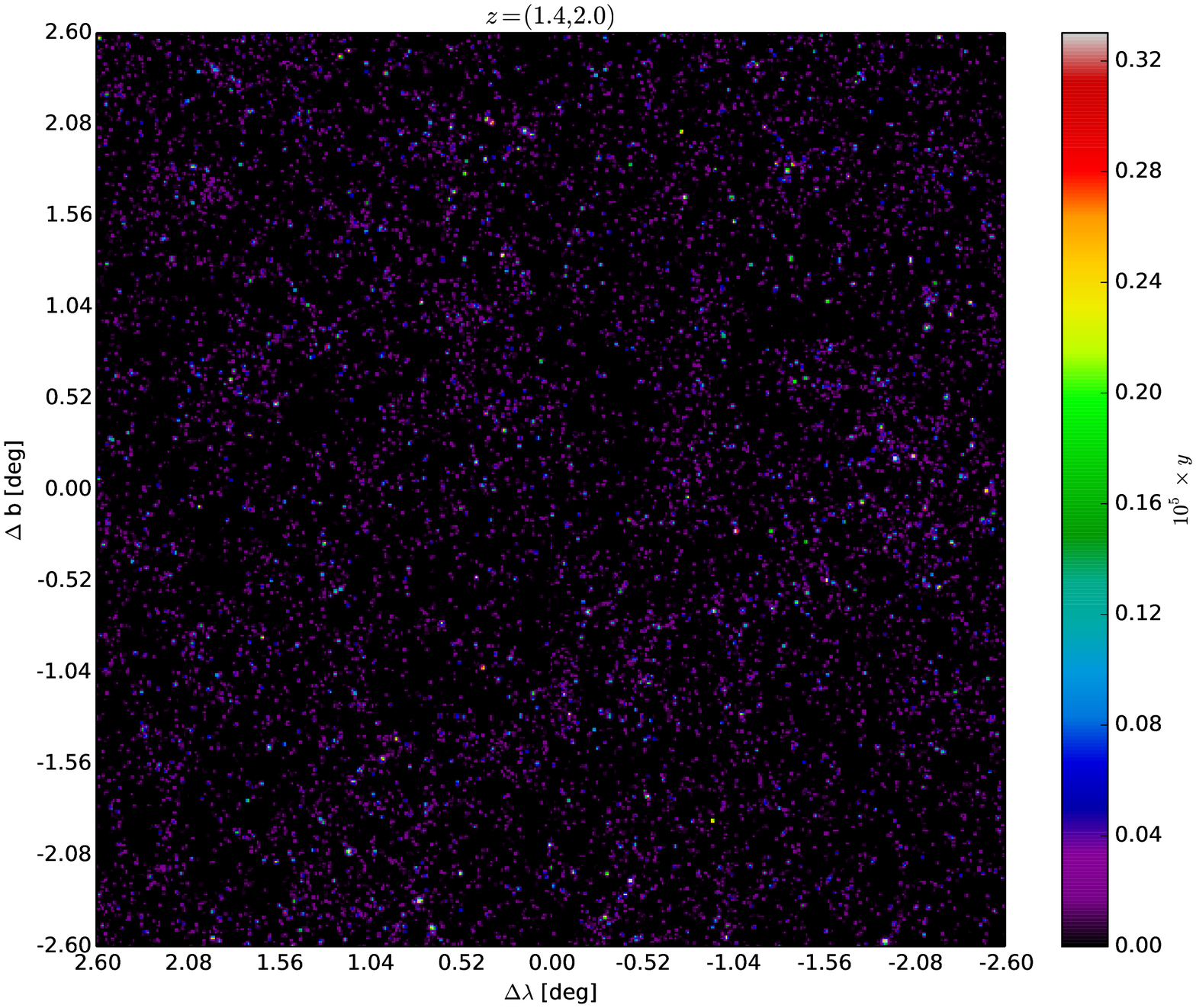}
\includegraphics[width=0.35\textwidth]{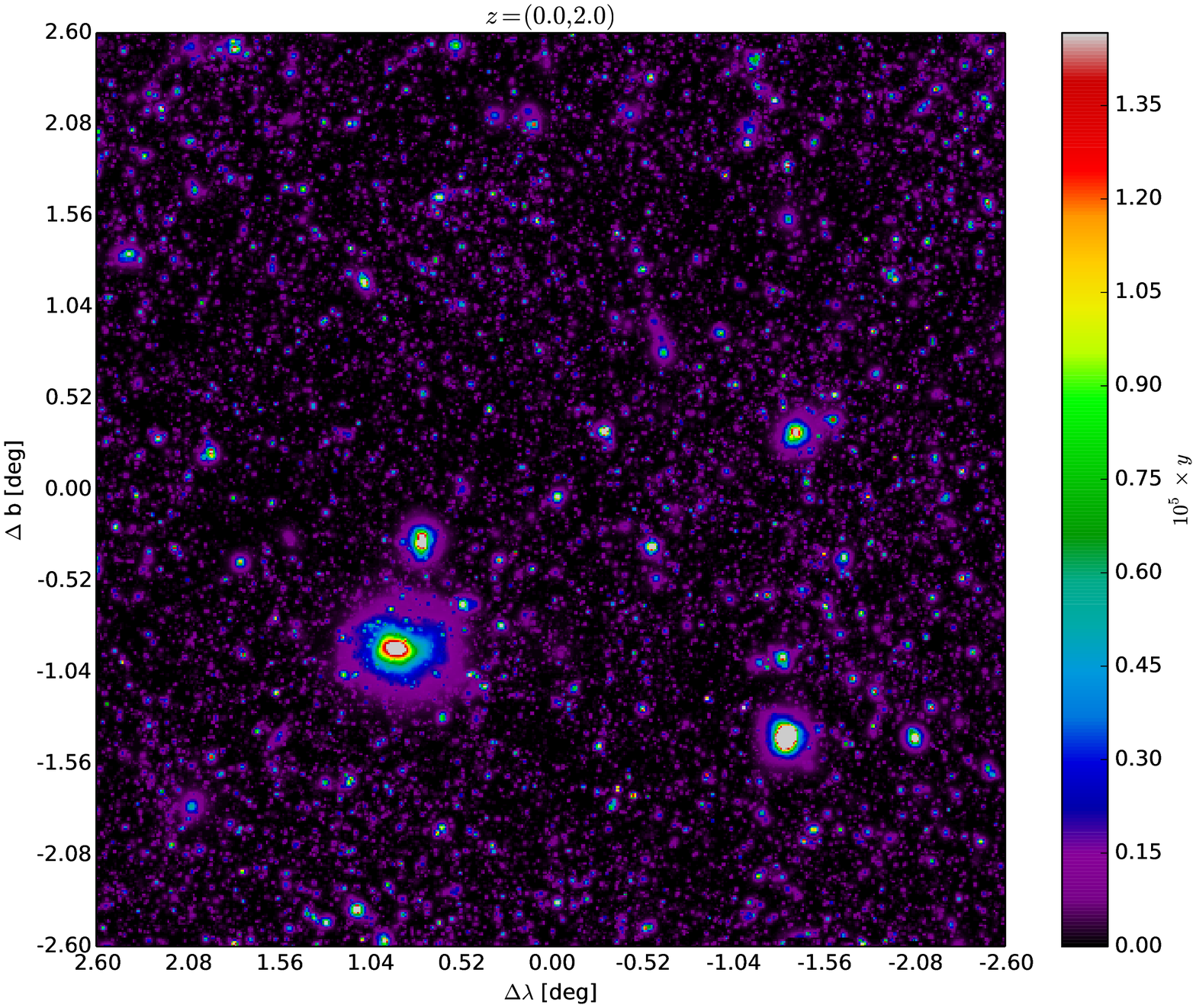}
\caption{Simulated deep field maps, rebinned for display.
{\it Top row}: CMB fluctuations without ({\it left}) and
with ({\it right}) the dipole component for field centre \lb{135}{40}.
The axes are $\pm 2.6^\circ$ 
about the field centroid, in
Galactic coordinates. 
{\it Second to fourth rows}: simulated realisation of the TSZE
signal from different redshift ranges, with the colour map scaled to
emphasise the lowest 25\% of the signal, with stronger signals
saturated in the white. The maximum Compton $y$-parameter value in
this realisation is $y \approx 5.9\times 10^{-5}$. The pixel scale,
of $\sim 1.9$~arcsec is well below the FWHM of the telescope beams
(table~\ref{tab:RTspecs}). The high-resolution simulation has been
rebinned for display preserving the maximum $y$ per bin, rather than
the bin average, in order to display the thermal energy content per
cluster in each redshift range.  }
\label{fig:CMBSZmaps}
\end{figure}

It is crucial to know the properties of the signals sought before
designing the software that will extract astrophysical signals from
noisy and incomplete radio-survey data. 
Therefore, one of the basic results of this work is a set of
high-resolution maps ($10^4\times 10^4$ pixels at about
1.9~arcsec/pixel) that include 
contributions from CMB fluctuations, cluster TSZEs, and point sources.
Sample simulated CMB and TSZE fields contributing to the frequency maps
of a deep field are shown in figure~\ref{fig:CMBSZmaps}.
The top row shows the simulated CMB field with and without the CMB
dipole, which takes the form
$T_d(\mathbf{\hat n}) = A \cos(\mathbf{\hat d} \cdot \mathbf{\hat n})$
with $A=3.346$ mK, and $\mathbf{\hat{d}}=$\lb{264.26}{48.22} 
\citep{Bennett2003}. The significant difference between these two
panels arises from the gradient in intensity across the field caused
by the dipole, and the choice of field direction --- \lb{135}{40}, chosen  
to lie at high galactic latitude and declination, to ensure year-round
visibility and high elevation from the vicinity of Toru{\'n}.

In a small survey it is unlikely that an extremely large TSZE, from a
very massive cluster, will be found. Using the eleven stacked
simulations we generated seven TSZE maps for each of the three 
frequencies of interest (table~\ref{tab:RTspecs}) by random
permutations of the simulation box order, particle shifts, and
coordinate transformations (Section~\ref{sec:lss}). The example shown
in figure~\ref{fig:CMBSZmaps} is typical of the set, and shows
the typical maximum TSZE that should be expected in a blind survey of
this size, $y \lesssim 1.0\times 10^{-4}$. 

\begin{figure}[!t]
\centering
\includegraphics[width=0.49\textwidth]{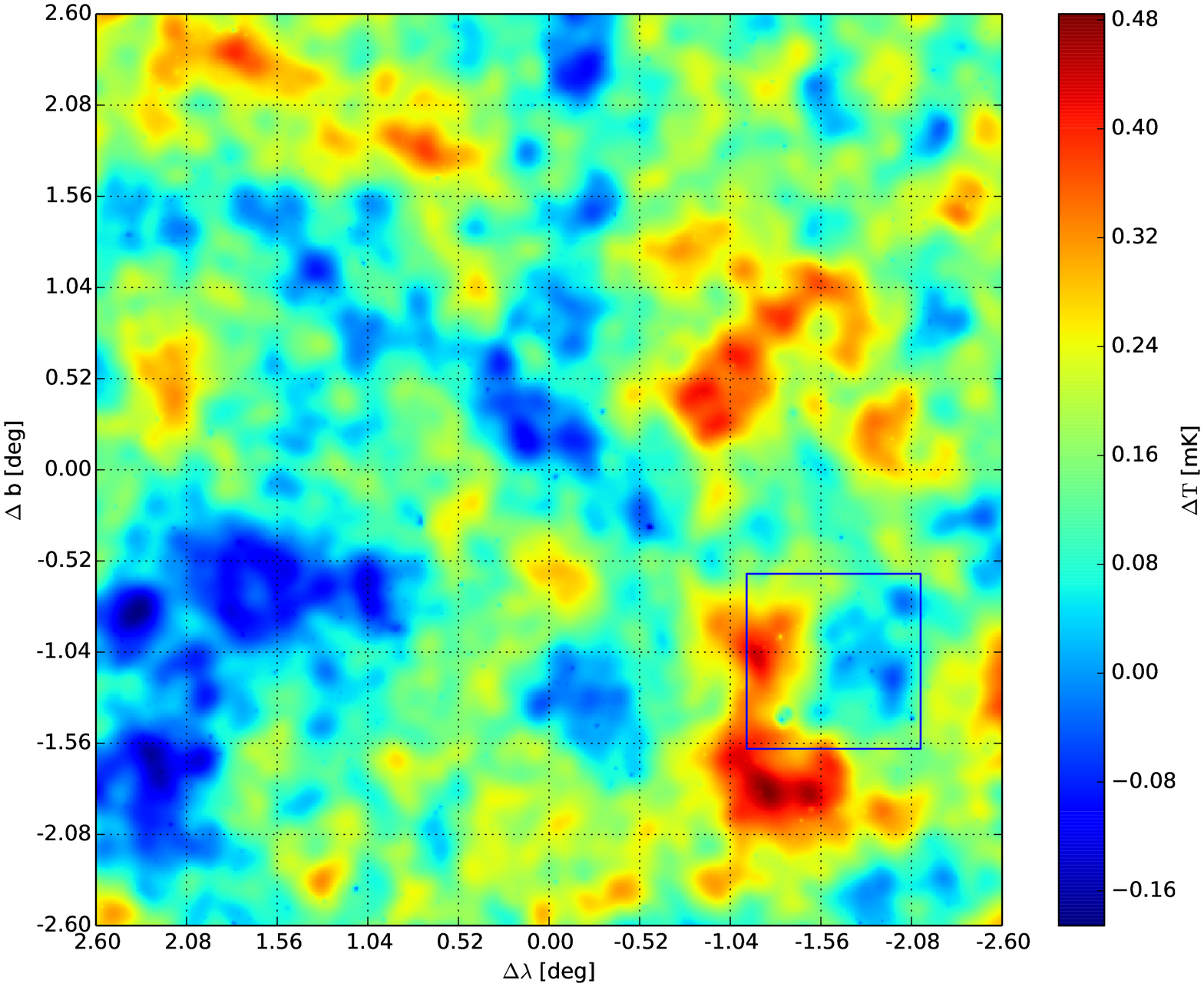}
\includegraphics[width=0.49\textwidth]{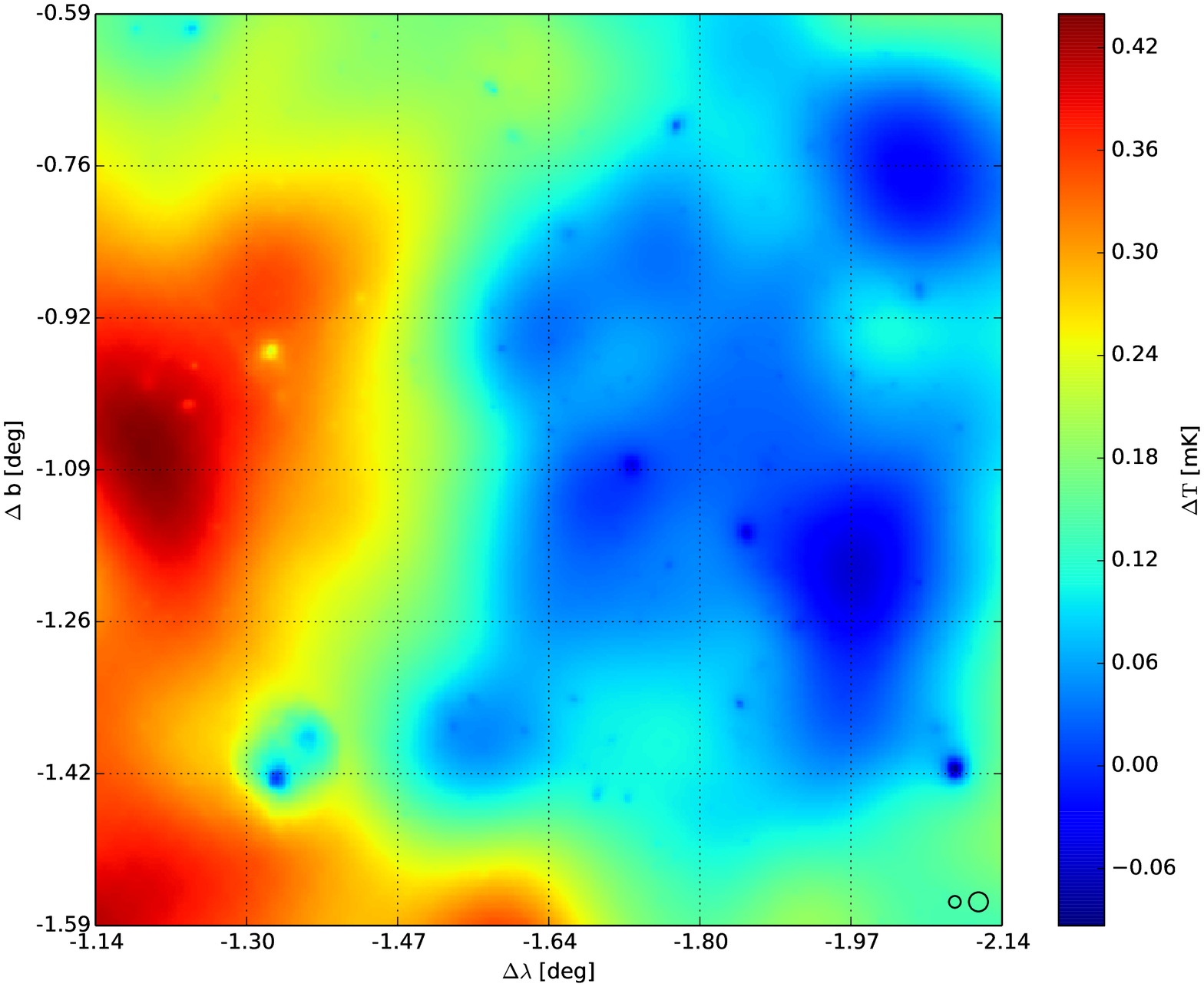}
\includegraphics[width=0.49\textwidth]{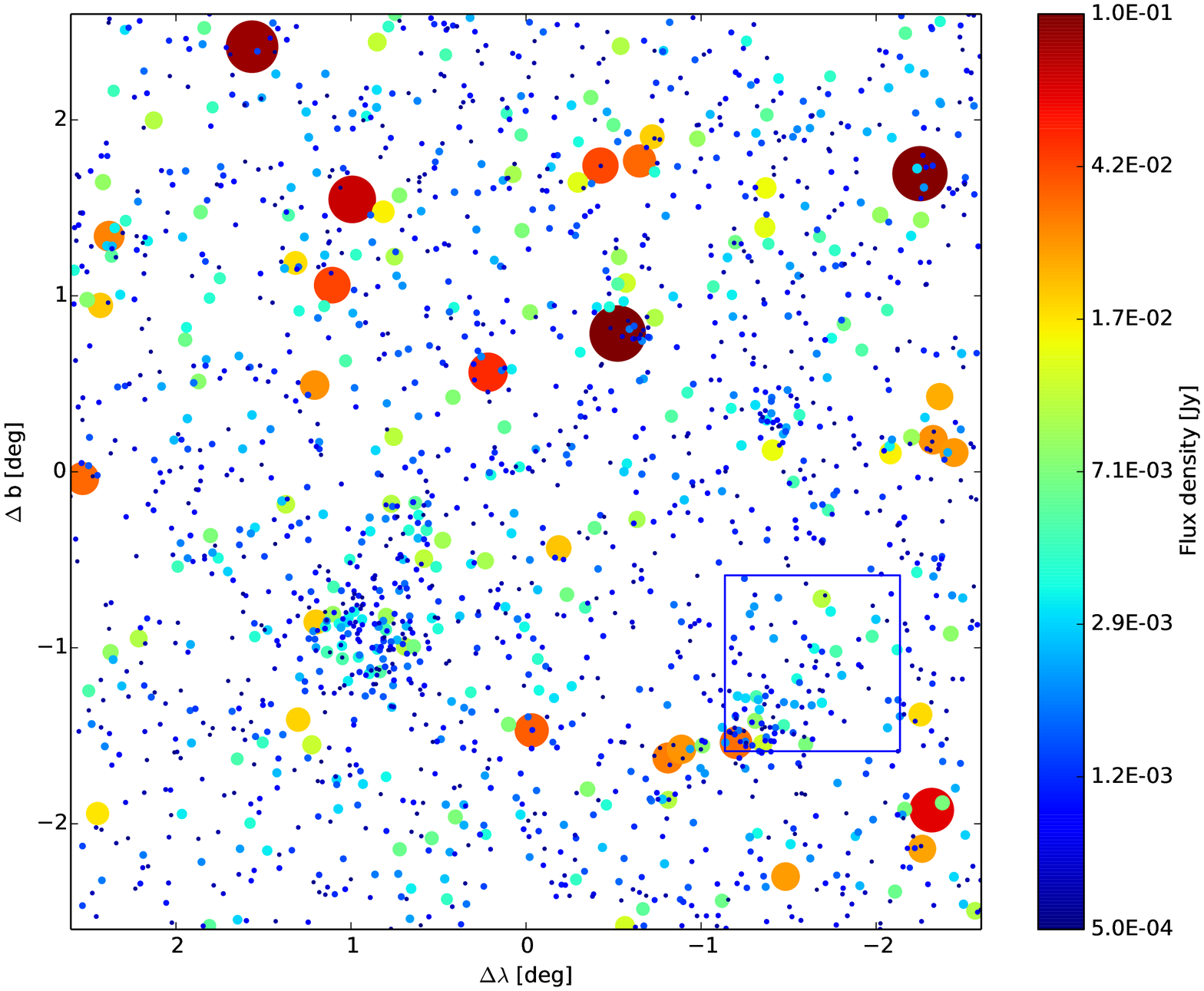}
\includegraphics[width=0.49\textwidth]{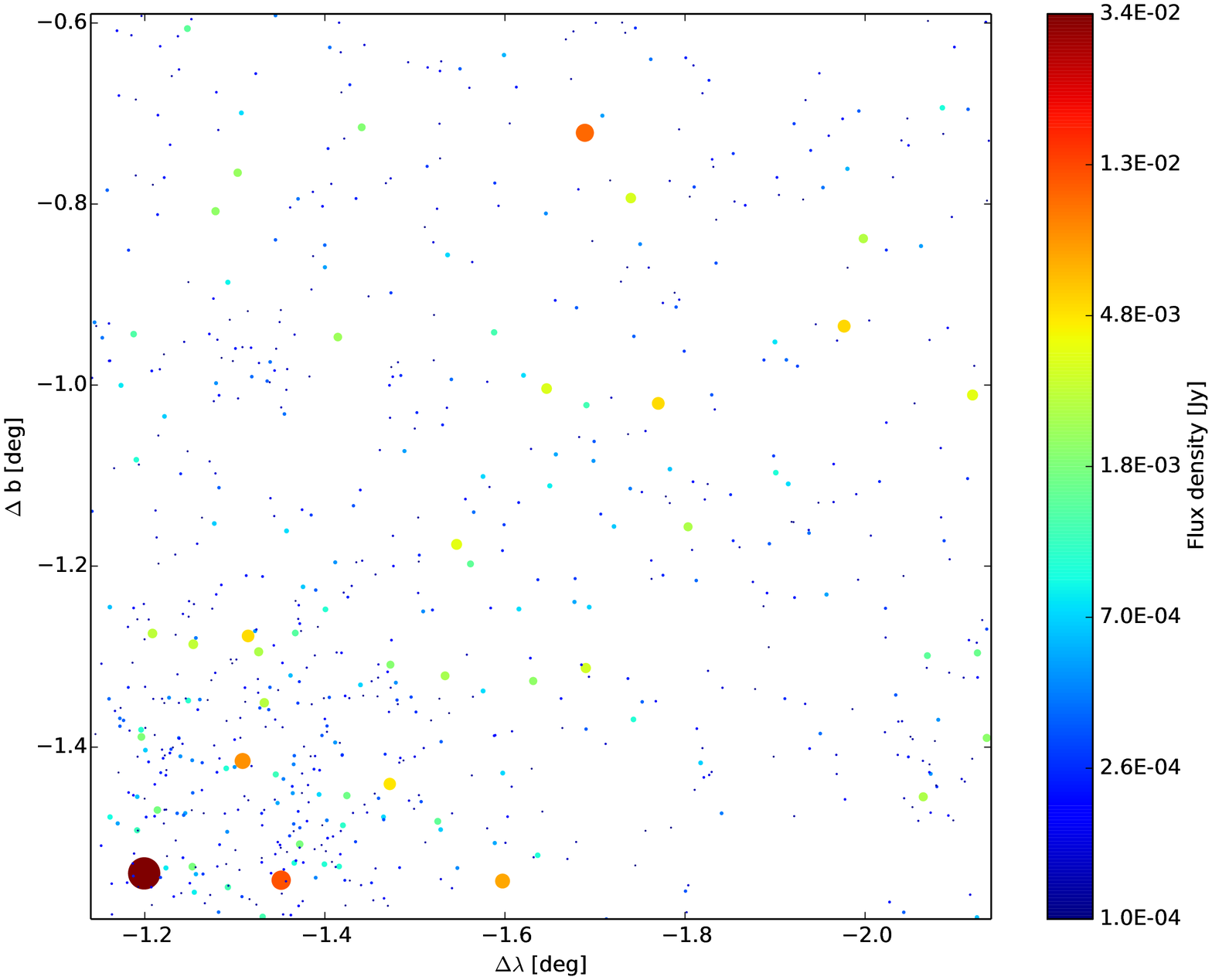}
\caption{Simulated deep field maps, rebinned for display. 
{\it Upper row}: a realisation of CMB fluctuations and the 30-GHz TSZE
over a $5.2^\circ \times 5.2^\circ$ field ({\it left}), with a zoom
view of the $1^\circ \times 1^\circ$ region marked by a square
({\it right}). In the right-hand panel the large and small circles
indicate the FWHM beam-sizes of the 30-GHz OCRA-f receiver 
on the 32-metre telescope and the 15-GHz RTH array on the RTH.
The axes are as in figure~\ref{fig:CMBSZmaps}.
{\it Lower row}: a realisation of the 30-GHz point sources in the
field, taking account of the correlations with the galaxy
clusters. The left and right panels correspond to the same fields
shown in the left and right panels above. 
The range of flux densities is indicated by the symbol size (linearly
mapped to flux density) and the colour scale (logarithmically
mapped). 
Only sources with flux density larger than $0.5\, \mathrm{mJy}$ are
shown in the lower-left panel, but the symbol size scale is 
the same as in the lower-right panel.}
\label{fig:CMBSZsrcMaps}
\end{figure}

Figure~\ref{fig:CMBSZsrcMaps} (top row) shows the combined CMB and TSZE
signals. TSZEs are seen in these images because they typically appear
on small angular scales, where the CMB fluctuations tend to be weak. 
The RT32/OCRA-f and RTH/15 GHz beam sizes are much smaller than the 
angular scale scale of strong CMB background fluctuations
(figure~\ref{fig:CMBSZsrcMaps} right panel). An example point source
realisation generated for this particular deep field is shown in the 
second row of the figure: the flux densities of simulated sources
span more than three orders of magnitude. The spatial correlations due
to point source clustering around heavy halos is evident (see the
lower-right panel), even though we show only sources stronger than
0.5~mJy.

The number of galaxy clusters that can be detected in a blind survey
depends on two main factors. The first is the angular size of the
telescope beam: the presence of clusters with smaller scales will tend
to be suppressed by the noise. This translates into
a high-redshift cut-off in the survey. The same effect is
intrinsically present in the cluster counts: cluster evolution means
that at high redshifts clusters are less evolved, smaller, less
massive, and provide weaker TSZE signals. The second factor is the
survey area, and this is limited by the total integration time,
receiver noise, and the size of the radio camera. Given the
high-redshift cut-off from angular resolution effects, the number of
detected clusters can be increased by increasing the survey area. 
This will pick up more low-redshift clusters, which are 
more massive, well resolved by the beam, and yield larger total TSZE
flux densities. Larger telescopes provide better sensitivity, but in
smaller beams, and so can take more time to cover the same sky area as
smaller telescopes. It is for this reason that large radio cameras are
highly effective, and even necessary. Larger telescopes could also use
lower frequencies, and hence produce larger beams, usually with lower
atmospheric noise. However, using a lower frequency comes at the cost 
of decreasing the TSZE intensity.

For example, a change of observation frequency from 30 GHz to 15 GHz
and a dish size from 32 meters to 100 meters results nearly in 10-fold
decrease in TSZE flux density, where a factor of $\frac{\theta_b(15
\mathrm{GHz})}{\theta_b(30 \mathrm{GHz})}\approx 1.56^2\approx 2.4$ is
due to the beam solid angle decrease (even though the observation
frequency decreased) and a factor of $\frac{f(30 \mathrm{GHz}) g(30
\mathrm{GHz})}{f(15 \mathrm{GHz}) g(15 \mathrm{GHz})}\approx 3.86$ is
due to TSZE intensity change.  This is clearly seen in the
figure~\ref{fig:SZthetaSz} where the TSZE flux density in RTH case is
statistically about an order of magnitude smaller than in the case of
the 32-metre telescope. However, in case of a small-dish survey the
effective number of detected TSZE clusters is strongly limited by
point source confusion.
This is quantified in Figure~\ref{fig:TSZcounts} and the following
sections.  In the figure the effect of beamwidth change within the
same frequency is clearly seen: the ratio of RTH to RT32 beam solid
angles (Table~\ref{tab:RTspecs}) at 30 GHz is $\sim 10$ which directly
translates onto the integrated flux density distributions.

\subsection{Sensitivity limits}
\label{sec:surveys}
Table~\ref{tab:surveys} summarises the theoretical (radiometer
equation) sensitivity levels reached by
six combinations of telescopes and receivers (see
table~\ref{tab:RTspecs}) for several distinct surveys: the SQuare
Degree Field survey (\onedegfield), $\pi$-steradian SKY survey
({\PIsteradians}), and the RTH Deep Field survey ({\RTHdeepfield}).

A SQuare Degree Field survey is planned for winter 2014/2015 using the
30-GHz (OCRA-f) and 22-GHz radiometers on the RT32 in Poland. Such
a survey is a logical first component of many survey campaigns, which
may later contain a number of $1-\rm deg^2$ tiles, and 
is likely to be dominated by low flux density sources. Strong TSZE
clusters and radio sources are unlikely to appear in small sky 
areas, and rare objects are subject to strong Poisson noise. 
We calculate predictions for two scan line densities which are
expressed in number of beams per output map pixel. This can be thought
of as an average number of times that a single beam centre visits a
beamwidth-sized sky stripe. The same effective number of visits is
assumed in cross(angle)-scans (i.e., $2^2$~beams/pixel implies a total
of 4~visits).
For comparison, we also provide predictions for a
15-GHz survey with the 100-m \RTH telescope (RTH) equipped with a
large radio camera. We emphasise the advantages of the increase in
antenna size in table~\ref{tab:surveys} by also predicting the 
sensitivities and number counts that would be obtained when the
receivers are swapped between telescopes. The sensitivity limits
in terms of cluster counts resulting from a {\onedegfield} survey are
shown in figure~\ref{fig:TSZcounts}, and in terms of radio sources in
figure~\ref{fig:ptSrcCounts}. The $3\sigma$ confidence ranges for TSZE
counts, and the $5\sigma$ range for source counts, are marked by the
coloured shaded areas. 

The {\PIsteradians} and the {\RTHdeepfield} are the planned
RTH 15-GHz wide and deep surveys respectively. The wide survey is
designed to reach mJy flux density detection limits within one year of
typical observation (see table~\ref{tab:surveys} for time efficiency
estimates). The {\RTHdeepfield} should reach $100\mu\mathrm{Jy}$ flux
density at SNR $\approx 5$ in the same time. The $5\sigma$ flux
density thresholds, with the corresponding confidence regions 
are plotted in figure~\ref{fig:ptSrcCounts} for the 
with green and blue shaded regions for the {\PIsteradians} and
{\RTHdeepfield} surveys, respectively.

In practice the theoretical sensitivity limits provided in
table~\ref{tab:surveys} are likely to be optimistic, since they ignore
the effects of receiver gain instability and sources of systematic
error.

\begin{table}
\caption{\label{tab:surveys} Sensitivity limits and
predictions for radio source and TSZE counts for the
instrumental configurations in table~\ref{tab:RTspecs} for
three surveys. Each survey involves one year of 
non-continuous observation with realistic
efficiency (see table description details). }
\Beginruledtabular
\begin{tabular}{rcccccc}
& \multicolumn{3}{c}{32-metre radio telescope (RT32)}& \multicolumn{3}{c}{100-metre {\it Hevelius} radio telescope (RTH)}  \\
Band                    &Ku & K & Ka &Ku & K & Ka\\
Central frequency [GHz] &15 &22 & 30 &15 &22 & 30\\
Number of receivers     &49 &1  &  4 &49 &1  & 4 \\
Time efficiency\footnotemark[10] [\%]& $\sim 16$& $\sim 16$& $\sim 16$& $\sim 16$& $\sim 16$& $\sim 16$\\
\hline
&\multicolumn{6}{c}{\onedegfield\footnotemark[1]} \\
Radiometric flux-density limit ($3\times \mathrm{RMS}$)\footnotemark[1]$^,$\footnotemark[2] [mJy] 		
&$0.019_{-0.005}^{+0.007}$	&$0.63_{-0.32}^{+0.33}$	&$0.52_{-0.16}^{+0.20}$	
&$0.006_{-0.002}^{+0.002}$	&$0.18_{-0.09}^{+0.010}$	&$0.13_{-0.04}^{+0.05}$\\
Clusters count (TSZE only) ($3\times  \mathrm{RMS}$)\footnotemark[3] [$\mathrm{deg^{-2}}$]
& $8_{-3}^{+4}$ &  $0.04_{-0.02}^{+0.13}$ & $0.08_{-0.04}^{+0.09}$ 
& $9_{-4}^{+5}$ & $<0.06$ & $0.03_{-0.03}^{+0.03}$\\
Effective clusters count ($3\times \mathrm{RMS}$)\footnotemark[4]  [$\mathrm{deg^{-2}}$]
& $<0.03$ & $< 0.03$ & $0.03_{-0.01}^{+0.02}$ 
& $<4$ & $< 0.05$ & $0.02_{-0.02}^{+0.05}$ \\
Point source confusion limit ($95\%$ CL)\footnotemark[5] [mJy] 		
&$26\pm 6$	&$10\pm 3$	&$4\pm 1$	&$2.1\pm 0.3$	&$0.8\pm 0.2$	&$0.22\pm 0.06$\\
Point source confusion limit ($68\%$ CL)(\footnotemark[5]) [mJy] 		
&$4.9\pm 0.5$	&$1.3\pm 0.1$	&$0.39\pm 0.06$	&$0.11\pm 0.02$	&$<0.1$	&$<0.1$\\
Radio source count ($5\times \mathrm{RMS}$)\footnotemark[6] 
& $\sim 2180$\footnotemark[7] (6) & $29(22)_{-13}^{+43}$ & $33_{-11}^{+17}$ 
& $\sim 9100$\footnotemark[7] (571) & $123_{-54}^{+192}$ & $164_{-55}^{+84}$\\
&\multicolumn{6}{c}{{{\RTHdeepfield} (60 $\mathrm{deg}^2$)\footnotemark[6]}} \\
Radiometric flux-density limit ($3\times \mathrm{RMS}$)\footnotemark[2]$^,$\footnotemark[6] [mJy] 		&	&	&	&$0.06_{-0.02}^{+0.02}$	& &\\
Cluster count (TSZE only) ($3\times  \mathrm{RMS}$)\footnotemark[3]$^,$\footnotemark[9] & &  &  & $7_{-4}^{+5}$ &  & \\
Effective cluster count ($3\times \mathrm{RMS}$)\footnotemark[4]$^,$\footnotemark[9] &  &  &  & $< 1.5$ &  &  \\
Radio source count ($5\times \mathrm{RMS}$) [$10^3$] &&&&$34_{-13}^{+19}$ && \\
&\multicolumn{6}{c}{{\PIsteradians\footnotemark[6]}} \\
Radiometric flux-density limit ($3\times \mathrm{RMS}$)\footnotemark[2]$^,$\footnotemark[6] [mJy] 		&	&	&	&$0.79_{-0.20}^{+0.31}$	& &\\
Cluster count (TSZE only) ($3\times \mathrm{RMS}$)\footnotemark[3] & & & &$\gtrsim 676^{+612}_{-321}$\footnotemark[8]\\
Effective cluster count ($3\times \mathrm{RMS}$)\footnotemark[4] &  &  &  & $ O(10)$ &  &  \\
Radio source count ($5\times \mathrm{RMS}$) [$10^3$]&&&&$294_{-105}^{+146}$&& \\

\end{tabular}
\Endruledtabular
\footnotetext[1]{Predictions for a one-year survey with coverage
$1.5^2=2.25$ beams per pixel.} 
\footnotetext[2]{The flux density sensitivity threshold
($S_{\mathrm{min}}^{\mathrm{th}}$) for a one-year survey using
radiometer parameters as specified in table~\ref{tab:RTspecs}.  The
uncertainties correspond to $3\sigma$ confidence ranges and include
seasonal variations of antenna sensitivity and seasonal and elevation variations
in $T_{\mathrm{sys}}$. TSZE results are given for  $3\times
\mathrm{RMS}$ sensitivity limits, and ignore point source confusion.  
Radio source survey results are quoted at $5\times \mathrm{RMS}$
sensitivity, but the predictions again do not account for
confusion.
}
\footnotetext[3]{Naive cluster count above the sensitivity limit
($S_{\mathrm{min}}^{\mathrm{th}}$) due to TSZE from a single halo
(i.e., neglecting radio sources and halo-halo LOS alignment).} 
\footnotetext[4]{The effective cluster count above the sensitivity
limit ($S_{\mathrm{min}}^{\mathrm{th}}$) including the effects of
limited angular resolution, LOS alignment and radio
source confusions. For the cases where the $S_{\mathrm{min}}^{\mathrm{th}}<100\mu$Jy 
the reported upper limits on TSZE counts prediction do not account for the extra effects 
of radio sources below $100\mu$Jy (section.~\ref{sec:flux_density}).}
\footnotetext[5]{The 95\% (68\%) upper tail sensitivity limits due to
flux density confusion for $\delta_{\mathrm{conf}}=0.1$ (see
equation~\ref{eq:radioSrcConfusionDelta} and
figure~\ref{fig:ptSrcFluxConfusion}) along with bootstrap errors. }
\footnotetext[6]{Predictions for a one-year survey with a coverage of 4 beams per pixel. The errors include
the extent of possible changes in the overall system performance due to elevation and seasonal variations. 
If the confusion limit is higher than the derived TSZE survey limit then the effective (68\% CL)
confusion-limited source count is given in brackets. }
\footnotetext[7]{This exceeds the number of beams in the
survey area. The source count at such low flux densities is
uncertain (section~\ref{sec:PtSrcSurveys}). }
\footnotetext[8]{The wide field prediction resulting from FOF mass function estimates for the clusters
massive enough to yield a detectable TSZE (see figure~\ref{fig:SZthetaSz} bottom-right panel). It should be regarded as a lower
bound due to assumed survey redshift depth $z<1$
(see section~\ref{sec:PIsteradiansFieldPrediction}). The $1\sigma$
uncertainties result from sample variance. The effective cluster count is however significantly smaller 
if point source impact is taken into account (see Section~\ref{sec:PIsteradiansFieldPrediction}).}

\footnotetext[9]{Predictions from the \onedegfield. The
numbers resulting from the statistics of the FOV of size $\sim
5.2^\circ \times 5.2^\circ$ are scaled for the size of the
{\RTHdeepfield}.} 
\footnotetext[10]{The estimated overall time efficiency $\eta_t =
\eta_w\, \eta_s\, \eta_{\mathrm{RT}}\, \eta_v $ includes (i) a weather
efficiency $\eta_w\approx 0.25$ for the fraction of time with
cloudless or near-cloudless sky conditions; (ii) a scan efficiency
$\eta_s\approx 0.7$ that accounts for the fraction of time used
in manoeuvring and depends on the scan strategy; (iii) a telescope
efficiency $\eta_{\mathrm{RT}}=0.9$ excluding service time; and  
(iv) a visibility efficiency $\eta_v=1$ giving the
fraction of time the target lies within the preferred range of
elevations. We assume circumpolar fields.} 
\end{table}

\subsection{Predictions for blind galaxy cluster surveys}
\label{sec:TSZsurveys}
\begin{figure}[!t]
\centering
\includegraphics[width=\textwidth]{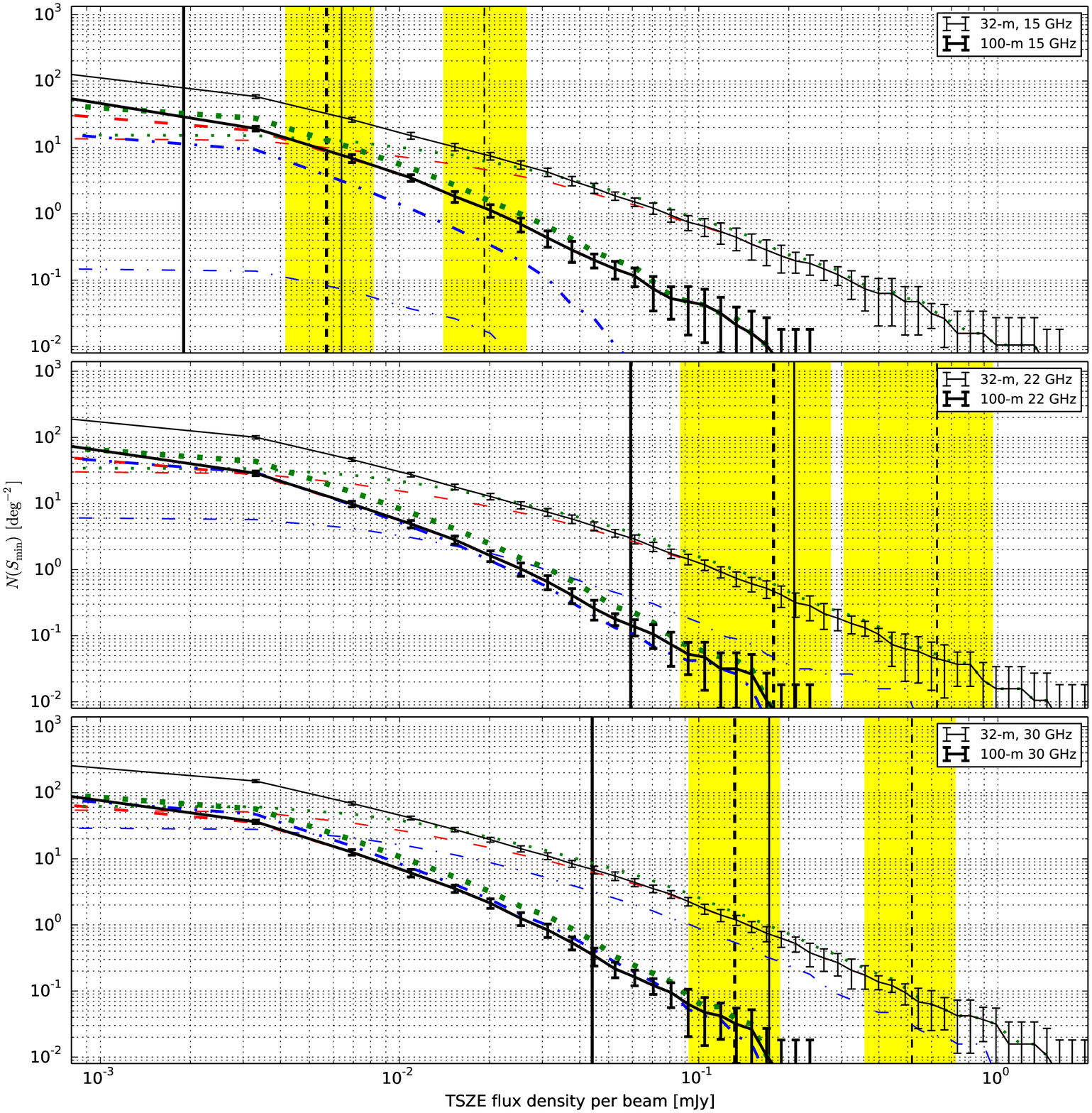}
\caption{Predictions for the number, $N(S_{\mathrm{min}})$, of galaxy
clusters per $\rm deg^2$ detectable through their TSZEs above a
given flux density limit, $S_{\mathrm{min}}$ (eq.~\ref{eq:NSmin}).
The three panels correspond to operation at $15$, $22$, and $30$
GHz. In each panel the thick lines correspond to the cluster count
expected for the 100-m RTH telescope and thin lines correspond
to the 32-m telescope. Solid (black) lines with error bars show
$N(S_{\mathrm{min}})$ due to TSZE only calculated for individual halos, with the $\pm 1\sigma$
sample variance -- ignoring halo alignments and radio source contamination. 
Dashed (red) lines trace the counts after filtering out halos with
angular virial diameter less than the HPBW of the corresponding
telescope/receiver combination. Dotted (green) lines show the
further change caused by TSZE flux density confusion, from the
superposition of clusters near the line of sight. 
Dash-dotted (blue) lines show the additional impact of the
synchrotron radio source population with flux-densities above 100$\mu$Jy, which effectively cancels 
part of the TSZE flux density. The vertical solid lines indicate
the RMS noise level for the telescope/receiver combination, and the   
vertical dashed lines mark the $3\,\times$~RMS flux density level
for candidate TSZE detections. The shaded (yellow) regions
encompass the $\pm 3\sigma$ confidence ranges in the theoretically achievable
$S_{\mathrm{min}}$ corresponding to seasonal and elevation-dependent variations of system
performance. 
}
\label{fig:TSZcounts}
\end{figure}

Figure~\ref{fig:TSZcounts} shows predictions for the {\onedegfield}
survey. The cumulative galaxy cluster count $N(S_{\mathrm{min}})$ 
is defined as
\begin{equation}
N(S_{\mathrm{min}}) = \int_{S_\mathrm{min}}^{\infty} \frac{\d n(S)}{\d S}\d S,
\label{eq:NSmin}
\end{equation}
and is calculated from simulated maps covering a field of about $5.2
\times 5.2\, \mathrm{deg}^2$, recalibrated to a $1\, \mathrm{deg^2}$
survey. We use seven distinct realisations of the field to make a rough
estimate of the variance in expected counts. 
The predictions for TSZE detections for each individual telescope and
receiver configuration are summarised in table~\ref{tab:surveys}. 

Our simulations allow us to investigate the detectability of TSZEs (i)
for individual halos; (ii) for the individual halos in the presence of
the neighbouring halos; and (iii) in the presence of point
sources. Thus we can gauge the effects that halo correlations
and radio source contamination have on the TSZE count from a survey. 
Figure~\ref{fig:TSZcounts} shows how much these effects matter. 

It is clear that using a small dish with a large beam will
preferentially detect clusters with a high TSZE flux density (upper
thin solid lines in figure~\ref{fig:TSZcounts}). However, the smaller
collecting area of a smaller dish also leads to lower flux-density
sensitivity ($S_{\mathrm{min}}$). For our study cases, $S_{\mathrm{min}}^{\mathrm{RT32}} >
S_{\mathrm{min}}^{\mathrm{RTH}}$ (table~\ref{tab:surveys}). The
combination of these effects means that the overall number of TSZEs
that can be detected is similar, if one only counts halos, and
neglects the impact of finite angular resolution and unresolved radio
sources.

The effects of the finite beamwidth on the number of sources detected,
$N(S_{\mathrm{min}})$, are insignificant down to the sensitivity
threshold for RTH, but are important at the lower resolution provided
by RT32. The effects of finite angular resolution (thick and thin red
dashed lines on figure~\ref{fig:TSZcounts}) limit
$N(S_{\mathrm{min}})$ more severely at lower frequencies, as expected. 
These lines result from filtering out all TSZE halos with
angular virial extent smaller than the beamwidth of the 
receiver--telescope configuration.
For all combinations of telescope and receiver, the effects of finite
angular resolution are less than a factor 2, and are usually only a
few percent. 

The impact of TSZE confusion due to LOS halo alignment
is noticeable at all frequencies, but is largely lost in the 
Poisson noise of the various realisations (compare the green
dotted lines and the dashed red lines in figure~\ref{fig:TSZcounts}).  
The green dotted confusion-corrected lines trace the TSZE halo
counts filtered by the angular size as before, but with TSZE
flux-densities integrated from high-resolution maps of specific
intensity at the pixel of individual halo TSZE peak -- hence
accounting for halo-halo LOS alignments.
The slight
apparent increase in the TSZEs count is due to the SZ flux-density redistribution caused by 
integrations  of multiple halos falling into the telescope beam. The increase is 
largest at smallest flux-densities 
and $\lesssim 40$\% ($\lesssim 30$\%) at RT32 (RTH) sensitivity threshold (compare the dashed and dotted lines in figure~\ref{fig:TSZcounts}).

Finally, the impact of the unresolved but spatially-correlated point
sources 
-- the dash-dotted (blue) lines in figure~\ref{fig:TSZcounts} --
is important and strongly frequency dependent. 
As for the dotted green lines,
these counts result from integrating specific intensity maps that
contain both TSZE and radio sources 
above 100$\mu$Jy (section~\ref{sec:unresolved_radio_sources}).

The source-induced TSZE flux density cancellation is
least problematic at high radio frequencies (30~GHz), and is
increasingly important at lower frequencies, as would be expected
due to beam size frequency dependence and
because of the difference between the spectrum of synchrotron
radiation and the TSZE. For example, for the RT32 with a large 15 GHz
radio camera the TSZE flux density in 
all considered surveys
would 
be almost completely cancelled out (thin dash-dotted blue line in the upper
panel of figure~\ref{fig:TSZcounts}). The same camera installed on a
100-m telescope is still significantly affected by the radio sources
(the thick dash-dotted blue line), with at least half the TSZEs lost to
radio-source contamination. 
A TSZE survey at a relatively low frequency, 
such as 15 GHz, is feasible only with a large radio camera
on a large telescope. 
For small telescopes the flux cancellation problem can be
alleviated and TSZE detection rate improved by observing at higher
frequencies as is done in case of OCRA-f instrument.  OCRA-f (30-GHz)
operation on the RT32 loses a similar fraction of TSZEs to
radio-source contamination as the 49-beam, 15-GHz, radio camera on the
RTH. A higher-frequency radio camera on the RTH would, of course, be
more efficient in these terms, but would require more time to complete
a survey to the same depth.

In case of large telescopes -- for a duration-fixed survey --
once the flux-cancellation problem is lifted by using smaller
beamwidths,
an increase of observing frequency and consequently the number of
required beam pointings at the cost of decreasing survey depth may
still be a safer way to pursue than choosing small and deep fields
that can be reached at lower frequencies and with larger beams because in
the latter case TSZE detections are dominated by very low flux
densities ($O(10)\mu$ Jy) where systematical effects induced by gain
instability and atmospheric turbulence (which we do not consider) are
likely to preclude stable flux-density measurements (we will return to
this issue in Section~\ref{sec:freqRTrec}).
Even though TSZEs are stronger at higher frequencies and the
radio source contamination is lower, for a duration-fixed survey,
increasing survey frequency leads to lowering TSZE detection rate
because integration time per beam scales with frequency as:
$t_{\mathrm{int}} \sim \theta_b^{2} \sim \nu^{-2}$ and also the
receiver and atmospheric noises typically increase with
frequency. Hence although the TSZE intensity increases with frequency
by similar amount as the beam solid angle decreases in K and Q bands
(Figure~\ref{fig:SZfluxDiag}), the survey necessarily becomes
shallower leading to lower SZE detection rate. We estimate that a
49-beam, 8-GHz wide bandwidth receiver operating at 30-GHz and 
installed on RTH would
have TSZE survey limit $\sim 4.5$ times larger than the 15-GHz receiver
(see Table~\ref{tab:surveys}) and would effectively detect $\sim 1.3$
clusters per square degree per year -- about 3 times less than 15-GHz
camera -- but the survey would operate at larger flux density levels
(tens and hundreds of $\mu$Jy for {\onedegfield} and {\RTHdeepfield}
respectively).  Widening the survey by means of a multi-year campaign
would increase the detection count while avoiding the risk of
searching weak TSZEs in very small and deep fields.

It is instructive to examine the competing effects on TSZE
detectability in more detail.

\subsubsection{Scale dependence and redshift coverage}
\begin{figure}[!t]
\centering
\includegraphics[width=.49\textwidth]{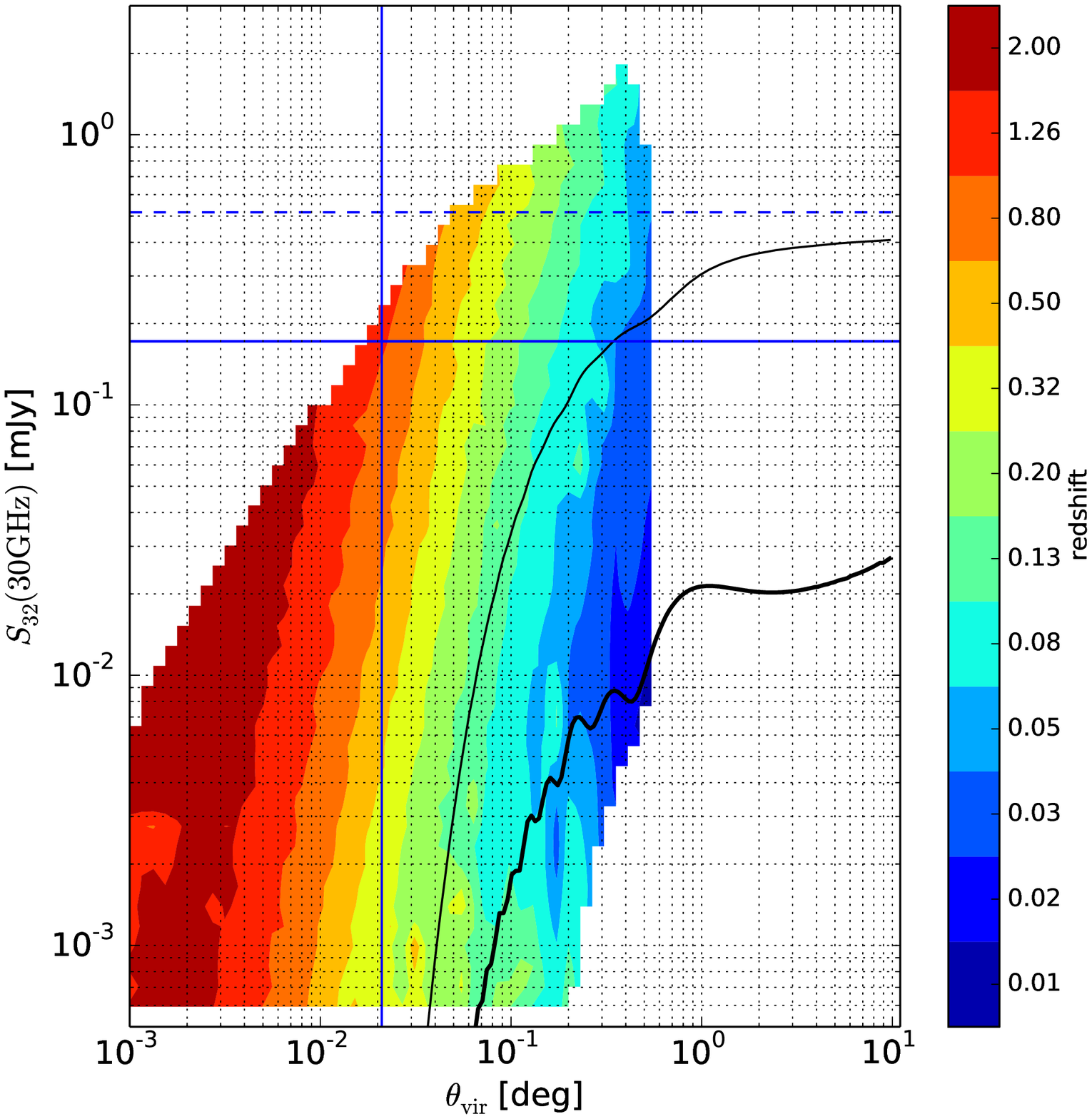}
\includegraphics[width=.49\textwidth]{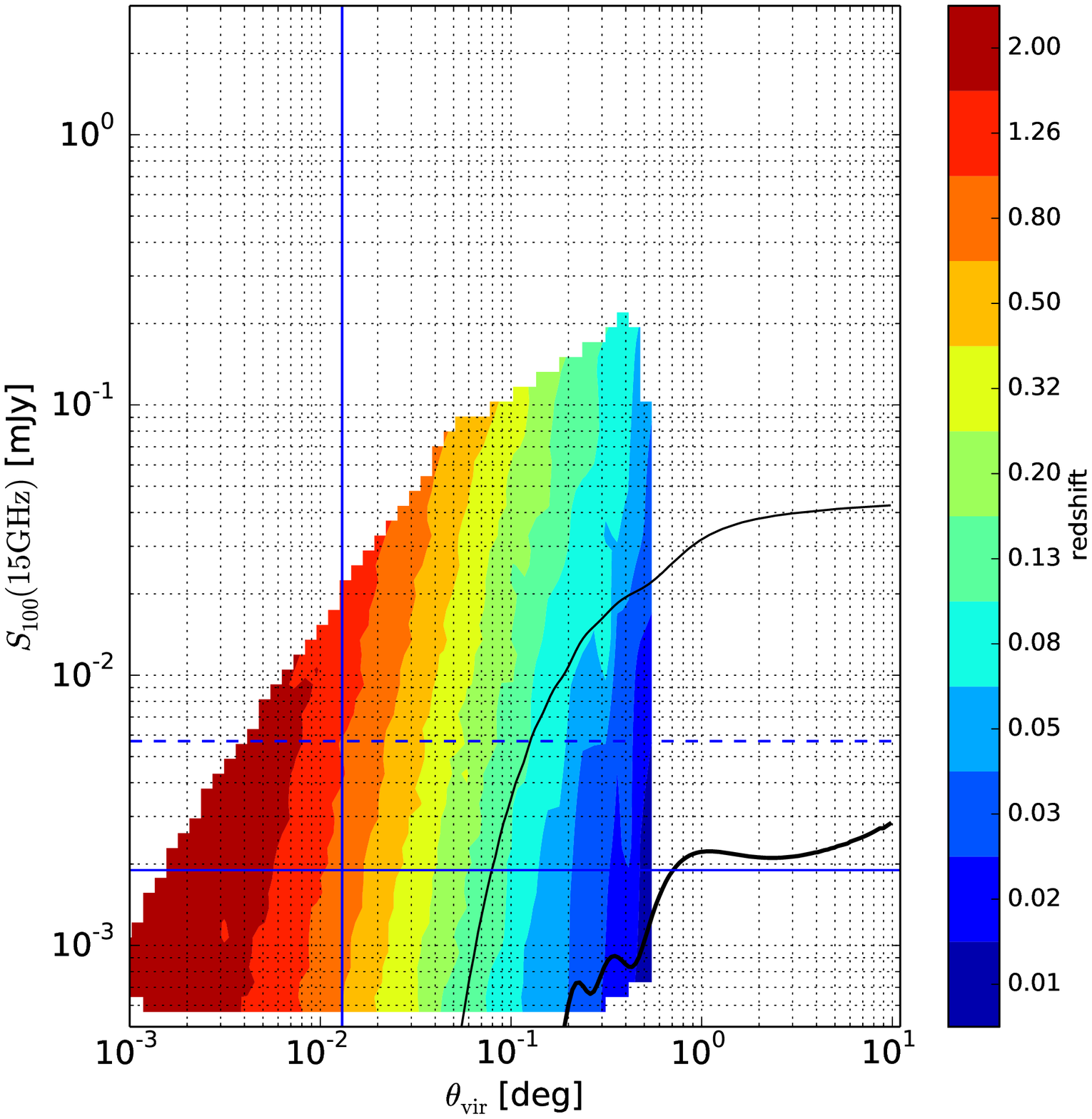}
\includegraphics[width=.49\textwidth]{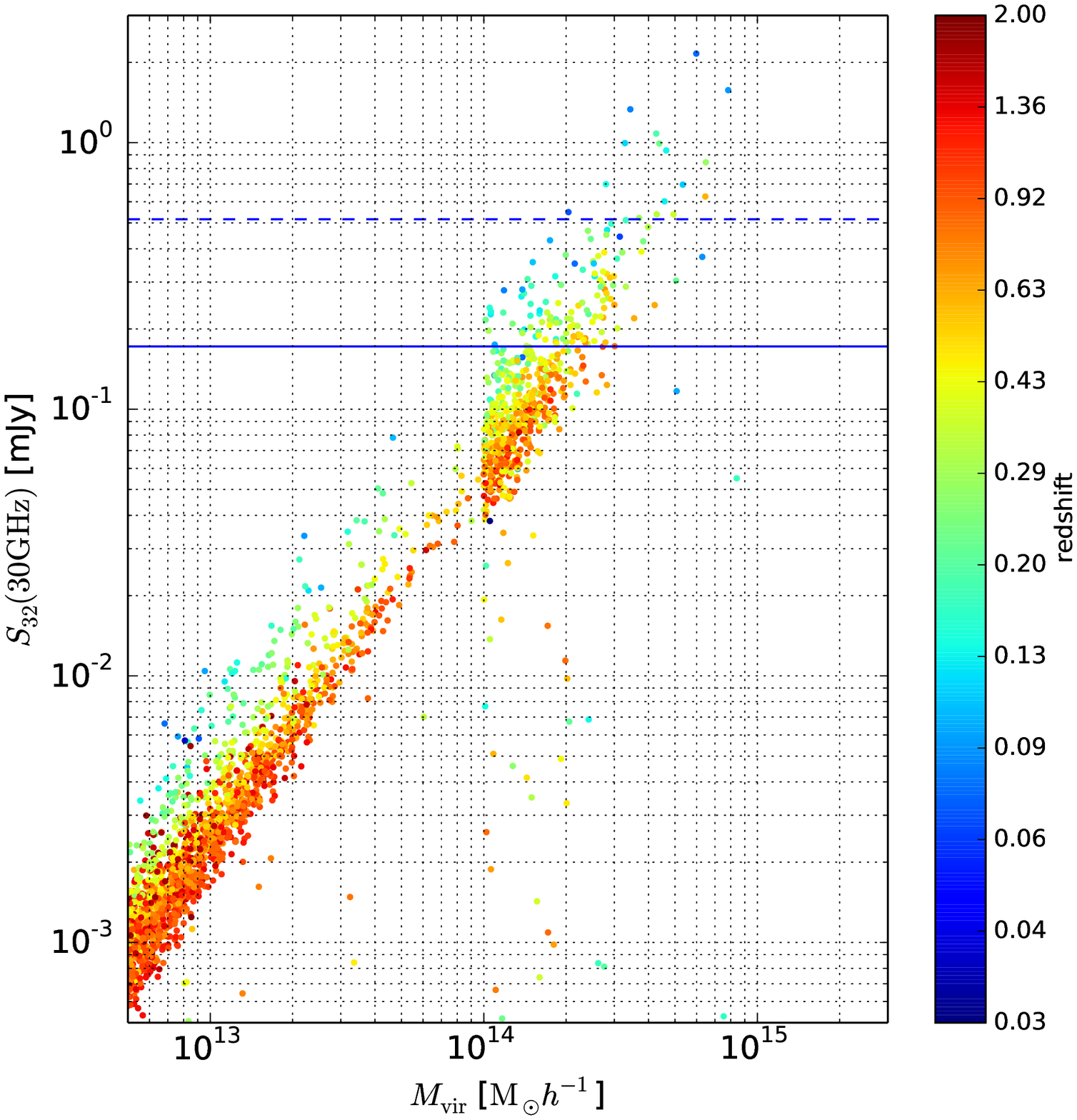}
\includegraphics[width=.49\textwidth]{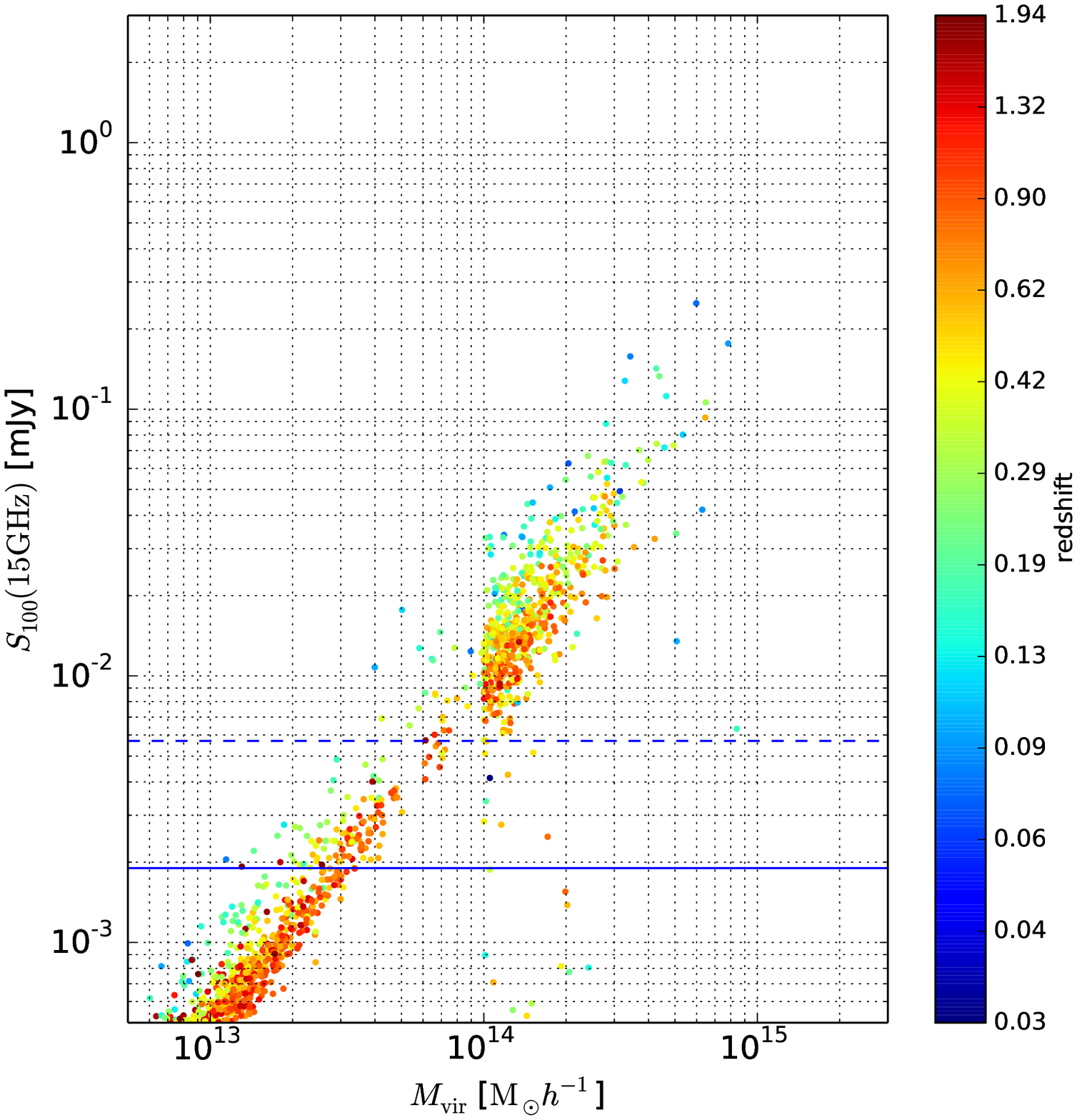}
\caption{Distribution of TSZE flux density per beam from single halos
as a function of halo virial angular size (upper row) and virial
mass (lower row), with redshift indicated by the colour scale. 
The left column shows the halos that could be detected by a 32-m
telescope at 30~GHz. The right column is for a 100-m telescope at
15~GHz. 
Horizontal lines indicate the same flux density thresholds as in
figure~\ref{fig:TSZcounts}. The thin black curves in the top row
indicate the $1\sigma$ rms level of the flux density (per beam)
variation from CMB fluctuations as a function of angular scale (see
text for details). The thick lines indicate the CMB flux density
per beam from all angular scales $< \theta_{\mathrm{vir}}$.
The effects of the halo superposition and point source confusion
are not included.
The filled contours in the upper row show the statistics of the halo
redshift distribution. The lower plots show individual halos.
The distribution of halos in the $\theta_{\mathrm{vir}}-S$ plane is
not uniform: there are more halos with low TSZE flux densities than
high flux densities. For clarity, only 2\% of the halos 
with virial mass of $< 10^{14} M_\odot/h$ are shown in the lower
plots. 
}
\label{fig:SZthetaSz}
\end{figure}

It is of interest to investigate the range of redshifts of clusters
from TSZE surveys, since clusters selected in this way are expected to
cover a wide range of redshifts, and so to provide excellent objects
for testing our understanding of the growth of structure. 
In figure~\ref{fig:SZthetaSz} we show the population of clusters
expected from 32-m/30-GHz and 100-m/15-GHz surveys as functions of
halo mass (and virial angular size), and redshift, based on seven
deep-field realisations. Halo TSZEs are placed on the figure ignoring
confusion from the primordial CMB and halo alignments, but the level
of CMB confusion is indicated as a curve of flux density noise per
beam, calculated from
\begin{equation}
\Delta S(\nu,\theta_b) = [B_\nu(T_{\mathrm{CMB}})- B_\nu(T_{\mathrm{CMB}}+\Delta T)] \int P_b(\hat{\mathbf n}) \d\Omega
\label{eq:fluxDensity}
\end{equation}
where $B_\nu(T_{\mathrm{CMB}})$ is given by eq.~\ref{eq:BBraiance} and
$P_b(\mathbf{\hat{n}})$ is the aperture- and frequency-dependent
instrumental beam profile. The angular dependence of CMB intensity is
approximated by a simple 1:1 mapping between multipole component
$\ell$ and angular scale, $\theta(\ell) = 2 \pi (2\ell+1)^{-1}$, with
the temperature contribution on that scale
$\Delta T_\ell = \left[ \frac{1}{4\pi} (2\ell +1) C_\ell \right]^{1/2}$
where $C_\ell$ is the amplitude of the power spectrum at multipole $\ell$. 
By analogy, the cumulative contribution from all scales less than
$\theta$ is calculated as $\sigma^2_{\ell_\mathrm{min}} =
\sum_{\ell=\ell_\mathrm{min}}^{\ell_{\mathrm{max}}}\sigma^2_\ell$
where $\ell_\mathrm{max}=6000$ is the highest multipole in the
calculated power spectrum. The corresponding CMB noise is plotted in
figure~\ref{fig:SZthetaSz} as black solid lines: for most halos
detectable in the planned surveys this noise is small because most
of the halos lie at high redshift, and so have small angular scales,
where the primary CMB fluctuations become exponentially damped. 

For the RT32/OCRA-f configuration halos with virial angular scales
between 3~and 30~arcmin are the most likely to be detected (at
$3\sigma$ confidence), and these lie in redshift range
$0.03 \lesssim z \lesssim 0.5$. The flux density peak is found at
redshift $z \approx 0.1$, and none of the halos with detectable flux
density has an angular scale smaller than the beam size.

In the case of RTH with the 49-beam Ku-band array, the most likely
virial angular scales are between $0.3$~and 30~arcmin, and clusters
out to $z \approx 2$ could be detected. 
The predictions of the telescope selection functions 
in terms of the detectable
cluster mass and redshift (figure~\ref{fig:SZthetaSz}) are based on a theoretical sensitivity 
limit (section~\ref{sec:surveys}) and the integrated
TSZE flux-densities but without the impact of point sources. 
We did not investigate the impact from individual effects jointly as it would
require extrapolations below the experimentally probed range of
radio source flux-densities (section~\ref{sec:flux_density}).
Note that the bottom panels of figure~\ref{fig:SZthetaSz} do not readily show
the number of clusters above certain flux-density limit because the
plotted $S-M_{\mathrm{vir}}$ scaling relation combines several FOV realisations.

The number of galaxy clusters detected in a blind survey depends
mostly on the observing frequency, the beam size, and the system
sensitivity. At centimetre wavelengths, increasing the observing
frequency makes the TSZE specific intensity larger, but the telescope
beam smaller, on any given telescope. The first factor causes only a
weak redshift dependence (related to cluster temperature evolution),
but the second factor scales with the square of the angular diameter
distance, and hence is a rapid function of redshift. 

Figure~\ref{fig:SZthetaSz} shows that at the virial overdensity,
clusters larger than $\sim 0.5^\circ$ are very rare, and although they
may be very massive, with large total flux densities, they do not
necessarily yield the largest flux density per beam as the TSZE signal
is strongly non-uniform -- peaking at the cluster core and decreasing
towards the cluster outskirts. Such rare objects are not well
represented in our simulations, and more precise predictions of their numbers in
the widest-area surveys requires a much larger simulated field --- a
calculation which we defer to the future. For the present, we note
only that the sampling noise at the lowest redshifts is significant,
as is apparent from the small number of clusters with $M_{\rm vir} > 4 
\times 10^{15} M_\odot h^{-1}$. 

The statistics of the halo distribution plotted in Figure~\ref{fig:SZthetaSz} is based on the selected FOV but the we argue that the result
is quite general.
While the flux density and redshift distribution are independent from the FOV, the total number of clusters in the light-cone would change
when the FOV was chosen to be slightly smaller or larger from the one in our simulation,
but the relative proportions of cluster counts at different redshifts should be conserved yielding similar statistics 
under assumption of statistical isotropy.

\subsubsection{The effects of limited angular resolution}
Finite beamwidth leads effectively to a cut-off in the distribution  
of detectable TSZE galaxy clusters and causes flattening in the
cumulative $N(S_{\mathrm{min}})$ distribution in the limit of high
redshift and so small
flux density (red dashed lines in figure~\ref{fig:TSZcounts}). 
The halo filtering criterion depends on the observing strategy used:
beam-switching techniques, for example, also imply a cutoff at low
redshift and so high flux density (\cite{Birkinshaw1999}). Detailed planning
of the survey methodology is required to investigate these effects for
the planned RT32 and RTH surveys, but a fair approximation for the
total number of detectable halos can be obtained by adopting the
criterion that $\theta_{\mathrm{vir}}>\theta_b$: i.e., that the
angular virial size of a halo should exceed the observing HPBW, and
that criterion was used in figure~\ref{fig:TSZcounts}). 

An objection to this procedure may be found from the results of the
\textit{Planck} satellite (\cite{PlanckCollaboration2011}). Most of the cluster
candidates detected by \textit{Planck} through their TSZEs correspond to
halos smaller than the \textit{Planck} beamwidth. However, ground-based 
observations are limited by strongly structured and powerful
atmospheric noise, and we expect it to be difficult to detect clusters
smaller than the beam size.

\subsubsection{The effects of halo LOS alignment }
A second effect caused (in part) by limited angular resolution is
confusion of halos by superposition on the line of sight. 

The additive nature of TSZEs from multiple halos on the line of sight 
is evident in figure~\ref{fig:CMBSZmaps}: the bottom-right panel
displays many examples of small angular size (distant) TSZEs
superposed on more extended (lower-redshift) clusters. 
Halo alignments lead to boosted TSZE flux density per beam, with the
flux density of maximum boosting being a function of the beam size and
the distribution of halos in redshift. 
Effectively, the TSZE flux density detected from a halo is increased
while the number of halos is unchanged.
Thus in figure~\ref{fig:TSZcounts} the green (dotted) lines that display
the effects of halo alignment lie above the red (dashed) lines that
show the effect of beam size selection of individual halos.
The two sets of curves converge at low flux densities because of the
halo filtering criterion.

Halo LOS alignments change the apparent count 
of detectable clusters in our statistics by an amount smaller than
sample variance for 22-
and 30-GHz surveys with either the 32-m or 100-m telescopes. However,
at the lower flux density threshold achievable with the 49-beam radio
camera on the 100-m telescope there is a significant impact, and a 
TSZE flux density bias 
should be 
taken into account when reconstructing scaling relations at the
faintest flux density levels, of order $100 \ \rm \mu Jy$ or less. 

\subsubsection{The effects of unresolved radio sources}
\label{sec:unresolved_radio_sources}
Unresolved radio sources appearing near the line of sight to a halo
will decrease the integrated TSZE flux density of that halo. This
effect is shown in figure~\ref{fig:TSZcounts} by the dash-dotted
(blue) lines. These lines result from integrating specific
intensity FOV  simulations that include TSZEs from all identified and
reconstructed halos (section~\ref{sec:simulation}) 
and from simulated radio sources brighter than $100\mu$Jy
(section.~\ref{sec:flux_density}), using the instrumental 
beams given in table~\ref{tab:RTspecs}. The statistics obtained from
different light-cone realisations account for TSZ--radio-source
cross-correlations, but in this calculation we ignore accidental
correlations with the primordial CMB.

Although massive halos tend to have more contaminating sources (see Section~\ref{sec:spatial_correlations}), 
a low redshift survey with a small beam makes it easier to clip out the affected source regions, 
and so reduce the confusion that would be obtained with a smaller telescope or lower frequency.

The strength of the clustering of radio sources towards massive halos is
key in quantifying the flux cancellation. In the present work we use 
a free parameter, calibrated by point-source/galaxy-cluster
associations. However, the source distribution is complicated, and the
description of the source properties that we use is a relatively
simple approximation (see section~\ref{sec:spatial_correlations}). While
the mean overdensity of radio sources towards known clusters can be
assessed from our cross-correlation analysis, 
the radial density profile, richness-mass relation, and radial
spectral variations in the source population should also be taken into
account. Our current approach weights the spatial radio source PDF by
cluster mass, but the masses of the clusters used in the 
cross-correlation analysis are not known. Instead we assign
2D PDF contributions based on the masses of halos that happen to be in
the simulated FOV. The contribution of clustered sources relative 
to the uniformly-distributed source component may be over- or
under-estimated depending on the extent to which the halos in the
simulated FOV correspond to the masses of the clusters  
used for the cross-correlation analysis. A more detailed analysis of
the clustering properties of radio sources will be possible as a
consequence of the surveys that we plan, and will allow better-tuned
modelling of the source contents of clusters. 
Although there is a possibility that, at the lowest radio-source
flux densities, a uniformly-distributed radio-source population with a
central void could mimic the TSZE, in the present work we do not seek
such configurations as we only measure TSZE flux density towards
known halos. These cases however, along with a possibility of spurious
detections due to presence of various noise components, will be
quantified when the final source extraction algorithms are
developed. However, this is beyond the scope of the present work.

Radio sources have most impact on TSZE flux density measurements at
lower frequencies, for a given radio telescope, due to the larger beam
as well as the brighter synchrotron radiation. At a fixed frequency,
the impact of radio sources can be mitigated only by observing with a
larger telescope.

\subsubsection{Low-mass halo contributions to the LOS integrated Compton $y$-parameter}
\label{app:lightHaloContrib}
Typically there are about $4\times 10^4$ halos (with at least
$N_h=600$ SPH particles per halo; see Section~\ref{sec:lss}) in the
simulated deep field. Such a value of $N_h$ 
could lead to an
underestimation of the calculated TSZE flux density distribution
because the lighter halos are neglected, but are abundant because of
the steepness of the halo mass function, and so could contribute
significantly to the sky distribution of the Compton $y$ parameter.

We have investigated the LOS contribution of halos up to a factor $10$ in
mass below our normal limit by decreasing $N_h$ to $60$ (corresponding
to mass scale $M_{h,\rm min} \approx 1.3 \cdot 10^{12} M_\odot
h^{-1}$, or about the mass of a large spiral galaxy), 
recalculating the Compton $y$-parameter map, and then examining the strongest
peaks in difference from our basic map. 
The reduction in $N_h$ raises the
typical number of halos to more than $6\times 10^5$, but the
$y$-parameter is increased by few per-cent at the most,
towards the halos selected using $N_h\geq 600$ condition.

We conclude that the LOS
contribution of light halos is statistically unimportant and typically $\ll 10\%$. 
The underlying reason stems from chances of LOS alignment and also from the fact that
these lighter halos have lower temperatures and densities, and
smaller physical sizes, than more massive halos. All three factors
work in the same sense, to reduce the TSZE from a halo, and the
reduction in TSZE per halo is more significant than the increase in
halo numbers at low masses. 

Since $N_h\geq 60$ halos constitute a superset of $N_h\geq 600$ halos there exist
multiple Compton $y$-parameter peaks from halos just below the $N_h=600$ threshold
which build structures distributed around heavier halos. Depending on the halo redshift and
telescope beamwidth these may also occasionally contribute at the level of up to several per-cent.

\subsubsection{Predictions for a wide field survey and impact of point source clustering}
\label{sec:PIsteradiansFieldPrediction}
It is difficult to infer directly the number of detectable TSZE galaxy
clusters in the {\PIsteradians} field from the simulated deep field,
because of high Poisson noise on the abundance of the most massive
halos. However, figure~\ref{fig:SZthetaSz} (bottom-right panel) shows that for the
$3\,\times$ RMS flux density threshold (0.79 mJy) 
derived for the RTH's {\PIsteradians} survey (table~\ref{tab:surveys}),
a detectable TSZE signal will arise from a cluster mass $\sim
2.2 \times 10^{15}M_\odot/h$. Assuming that only the nearby (low
redshift) galaxy clusters will be massive enough to yield that
requirement, and using the reconstructed FOF mass function depicted in
the figure~\ref{fig:massFn} we estimate that there are 
$17^{+15}_{-8}$ $\mathrm{Gpc^{-3}}$ (68\% CL)
galaxy clusters within the FOF mass range  $(1.8, 3.2) \times 10^{15}$
$\mathrm{M_\odot}/h$. 
For the cone-like {\PIsteradians} field survey with the assumed
opening angle of 120 deg, and depth reaching redshift $z=1.0$, the
corresponding comoving volume $V=39.2$ Gpc$^3$ by a simple rescaling
would contain $N_{\PIsteradiansStr}=676^{+612}_{-321}$ clusters (68\%
CL).\footnote{Higher-redshift clusters are not accounted for, because
the modelling would need to include the mass-function
redshift-dependence.}  Thus, given the mass-TSZE flux-density relation
(figure~\ref{fig:SZthetaSz}) it should be expected that deepening this
survey in a multi-year campaign would quickly increase the detectable
cluster's count.
This result, however, still needs to be corrected for the
presence of point sources. Accurate predictions for the wide and shallow
survey are difficult, as the degree of clustering has a strong effect,
and Figure~\ref{fig:TSZcounts} shows that, at 15-GHz, point source
confusion is increasingly significant at higher halo masses 
(higher flux-densities), 
and that point sources effectively decrease the halo counts by a
factor of $> 10$ at 0.1 mJy 
for the assumed clustering level. 
We recalculated the statistic for the case of non-clustered
sources and found that the effective TSZE counts are less suppressed,
as expected. 
In our simulations the clustering is frequency independent, and the
impact on the counts results only from frequency-dependent
beamwidths. The suppression in the total TSZEs counts due to switching
from the non-clustered to the clustered case is largest 
at low frequencies giving a ratio of $\sim 10$ ($\sim 1.6$) at 15 GHz
for RT32 (RTH). At 30~GHz the overall impact on the count ratio is
small: $<1.2$ for both RT32 and RTH. 
However, the angular clustering PDF peak amplitudes are proportional
to halo masses and so perhaps a more significant impact of clustering
is seen at the highest masses --- which is the most interesting from
the observational point of view. In the high mass limit, and in
the non-clustered case, the effective cluster count curves for RTH 
approach the green (dotted) lines in Fig.~\ref{fig:TSZcounts} -- the counts without the effects of point sources
for all frequencies. RTH at 30 GHz would be, in principle, immune to
variations of point source clustering.

Details of the point source clustering therefore do have impact
on detectable cluster counts predictions in the {\PIsteradians}
survey. We estimate the effective cluster count 
as $N_{\PIsteradiansStr}*f$ up to  a multiplicative factor $f$.
$f$ ranges from  $\sim 1$ in the non-clustered case
to $<0.1$ for the case of clustering assumed in our simulations 
hence, the resulting effective cluster count will be of order $O(10)$.

Less confusion is observed at higher frequencies (smaller beamwidths),
but the penalty is a decreased survey depth (for a duration-fixed survey). 
A 4-year {\PIsteradians} survey would have TSZE
flux density limit 1.8 mJy ($3\times$ RMS) if the 49-beam camera of RTH was
operating at 30-GHz. As pointed out in section~\ref{sec:mockMaps} the
integrated TSZE flux density changes as square of beamwidth which, in
turn, shifts the $N(S_{\mathrm{min}})$ distribution towards smaller flux
densities and effectively increases the cluster mass required for
successful detection to $\sim 10^{16}M_\odot/h$ for a 30-GHz
{\PIsteradians} survey with RTH. Since galaxy clusters generally have 
mass less than $10^{16}M_\odot/h$,it is clear that large apertures
surveying fields that are very wide and very shallow and at an
increased frequency will not guarantee many detections. This
makes SZ observations challenging, since at larger beamwidths the only
way to effectively detect TSZE is to either use very large cameras
and/or have a better receiver performance and stable atmospheric
conditions.

\subsubsection{Frequency and receiver-telescope dependence}
\label{sec:freqRTrec}
\begin{figure}[!t]
\centering
\includegraphics[width=0.75\textwidth]{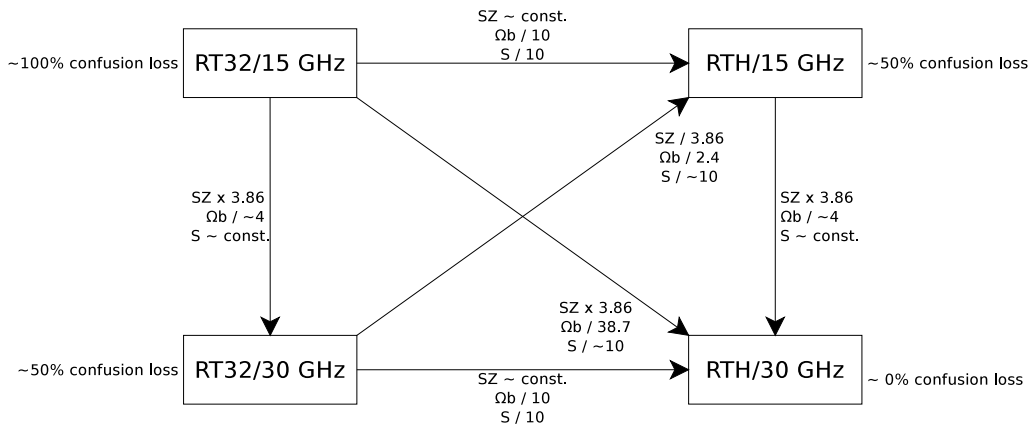}
\caption{A diagrammatic and simplified representation of
approximate relations between the TSZE intensity (marked as SZ),
telescope beam size $\Omega_b$ and TSZE flux per beam (S) under
changes of observing frequency and telescope aperture. Within the
same frequency the amplitude of TSZE remains constant. The '$\times
f$' ('$/f$') indicates an increase (drop) of a quantity by a factor
$f$ under the configuration change indicated by the corresponding
arrow. The diagram explains the mutual relations of
$N(S_{\mathrm{min}})$ distributions in Figure.~\ref{fig:TSZcounts}.
For each telescope--observing-frequency combination an approximate
fraction of TSZEs lost to radio-source contamination is also
indicated.}
\label{fig:SZfluxDiag}
\end{figure}

In Table~\ref{tab:surveys} we report that at 15 GHz RTH will
statistically detect $4^{+2}_{-2}$ clusters per year in a
{\onedegfield} above $3\times \mathrm{RMS}=6\mu$
Jy. Figure~\ref{fig:SZfluxDiag} summarises the flux density relations
under change of observational frequency and telescope-receiver
configurations. In this section we investigate the case of 30-GHz (no
beam-switched) 49-beam receiver with the same receiver noise level and bandwidth
as OCRA-f if installed on RT32. The TSZE
survey limit after 1-year (4-year) of operation would be $\sim 0.1$
mJy (0.05 mJy) and within the {\onedegfield} survey it would effectively
detect 0.9 ($\sim 2.5$) clusters per square degree.  Within this
regime this shows that it is slightly more effective to keep constant
survey depth and increase survey area after reaching the initial depth.  
The effective cluster detection
rate would be then  $\sim 4.2$ times lower than in the case of 15-GHz camera
on RTH, the TSZEs loss due to point sources would be similar, but the
survey would find much stronger TSZE flux densities -- tens-to-hundreds
of times stronger -- which could turn out to be more feasible than the low
frequency survey.

If the RTH camera were to operate at the higher frequency of 30 GHz it
would provide a better signal/noise ratio and would detect stronger
TSZEs with superior angular resolution. As pointed out in earlier
sections, such a configuration results in a TSZE distribution shifted
towards smaller flux densities per beam, but the advantages of a
larger beam in terms of integrated signal should be recovered with
data smoothing, and point source excision, during data analysis.

\subsection{Predictions for blind point source surveys}
\label{sec:PtSrcSurveys}

\begin{figure}[!t]
\centering
\includegraphics[width=\textwidth]{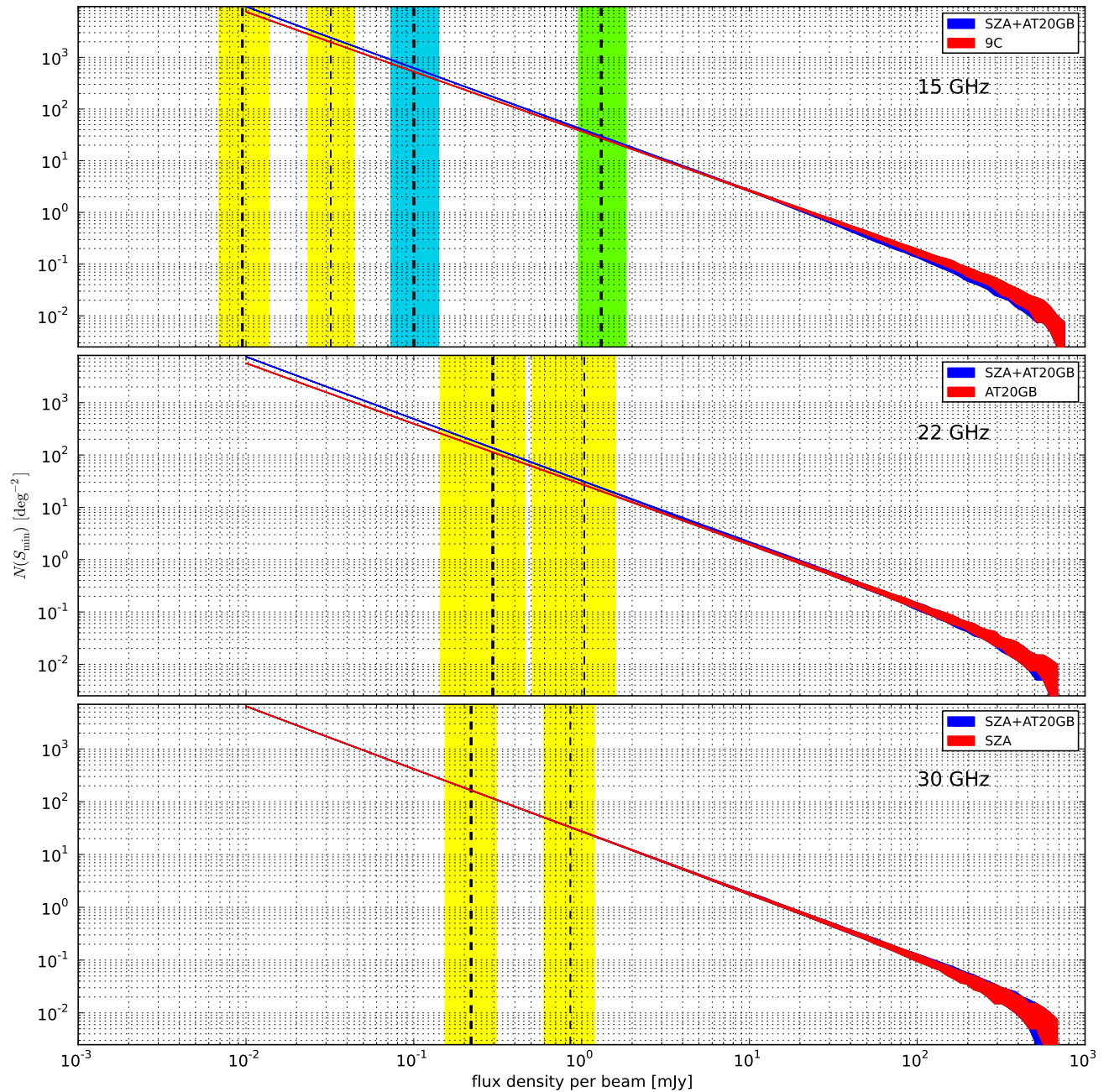}
\caption{Predictions for the total count of radio sources,
$N(S_{\mathrm{min}})$, above flux density threshold
$S_{\mathrm{min}}$ (eq.~\ref{eq:NSmin}). 
The three panels correspond to frequencies $15$, $22$, and $30$
GHz. In each panel thick vertical lines relate to surveys with the
100-m RTH telescope and thin lines to surveys with the 32-m
telescope. 
The shaded regions represent 68\% confidence limits based on 100
Monte-Carlo realisations of a large sky area. 
For each frequency we show two sets of predictions: one based on a
combination of the SZA and AT20B catalogues, and one based on the
single catalogue that most closely matches the planned survey
frequency. 
Vertical dashed lines indicate the $5\,\times$ RMS flux density
level expected for the {\onedegfield} surveys. Yellow shaded regions
show the variation in the $5\sigma$ confidence range corresponding
seasonal variations of the system performance. 
Flux density thresholds for the {\PIsteradians} and {\RTHdeepfield} surveys,
as defined in table~\ref{tab:surveys}, are shown by the green and
cyan shaded regions in the top panel.
}
\label{fig:ptSrcCounts}
\end{figure}

In this section we calculate the total number of radio sources that we
expect to detect in blind surveys to flux density threshold $S_{\mathrm{min}}$.
Our results are summarised in figure~\ref{fig:ptSrcCounts}, where the
$N(S_{\mathrm{min}})$ distributions are shown for each of the surveys
described in table~\ref{tab:surveys}.

At each of the survey frequencies we compare the predicted source
counts based on two different datasets. The first prediction is
derived from 
the SZA and AT20GB surveys, combined and processed as described in
section~\ref{sec:flux_density}. For the second we choose the survey
performed at the frequency closest to the frequency of interest. 
In either case the predictions were made at the relevant survey
frequency. Figure~\ref{fig:ptSrcCounts} shows that the
different predictions are in good agreement, except at low flux
densities, below the range probed by the 9C and AT20GB surveys.
One of the aim of the planned surveys is to provide a
large sample of sources in this sub-mJy flux density range, where the
differences in the predicted source counts reflect uncertainties in
the contributing source populations.

\subsubsection{The effects of point source confusion }
The deepest surveys with a given instrumental configuration are
limited by confusion.
The confusion limit is usually discussed based on some limiting number
of sources per beam: at some assumed source density, such as 
three beams per source, the corresponding flux density
is taken as the effective confusion level. This approach ignores the
spatial correlations of radio sources and
the contribution of sources below the adopted flux density limit. Both
shortcomings underestimate the level of confusion.

\begin{figure}[!t]
\centering
\includegraphics[width=\textwidth]{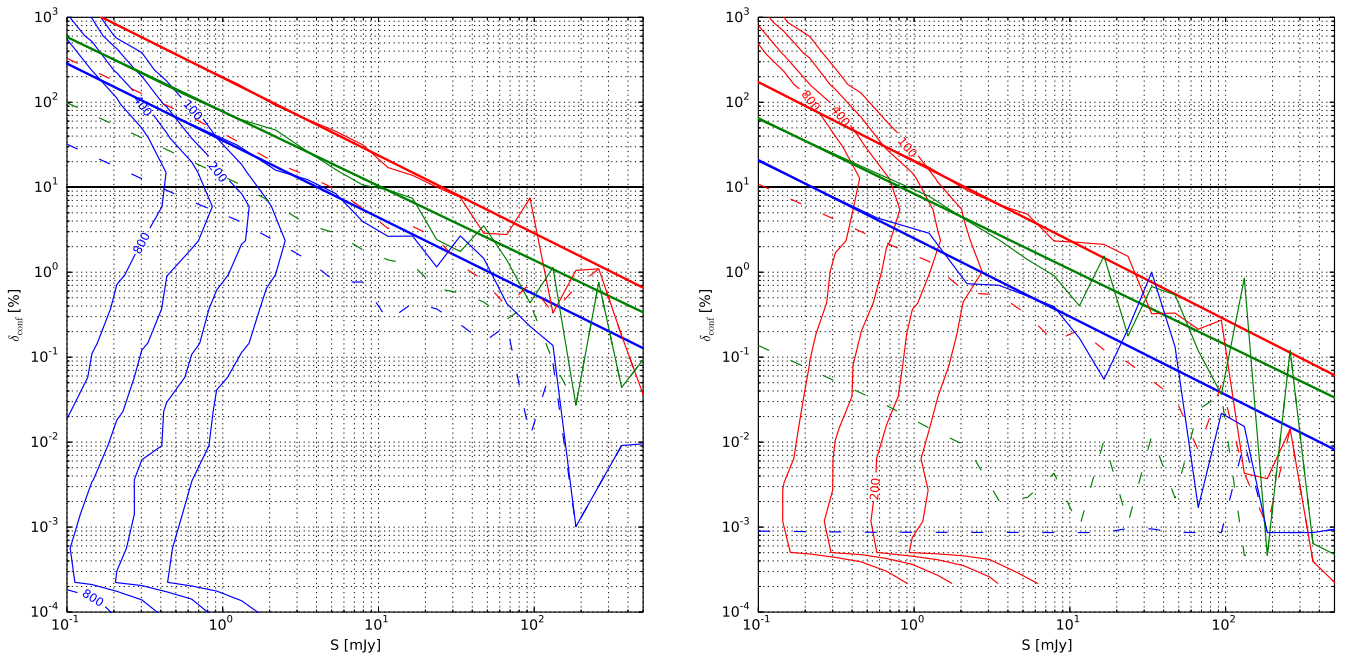}
\caption{Radio source flux density confusion for RT32 (left) and RTH (right). 
Solid lines and dashed lines connect the 95\% and 68\% upper-tail
percentiles, respectively, of the $\delta_{\mathrm{conf}}$
distributions. 
Linear fits (in log-log space) to the 95\% distributions are shown:
these approximately follow power-laws of the form $\delta_{\rm conf}
\propto S_g^{-0.9}$. From top to bottom, the red, green, blue lines
correspond to frequencies of 15, 22, and 30~GHz.
The intersection of the horizontal line at
$\delta_{\mathrm{conf}}=10$\% with the fitted lines defines the flux
density at which less than 5\% of sources in the field  have a
systematic flux density measurement error larger than 10\%. We use
this as our definition of the confusion level, and summarise the
values for the different telescopes and operating frequencies in
table~\ref{tab:surveys}.
The contours show the densities of radio sources in the $\delta_{\mathrm{conf}}-S$ plane (in the same, but
arbitrary, units in both plots) for the RT32 at 30'~GHz (left) and
for the RTH at 15~GHz (right). }
\label{fig:ptSrcFluxConfusion}
\end{figure}

In this work we derive the confusion limits in an alternative way that
alleviates these problems. We calculate the confusion limit via
the relative flux density measurement error,
$\delta_{\mathrm{conf}}$, as
\begin{equation}
\delta_{\mathrm{conf}}(\nu) = \frac{S_g - S_m}{S_g}
\label{eq:radioSrcConfusionDelta}
\end{equation}
where $S_g$ is the flux density generated according to the recipe
outlined in Sect.~\ref{sec:flux_density}, and $S_m$ is the flux
density measured from the high-resolution map. $S_m$ therefore
includes the effect of confusion. 
In figure~\ref{fig:ptSrcFluxConfusion} we show
$\delta_{\mathrm{conf}}(\nu)$ for the two radio telescopes for all
three frequencies as a function of $S_g$.

Dashed and solid lines show
the upper 68\% (95\%) percentile points for the distribution of 
$\delta_{\mathrm{conf}}$ --- we base the confusion level as
given in table~\ref{tab:surveys} on the flux density at which the 68\% (95\%)
percentile contour crosses the line
$\delta_{\mathrm{conf}} = 10$\%. That is, at the confusion levels
quoted in the table, 32\% (5\%) of the flux densities will have errors $>10$\%.

In figure~\ref{fig:ptSrcFluxConfusion} we also plot contours of
constant point source density distribution in the
$\log\delta_{\mathrm{conf}}-\log S$ plane in order to visualise the
effects of confusion when approaching the low flux density levels.  In
case of no confusion the sources should group horizontally along the
$\delta=0$ line. In case of non-zero confusion the sources will occupy
$\delta>0$ regions with the amplitude of the deviation depending on
the initial flux-density and clustering properties.

\section{Discussion}
\label{sec:discussion}

Our simulations have neglected AGN and star
formation feedback and radiative losses, so that the M-T scaling 
relation (figure~\ref{fig:MT}) lies below the scaling relations from
observations or simulations that include more physics. Nevertheless,
the predictions for the detected TSZE galaxy cluster count do not
represent lower limits, but quite the opposite, as can be seen from 
the M-Y scaling relation (figure~\ref{fig:MY}).

We speculate that one of possible sources of scatter in the scaling 
relations is the assumption that our estimates for $r_{\delta}$, which
are well-defined for spherically-symmetric clusters, are only
approximate indications of mass in the typically-asymmetrical density
distributions of realistic halos at the usual values of $\delta$ (see
figures~\ref{fig:halo34582} and~\ref{fig:halo10801}). 
The implicit uncertainty propagates to quantities that depend on
overdensity which are extracted from the reconstructed temperature and
mass profiles. If a galaxy cluster is going through a merger event,
then its temperature and mass profiles may well be irregular, and we
should not even expect the
density profile to fall monotonically with distance from the cluster
centre. 
We mitigate against these problems somewhat, by fitting scaling
relations using only the most massive halos, which should have more
relaxed and well-defined density profiles.

The relative errors of the SPH interpolations, which we use heavily to
construct the 3D $T_\rho$ distributions, is of the order of 1\% in the
signal dominated regions and depends on the chosen number of SPH
neighbours. Although for the main results we use a fixed number of
neighbours, $N_{\mathrm{neigh}}=33$, we might be able to improve the
physical accuracy of the simulation by relating this number to 
the halo mass, and so to the number of SPH particles contained in the
halo. We have not investigated the impact of changing from a fixed
number of neighbours. For well-mixed CDM and baryon fluids we would
expect the effect to be unimportant, but some differences may arise in
cases where baryons are spatially separated from the CDM component,
such as when baryonic gas is concentrated towards the centre of an 
extended CDM halo. In that case one would expect larger smoothing
lengths in the central regions for a fixed $N_{\mathrm{neigh}}$ as
compared to the smoothing lengths calculated for a fixed mass enclosed
within the smoothing length.

The simulation mass resolution affects the amplitude of the $M$-$Y$
scaling relation (figure~\ref{fig:MY}), and a simulation with a larger
number of particles would result in better estimates of 
halo quantities at high overdensity, where most of the TSZE signal
arises. While our scaling relations show that the process of computing
Compton $y$-parameter profiles is relatively unaffected, we found
that the $M_{500}-Y_{500}$ scaling relation amplitude is sensitive to
the simulation mass resolution. Convergence at high density thresholds
requires better mass resolution than we were able to undertake, but we
quantified the effect by a sequence of simulations with $(N,L
[\mathrm{Mpc}])=\{(256^3,512), (512^3,512), (256^3,128)\}$, where $N$
is the number of particles and $L$ simulation box size. We found
that the $M_{500}-Y_{500}$ scaling amplitude decreases, from
$Y_{14}=6.62$ to $6.03$ and then $Y_{14}=5.61$ as the resolution
increases, for a fixed number of neighbours.

Our $M$-$T$ relation (figure~\ref{fig:MT}) shows that adiabatic
simulations yield the scaling relations amplitude, $T_{15}$, offset by
a few tens of percent from the observational amplitude. The value 
of amplitude depends on the smoothing length parameter used, or
equivalently to the number of neighbours used in the SPH density
calculations. For a fixed mass resolution, larger smoothing lengths
tend to increase the amplitudes of the $M$-$Y$ and $M$-$T$ scaling
relations, but we found that increasing $N_{\mathrm{neigh}}$ cannot
correct for this $\sim 30$ \% discrepancy. Increasing
$N_{\mathrm{neigh}}$ from 33 to 66 with $N_h = 600$ increases the
value of $T_{15}$ by only about 3\%. Decreasing $N_{\mathrm{neigh}}$
has a larger effect and exacerbates the discrepancy with observations.
Future improvements will investigate changes in the smoothing scheme
and how they might improve our results. 
Similarly, we showed that our results have converged well: doubling the grid resolution
(from the 50 kpc value assumed here) has negligible effect, of order 
1\% in the $M$-$Y$ scaling relation. Extra physics, describing cooling
and heating, is needed to reproduce the observed scaling relation more
closely. We therefore consider our predictions for TSZE
detections to be upper limits.

The most massive low-redshift clusters in the simulations have central
Compton $y$-parameters of order $10^{-4}$, which corresponds to 
TSZE decrements of order $\mathrm{mK}$ in the Raleigh-Jeans regime.
There are at few such clusters found in any $5.2 \times 5.2 \ \rm
deg^2$ simulation (figure~\ref{fig:CMBSZmaps}). Thus tracing the
distribution of $y$ in the cosmic population of halos requires wide
surveys, and hence large radio cameras if the survey is to be
conducted in a modest time. TSZE observations are therefore
challenging, and are further limited, for a blind survey at a single
frequency, by the lack of filtering methods that could reduce the
effect of confusion from primary CMB fluctuations
(figure~\ref{fig:SZthetaSz}) and, more importantly, from radio
sources. Once again, figure~\ref{fig:SZthetaSz} emphasises that wide
fields of view are required to find significant numbers of clusters
with strong TSZEs, reaching mJy flux densities at the survey
frequencies.

The vertical lines in the figure~\ref{fig:TSZcounts} do not
monotonically increase towards higher flux density levels with the
frequency. Rather, the worst sensitivity is seen at about 22~GHz. This
results from the strength of the atmospheric water vapour feature that
peaks at about 22~GHz and for the small number of beams available in
the receiver. The substantially improved sensitivity for the
15-GHz surveys stems mainly from the large number of channels in the
RTH camera.

Given the scaling relations exhibited by our mock maps, as
derived in sections~\ref{sec:MT} and~\ref{sec:MY}, it could be
possible to roughly correct the simulations for missing or imprecise 
physics by multiplying the integrated Compton $y$-parameters by a 
halo-mass dependent factor that relates the
temperatures and $Y^{\mathrm{INT}}$ from our simulations to those
containing more gas physics and assumptions about heating mechanisms. 
A simple scaling, however, would not suffice as non-adiabatic
processes alter both the amplitudes and the shapes of the
temperature and Compton $y$-parameter profiles. In the present paper
we choose not to attempt the correction, and accept systematic errors
in our predictions that we expect to be of order 10~per~cent
(sections~\ref{sec:MT} and~\ref{sec:MY}). 

As a result of the planned, tiered RTH surveys,
an interesting clue on 
AGN feedback activity in the most massive galaxies, 
and its relation to 
the state of the intra-cluster medium,
should partially come from measurements of redshift and flux-density
distributions of radio galaxies and radio-loud quasars
at sub-mJy flux-density levels
where spatial comoving number density decline is expected 
at $z\gtrsim 2$. Such a high-frequency radio survey 
will complement the low-frequency counts,
will help to quantify the flux-density and redshift dependent ratios between 
flat, steep and inverted spectrum populations, will track contributions
of starburst galaxies and will provide reliable flux-density   
extrapolations into mm-wavelengths 
\citep[][and references therein]{deZotti2010}.

According to recent {\it Planck} results, 
the low-redshift cluster counts data favour a 
lower normalisation of the matter power spectrum than the
high-redshift CMB(+lensing) data \citep{PlanckCollaboration2014,PlanckCollaboration2014a}. 
This tension can be reconciled by 
increasing the mass of active neutrinos or by introducing a massive sterile neutrino
that would suppress the small scale fluctuations with respect to the
standard $\Lambda$CDM model \citep{Wyman2014,PlanckCollaboration2014}.
The one-dimensional constraint on $\sigma_8=0.77\pm0.02$ reported
in \cite{PlanckCollaboration2014} from {\it Planck} SZ+BAO+BBN
can also be reconciled with the CMB data by introducing mass estimate
biases of about 40\% in the reconstructed Y-M scaling relation. 
Interestingly, \cite{Evrard2008} reports that 
high normalisation values -- that are now more compatible with the
current CMB data -- are also preferred based on analysis of the
$\Lambda$CDM model 
halo space density as a function of dark matter velocity dispersion.
In this context our chosen value could be a reasonable assumption. 
The currently available CMB data
combined with external priors yield 
maximum likelihood $\sigma_8$ estimates that can
differ between each other by as much as 
$\Delta\sigma_8\approx 0.1$ \citep{Hou2014,Hinshaw2013} 
in cosmologies with massive neutrinos.
Since for our main results we chose $\sigma_8=0.821$
(section~\ref{sec:simulation}) we test the impact of using a different
value from the one preferred by the CMB data in the concordance
$\Lambda$CDM model. We test other observationally motivated
normalisations using constrained,
lower mass resolution simulations and PS mass functions (figure~\ref{fig:massFn}).
Using PS mass function predictions for $\sigma_8=0.77$ ($0.90$) 
we estimate that the spatial halo density would be
decreased (increased) by a factor of about $\sim2.0$ ($\sim2.6$) 
at the cluster-size mass scale with respect to the abundances 
used in our analyses.
Using reconstructed mass functions from
constrained simulations we find that the impact is somewhat smaller (due
to the smaller steepness).
In particular, we estimate that a decrease (increase) of $\sigma_8$ to the value 0.77 (0.90) would result
in decreasing (increasing) the halo abundances by a factor of $\sim1.6$ ($\sim2.1$)
for the cluster-mass halos ($\sim 1.3 \times 10^{15} M_\odot/h$).
The difference due to Planck CMB preferred normalisations is insignificant,
but clearly the overall variation of the halo abundance is a quite 
rapid function of the chosen normalisation in this mass regime, 
thus making halo counts an interesting cosmological test.
Further progress in the field will certainly reveal
the nature of this interesting dichotomy in the observed normalisation
values favoured by different data, be it due to cluster mass biases, 
impact of massive neutrinos or yet another process.

\section{Conclusions}
\label{sec:conclusions}

We have used adiabatic hydrodynamical simulations to predict the
number of galaxy clusters detectable via the thermal Sunyaev-Zeldovich
effect (TSZE) in a set of blind simulated surveys to be performed with
the existing 8-beam, beam-switched, OCRA-f radiometer on the 32-m
radio telescope in Toru{\'n} (Poland), and with the planned 100-m
radio telescope Hevelius equipped with a 49-beam focal plane radio
camera. Based on existing radio source surveys, we have also
calculated the number of radio sources detectable in these surveys.

We analysed the effects of halo--halo and halo--radio-source
alignments on the integrated flux density and count of detected TSZEs,
and calculated the redshift distribution of the detectable clusters,
in the presence of radio source clustering in the halos.
By cross-correlating the NVSS survey with the {\it Planck} SZ cluster
candidate sample we found that statistically there is roughly a
ten-fold increase in the point source density towards massive clusters
as compared with directions away from the cluster centres, 
although the value is a function of survey depth, cluster mass, and
redshift range.

We visualised the improvements in the receiver sensitivity, detections
count and available survey area resulting from using a large dish, and
large radio cameras when carrying out blind surveys, by 
analysing a number of receiver-antenna combinations for frequencies
from 15~to 30~GHz, taking account of the weather effects.

We used a set of high-resolution mock maps of the signals sought (section~\ref{sec:mockMaps})
including (i) the CMB primary fluctuations; (ii) TSZEs; and (iii) the radio source population. 
In future these maps will be used to test the image reconstruction 
procedures from simulated observations, taking account of realistic
atmospheric brightness instabilities and receiver performance. This
process will also generate completeness functions for the
interpretation of the survey results.

Based on our simulations, we showed RT32/OCRA-f will be capable of detecting
$33^{+17}_{-11}$ radio sources above $0.87$ $\mathrm{mJy}$ at $5\sigma$
CL at 30 GHz in a square degree field every year, taking account of
the Polish climate. The relative flux density measurement error
due to background point sources is constrained to be at the level
$\lesssim 8\% (66\%)$ in 68\% (95\%) of the detected radio sources. 
It is unlikely that such a survey will detect any galaxy clusters,
even at $3\sigma$ CL, since the expected source density is
$0.03^{+0.02}_{-0.01}$ $\mathrm{deg}^{-2}$. The main limitations of
the survey arise from observing time constraints, the small number of
feeds, and confusion from radio sources. Reducing the detection
confidence to $2\sigma$ would result only in a modest increase of the
number of detectable clusters, to $\approx 0.07$ $\mathrm{deg}^{-2}$.
Therefore OCRA-f (apart from surveys for radio sources) is best-suited
to small-field mapping of candidate galaxy clusters from X-ray
surveys, optical/IR surveys, or \textit{Planck} selections.

On the other hand, the planned 100-m radio telescope Hevelius (RTH)
equipped with (initially) 49 beams and wide bandwidth horns and
operating up to $\sim 22$ GHz, should be suitable for
blind TSZE surveys centred around 15~GHz. 
Given the spectrum of TSZEs, this particular frequency is a compromise due to
climate conditions at
RTH's location  (giving a high frequency 
constraint) and radio frequency interference (giving a low
frequency constraint). Taking account of
confusion from radio sources, and ignoring systematical effects,
we would expect RTH to detect $<4$ galaxy clusters per year 
in a square-degree field at $3\sigma$ CL above $6$ $\mathrm{\mu Jy}$
TSZE flux density with a field coverage of $1.5^2$ beam per pixel,  
and $<1.5$  clusters per year (at $3\sigma$ CL) in the
$\approx 60$ $\mathrm{deg}^2$ field, above TSZE flux density of $60$
$\mathrm{\mu Jy}$ with field coverage of $2^2$ beam per pixel
(table~\ref{tab:surveys}). The reason for such a low detection
rate stems from radio source emission which dominates at low frequencies
and even though the beam size of a 100-m telescope at 15 GHz is
smaller than that of 32-m telescope at 30 GHz, in the latter case the
confusion is significantly smaller while TSZE is larger. Therefore
although increasing observational frequency would result in fewer
counts per square degree for a fixed duration survey, it would select
stronger TSZEs in the range of hundreds of $\mu$Jy where flux
measurements may be less prone to atmospheric (and gain instability)
systematic effects which we haven't included in our analysis. We
also find that the angular resolution of RTH operating at 30-GHz would
in principle make TSZEs detection counts statistics insensitive to
point source clustering properties within the limits constrained by
observations.  The $\approx 60$ $\mathrm{deg}^2$ survey would yield
nearly $34\,000$ point sources brighter than $1$ $\mathrm{mJy}$ at
$5\sigma$ CL, with a relative systematic flux density error due to
source confusion at the level $\lesssim 2\% (20\%)$ in 68\% (95\%) of
the detections.

A primary goal of the planned RTH will be a new, wide-area, radio source survey.
Within this survey RTH will detect nearly $3 \times 10^5$ radio sources
at $5\sigma$ CL at 15~GHz in a $\pi$-steradians ({\PIsteradians})
field
with flux density $> 1.3$ $\mathrm{mJy}$  in a one year of
non-continuous mapping (with a realistic time efficiency factor) and
with the relative systematic error on the flux density measurement due
to confusion with unresolved radio sources  at the level $\lesssim 1.5\%
(16\%)$ in 68\% (95\%) of cases of the detected radio sources
(section~\ref{sec:instruments} and~\ref{sec:PtSrcSurveys}). 
We estimate that with the {\PIsteradians} survey 
effectively tens of 
galaxy clusters within the mass range  $(1.8, 3.2)
\times 10^{15}$ $\mathrm{M_\odot}/h$ should be detected with $z \lesssim 1$
however the prediction depends on point source clustering properties.
Deepening this survey in a multi-year campaign would quickly increase
the detectable clusters count (potentially to hundreds of clusters)
even if operating at relatively low frequencies around
$15$ GHz.

\section*{Acknowledgements}
We thank the referee for useful comments that helped improving the original version of the manuscript.
BL would like to thank Gieniu Pazderski for useful discussions on technical aspects 
of the RT32 and RTH receivers. Thank you to Boud Roukema for reading the manuscript and providing
comments.

We acknowledge use of the FOF halo finder developed by the NASA HPCC ESS group at the University of Washington.
Use of a web service of the Astrophysics Science Division at NASA/GSFC and the High Energy Astrophysics Division of the Smithsonian Astrophysical Observatory (SAO) ({\protect\url{http://heasarc.gsfc.nasa.gov/docs/archive.html}}) is also acknowledged.
Use was made of the ``matplotlib'' plotting library \citep{Hunter2007} distributed on the GPL-compatible
Python Software Foundation license.

This work was financially supported by the Polish National Science Centre 
through grant DEC-2011/03/D/ST9/03373.
A part of this project has made use of ``Program Oblicze{\'n} WIElkich Wyzwa{\'n} nauki i techniki'' 
(POWIEW) computational resources (grant 87) at the Pozna{\'n}
Supercomputing and Networking Center (PSNC).
A part of this project also benefited from the 
EC RadioNet FP7 Joint Research Activity ``APRICOT''
(All Purpose Radio Imaging Cameras On Telescopes).

\bibliography{bibliography} 
\bibliographystyle{aaeprint}

\end{document}